\def\isarxiv{1} 

\ifdefined\isarxiv
\documentclass[11pt]{article}

\usepackage[numbers]{natbib}

\else
\documentclass{article}
\usepackage[final]{neurips_2024}
\fi

\usepackage{amsmath}
\usepackage{amsthm} 
\usepackage{amssymb}
\usepackage{algorithm}
\usepackage{enumitem}
\usepackage{subfig}
\usepackage{algpseudocode}
\usepackage{graphicx}
\usepackage{comment}
\usepackage{grffile}
\usepackage{wrapfig}
\usepackage{epsfig}
\usepackage{url}
\usepackage{xcolor}
\usepackage{epstopdf}

\usepackage{bbm}
\usepackage{dsfont}

\allowdisplaybreaks

\ifdefined\isarxiv

\usepackage{tikz}
\usepackage{hyperref}  
\hypersetup{colorlinks=true,citecolor=blue,linkcolor=blue} 
\usetikzlibrary{arrows}
\usepackage[margin=1in]{geometry}
\renewcommand{\cite}{\citep}
\else

\usepackage[utf8]{inputenc} 
\usepackage[T1]{fontenc}    
\usepackage{hyperref}       
\usepackage{booktabs}       
\usepackage{amsfonts}       
\usepackage{nicefrac}       
\usepackage{microtype}      

\definecolor{mydarkblue}{rgb}{0,0.08,0.45}
\hypersetup{colorlinks=true, citecolor=mydarkblue,linkcolor=mydarkblue}

\fi

\newtheorem{theorem}{Theorem}[section]
\newtheorem{lemma}[theorem]{Lemma}
\newtheorem{definition}[theorem]{Definition}

\newtheorem{assumption}[theorem]{Assumption}

\newtheorem{fact}[theorem]{Fact}
\newtheorem{remark}[theorem]{Remark}

\newcommand{\wh}{\widehat}
\newcommand{\wt}{\widetilde}
\newcommand{\ov}{\overline}
\newcommand{\N}{\mathcal{N}}
\newcommand{\K}{\ensuremath{\mathcal K}}
\newcommand{\R}{\mathbb{R}}

\renewcommand{\d}{\partial}

\renewcommand{\tilde}{\wt}
\renewcommand{\hat}{\wh}

\newcommand{\Tmat}{{\cal T}_{\mathrm{mat}}}

\DeclareMathOperator*{\E}{{\mathbb{E}}}

\DeclareMathOperator{\poly}{poly}

\DeclareMathOperator{\nnz}{nnz}

\DeclareMathOperator{\diag}{diag}

\DeclareMathOperator{\vect}{vec}
\DeclareMathOperator{\tr}{tr}

\DeclareMathOperator{\TV}{TV}

\makeatletter
\newcommand*{\RN}[1]{\expandafter\@slowromancap\romannumeral #1@}
\makeatother
\newcommand{\Zhao}[1]{{\color{red}[Zhao: #1]}}
\newcommand{\Lichen}[1]{{\color{purple}[Lichen: #1]}}

\usepackage{lineno}

\begin{document}

\ifdefined\isarxiv

\date{}

\title{Log-concave Sampling from a Convex Body with a Barrier: a Robust and Unified Dikin Walk\footnote{A preliminary version of this paper appeared in the proceedings of 38th Conference on Neural Informal Processing
Systems (NeurIPS 2024).}}
\author{
Yuzhou Gu\thanks{\texttt{yuzhougu@nyu.edu}. New York University.}
\and 
Nikki Lijing Kuang\thanks{\texttt{l1kuang@ucsd.edu}. University of California, San Diego.}
\and 
Yi-An Ma\thanks{\texttt{yianma@ucsd.edu}. University of California, San Diego.}
\and 
Zhao Song\thanks{\texttt{magic.linuxkde@gmail.com}. Simons Institute for the Theory of Computing, UC Berkeley.}
\and 
Lichen Zhang\thanks{\texttt{lichenz@mit.edu}. Massachusetts Institute of Technology.}
}

\else

\title{Log-concave Sampling from a Convex Body with a Barrier: a Robust and Unified Dikin Walk}
\author{Yuzhou Gu \\
New York University \\
\texttt{yuzhougu@nyu.edu} \And
Nikki Lijing Kuang \\
University of California, San Diego \\
\texttt{l1kuang@ucsd.edu} \And
Yi-An Ma \\
University of California, San Diego \\
\texttt{yianma@ucsd.edu} \And
Zhao Song \\
Simons Institute for the Theory of Computing, UC Berkeley\\
\texttt{magic.linuxkde@gmail.com} \And 
Lichen Zhang \\
MIT CSAIL \\
\texttt{lichenz@csail.mit.edu}
}
\maketitle 

\fi

\ifdefined\isarxiv
\begin{titlepage}
  \maketitle
  \begin{abstract}

We consider the problem of sampling from a $d$-dimensional log-concave distribution $\pi(\theta) \propto \exp(-f(\theta))$ for $L$-Lipschitz $f$, constrained to a convex body with an efficiently computable self-concordant barrier function, contained in a ball of radius $R$ with a $w$-warm start. 

We propose a \emph{robust} sampling framework that computes spectral approximations to the Hessian of the barrier functions in each iteration. We prove that for polytopes that are described by $n$ hyperplanes, sampling with the Lee-Sidford barrier function mixes within $\widetilde O((d^2+dL^2R^2)\log(w/\delta))$ steps with a per step cost of $\widetilde O(nd^{\omega-1})$, where $\omega\approx 2.37$ is the fast matrix multiplication exponent. Compared to the prior work of Mangoubi and Vishnoi, our approach gives faster mixing time as we are able to design a generalized soft-threshold Dikin walk beyond log-barrier.

We further extend our result to show how to sample from a $d$-dimensional spectrahedron, the constrained set of a semidefinite program, specified by the set $\{x\in \mathbb{R}^d: \sum_{i=1}^d x_i A_i \succeq C \}$ where $A_1,\ldots,A_d, C$ are $n\times n$ real symmetric matrices. We design a walk that mixes in $\widetilde O((nd+dL^2R^2)\log(w/\delta))$ steps with a per iteration cost of $\widetilde O(n^\omega+n^2d^{3\omega-5})$. We improve the mixing time bound of prior best Dikin walk due to Narayanan and Rakhlin that mixes in $\widetilde O((n^2d^3+n^2dL^2R^2)\log(w/\delta))$ steps.

  \end{abstract}
  \thispagestyle{empty}
\end{titlepage}

{\hypersetup{linkcolor=black}
}
\newpage

\else

\begin{abstract}

We consider the problem of sampling from a $d$-dimensional log-concave distribution $\pi(\theta) \propto \exp(-f(\theta))$ for $L$-Lipschitz $f$, constrained to a convex body with an efficiently computable self-concordant barrier function, contained in a ball of radius $R$ with a $w$-warm start. 

We propose a \emph{robust} sampling framework that computes spectral approximations to the Hessian of the barrier functions in each iteration. We prove that for polytopes that are described by $n$ hyperplanes, sampling with the Lee-Sidford barrier function mixes within $\widetilde O((d^2+dL^2R^2)\log(w/\delta))$ steps with a per step cost of $\widetilde O(nd^{\omega-1})$, where $\omega\approx 2.37$ is the fast matrix multiplication exponent. Compared to the prior work of Mangoubi and Vishnoi, our approach gives faster mixing time as we are able to design a generalized soft-threshold Dikin walk beyond log-barrier.

We further extend our result to show how to sample from a $d$-dimensional spectrahedron, the constrained set of a semidefinite program, specified by the set $\{x\in \mathbb{R}^d: \sum_{i=1}^d x_i A_i \succeq C \}$ where $A_1,\ldots,A_d, C$ are $n\times n$ real symmetric matrices. We design a walk that mixes in $\widetilde O((nd+dL^2R^2)\log(w/\delta))$ steps with a per iteration cost of $\widetilde O(n^\omega+n^2d^{3\omega-5})$. We improve the mixing time bound of prior best Dikin walk due to Narayanan and Rakhlin that mixes in $\widetilde O((n^2d^3+n^2dL^2R^2)\log(w/\delta))$ steps.

\end{abstract}

\fi

\section{Introduction}

Given a convex body, generate samples from the body according to structured densities is a fundamental problem in computer science and machine learning. It has extensive applications in constrained convex optimization~\citep{lv06,n16}, differentially private learning~\citep{wfs15,lmwrb24} and online optimization~\citep{nr17}. A central theme in the theory of sampling is to leverage stochasticity and reduce per iteration costs without having to proportionally increase the number of iterations.
This theme has played out in continuous optimization for both first and second order methods.
For the first order methods, focus is on reducing the variance of the gradient estimators \citep{jz13,sz13,dbl14}.
For the second order methods, matrix sketching is often used to reduce computation and storage related to the Hessian matrix \citep{lsz19,jlsw20,jswz21,sy21,qszz23}.

In Markov chain Monte Carlo (MCMC), much of the progress is on the first order methods.
Non-asymptotic analyses are performed for the stochastic gradient Langevin algorithms and their variance reduced extensions \citep{djw+16,rrt17,bdm18,cfmbj18,zxg18,dk19,lzc+19,dl21}. For second order methods, there has been a paucity when it comes to applying the sketching techniques.

In this paper, we focus on sampling from a log-concave distribution constrained to a $d$-dimensional convex body $\K$ that is described by $n$ constraints, where second order information is proven essential for capturing the geometry of the convex body and consequently for achieving fast convergence rates~\citep{n16,nr17,cdwy18,llv20,mv22,mv24}.
In particular, we associate the convex body with a \emph{self-concordant barrier function}~\citep{nn94,r88,v89_cp} and utilize the Hessian matrix $H(x)$ of the barrier function in the sampling algorithm.
We then use the Hessian matrix to propose samples and compute the probability to accept or reject the proposed samples. We note that this problem involves subtleties beyond the scope of constrained optimization.
In continuous optimization, which focuses on finding the descent directions, unbiased estimators with reasonable variance oftentimes suffice~\citep{v89_cp,lsw15,hjs+22}.
In MCMC, on the other hand, the target probability distribution needs to be maintained along the entire trajectory. This poses significant challenges to speed up the sampling algorithms. In the scenario of uniform sampling over a polytope,~\cite{llv20} shows that for a simple logarithmic barrier\footnote{Logarithmic barriers, or log-barriers have been extensively studied in the context of sampling and convex optimization. For a $d$-dimensional polytope with $n$ constraints, it typically leads to a convergence rate polynomially depends on $n$.}, an unbiased estimator in fact suffices. However, more complicated barriers such as the Lee-Sidford barrier~\citep{ls14,ls19} can only be approximately computed\footnote{Lee-Sidford barriers are the first polynomial-time computable barriers with complexity parameter depends only logarithmic on $n$ and in extension, the convergence rate.}, an unbiased estimator is extremely hard to be obtained. Moreover, for sampling from more sophisticated log-concave distributions and convex bodies, a more general approach is needed.

We therefore propose to obtain a high precision estimator to the desired acceptance rate with an improved per step running time and without sacrificing the rapid mixing time.
In particular, we ask the following question:
\begin{center}
    {\it Can we significantly reduce per iteration cost, while preserving the convergence rate of the log-concave sampling algorithms over various convex bodies?
    }
\end{center}

We answer the above question in the affirmative. To this end, we present a slew of results regarding log-concave sampling. For polytopes, we provide a walk that mixes in $\wt O(d^2+dL^2R^2)$ steps\footnote{We use $\wt O(\cdot)$ to hide polylogarithmic dependence on $n, d$ and $\delta$ where $\delta$ is the TV distance between the target distribution and our Markov chain $\mu$.} with per iteration cost $\wt O(nd^{\omega-1})$. Prior state-of-the-art results are due to~\cite{mv22,mv24}, for which their walks mix in $\wt O(nd+dL^2R^2)$ with a per iteration cost\footnote{We use $\nnz(A)$ to denote the number of nonzero entries in matrix $A$.} $\wt O(\nnz(A)+d^2)$. Our walk mixes faster whenever $n\geq d$. Our result partially resolves an open problem in~\cite{mv22} as they asked whether it's possible to design a Dikin walk whose mixing time is only depends on $d$ and independent of $L, R$. We remove the $L$ and $R$ dependence on the dominating term $d^2$, hence for the case of sampling from uniform distribution ($L=0$), we recover the state-of-art result of~\cite{llv20} which mixes in $\wt O(d^2)$ steps with a per iteration cost $\wt O(nd^{\omega-1})$.

We also consider sampling from convex bodies beyond polytopes. In semidefinite programming (SDP), one often focuses on the dual program of an SDP, where the constraint set is defined as a spectrahedron ${\cal K}=\{x\in \R^d: \sum_{i=1}^d x_iA_i\succeq C \}$\footnote{We use $A\succeq 0$ to denote $A$ is a positive semidefinite matrix, and $A\succeq B$ to denote $A-B\succeq 0$.} for $n\times n$ symmetric matrices $A_i$ and $C$~\citep{jkl+20,hjs+22}. We propose a walk that samples from a log-concave distribution over a spectrahedron ${\cal K}$ using the Hessian information of the log-barrier. The walk mixes in $\wt O(nd+dL^2R^2)$ steps. For log-barrier of an SDP, explicitly forming the Hessian matrix takes a prohibitively large $O(n^\omega d+n^2 d^{\omega-1})$ time. We utilize our robust sampling framework to approximately compute this Hessian via tensor-based sketching techniques, achieving a runtime of $\wt O(n^\omega+n^2 d^{3\omega-5})$. As long as $n\geq d^{\frac{3\omega-4}{\omega-2}}$ (which is a usual setting for SDP, where $d\ll n$), this provides a speedup. 

\subsection{Related works}
In this work, we focus on Dikin walk~\citep{kn12}, a refined variant of ball walk~\citep{ls93}. Roughly speaking, ball walk progresses by moving to a random point in the ball centered at the current point with the obvious downside that when the convex body is flat, ball walk progresses slowly. Dikin walk overcomes this problem by instead moving to a random point in a good ellipse centered at the current point, in an effort to better utilize the local geometry of the convex body.

For sampling over polytopes, a number of Dikin walks use the ellipse induced by the log-barrier functions~\citep{kn12}.
The work of~\cite{llv20} shows that uniform sampling over a polytope can mix in $O(\nu d)$ steps, where $\nu$ is the self-concordant parameter of the barrier function\footnote{Technically, what they have shown is that if the Hessian is a $\ov \nu$-symmetric self-concordant barrier function, then the walk mixes in $O(\ov \nu d)$ steps, and they subsequently prove for various barriers of interest, $\ov \nu=\nu$.}. Going beyond uniform distributions,~\cite{nr17} proposes a walk that samples from a log-concave distribution on a polytope in $\wt O(\nu^2 d)$ steps. This bound is later improved by~\cite{mv22}, in which they show that for logarithmic barrier, a mixing time of $\wt O(nd+dL^2R^2)$ is attainable, where $L$ is the Lipschitz constant of $f$ and $R$ is the radius of the ball that contains $\K$. A more recent work by~\cite{kv23} has obtained better bound when the function $f$ is $\alpha$-relative strongly-convex and the density $\pi$ is $\beta$-Lipschitz. They manage to obtain a mixing bound of $O(\nu d\beta\log(w/\delta))$. However, their algorithm requires stronger assumptions on $f$ and hence is incomparable to our result. In most previous works, the focus has been on improving the mixing time, rather than on per iteration costs.

For any convex body, it is well-known that a universal barrier with $\nu=d$ exists~\citep{nn89,ly18}, however it is computationally hard to construct the Hessian of the universal barrier. In short, the universal barrier requires to compute the volume of the polar set of a high dimensional body, which is even hard to approximate deterministically~\citep{fb86,nn94}. The seminal work of Lee and Sidford~\citep{ls14} presents a nearly-universal barrier with $\nu=O(d\log^5 n)$ and the Hessian can be \emph{approximated} in $O(nd^{\omega-1})$ time. The Lee-Sidford barrier was originally designed to solve linear programs in $O(\sqrt d)$ iterations, and it has been leveraged lately for Dikin walks with rapid mixing time. For uniform sampling,~\cite{cdwy18} gives an analysis with a walk that mixes in $\wt O(d^{2.5})$ steps. Subsequently,~\cite{llv20} improves the mixing to $\wt O(d^2)$ steps. In a pursuit to better leverage the local geometry,~\cite{lv17} proposes a walk relying on Riemannian metric that mixes in $\wt O(nd^{3/4})$ steps. The mixing rate is later improved to $\wt O(nd^{2/3})$ via Riemannian Hamiltonian Monte Carlo (RHMC) with log-barrier~\citep{lv18} and $\wt O(n^{1/3}d^{4/3})$ with a mixed Lee-Sidford and log-barrier~\citep{gkv24}. In this work, we show that log-concave sampling can also leverage Lee-Sidford barrier to obtain a mixing time of $\wt O(d^2+dL^2R^2)$. In comparison, the hit-and-run algorithm~\citep{lv07} mixes in $\wt O(d^2R^2/r^2)$ steps where $r$ is the radius of the inscribed ball inside $\K$. For the regime where $L=O(1)$ and $r=O(1)$, our walk mixes strictly faster than that of hit-and-run. 

We note that for sampling over more general convex bodies, a recent work~\citep{ce22} proves that for uniform sampling over an isotropic convex body, the mixing time bound is $\wt O(d^2/ \psi_d^2)$ where $\psi_d$ is the KLS constant~\citep{kls95}. However, it is unclear how to generalize their result to non-isotropic convex bodies or log-concave densities. 

\subsection{Our results}
Our results concern log-concave sampling from polytopes and spectrahedra. For polytopes, we state the result in its full generality.

\begin{theorem}[Robust sampling for log-concave distribution over polytopes]
\label{thm:main_polytope}
Let $\delta\in (0,1)$ and $R\geq 1$. Given a constraint matrix $A\in \R^{n\times d}$ with a vector $b\in \R^n$, let $\K:=\{x\in \R^d: Ax\leq b \}$ be the corresponding polytope. Suppose $\K$ is enclosed in a ball of radius $R$ with non-empty interior. Let $f:\K\rightarrow \R$ be an $L$-Lipschitz, convex function with an evaluation oracle. Finally, let $\pi$ be the distribution such that $\pi\propto e^{-f}$.

Suppose we are given an initial point from $\K$ that is sampled from a $w$-warm start distribution\footnote{We say a distribution $\rho$ is a $w$-warm start of with respect to a distribution $\pi$ if $\sup_{x\in \K} \frac{\rho(x)}{\pi(x)}\leq w$.} with respect to $\pi$ for some $w> 0$, then there is an algorithm (Algorithm~\ref{alg:informal}) that outputs a point from a distribution $\mu$ where $\TV( \mu, \pi)\leq \delta$.

Let $g:\K\rightarrow \R$ be a $\nu$-self-concordant barrier function such that in time ${\cal C}_g$, a spectral approximation $\wt H_g$ of the Hessian of $g$ denoted by $H_g$ can be computed satisfying
\begin{align*}
    (1-\epsilon)\cdot H_g\preceq \wt H_g \preceq (1+\epsilon)\cdot H_g
\end{align*}
for $\epsilon=\Theta(1/d)$. Then, Algorithm~\ref{alg:informal} takes at most 
\begin{align*}
    \wt O( ( \nu d + d L^2 R^2) \log(w/\delta))
\end{align*}
Markov chain steps. It uses $O(1)$ function evaluations and an extra ${\cal C}_g+d^\omega$ time at each step.
\end{theorem}

Let us pause and make some remarks regarding Theorem~\ref{thm:main_polytope}. As long as the Hessian matrix can be approximately generated with an error inversely depends on $d$, then our algorithm is guaranteed to converge. Moreover, if the approximate Hessian can be generated more efficiently, then this directly implies an improvement of our algorithm. On the mixing time side, Theorem~\ref{thm:main_polytope} is nearly-optimal up to polylogarithmic factors and the dependence on $L$ and $R$. In particular, the $\nu d$ mixing time bound is also achieved by~\cite{llv20} in the case of uniform sampling. We instantiate the meta result of Theorem~\ref{thm:main_polytope} into the following theorem.

\begin{theorem}[Robust sampling with nearly-universal barrier]
\label{thm:main_ls}
Under the conditions of Theorem~\ref{thm:main_polytope}, let $g$ be the Lee-Sidford barrier with $\nu=O(d\log^5 n)$. Then, we have
\begin{align*}
    {\cal C}_g = & ~ \wt O(nd^{\omega-1}).
\end{align*}
The algorithm takes at most 
\begin{align*}
    \wt O((d^2+dL^2R^2)\log(w/\delta))
\end{align*}
Markov chain steps.
\end{theorem}

The Lee-Sidford barrier~\citep{ls14,ls19} is the first polynomial-time computable barrier with a nearly-optimal self-concordance parameter. Several prior works~\citep{cdwy18,llv20} utilize this barrier for uniform sampling. For log-concave sampling, the walk of~\cite{kv23} requires strong-convexity-like assumption on $f$ in order to attain an $\wt O(d^2)$ mixing time. Our work is the first to obtain an $\wt O(d^2)$ mixing time for log-concave sampling over polytopes, when $L$ and $R$ are small.

\begin{table}[!ht]
    \centering
    \begin{tabular}{|l|l|l|}
    \hline
     {\bf References}    &  {\bf Mixing time} & {\bf Per iteration cost}\\ \hline
       \cite{lv06}  & $d^2R^2/r^2$ & $nd$\\ \hline
       \cite{nr17} & $d^5+d^3L^2R^2$ & $nd^{\omega-1}$ \\ \hline
       \cite{mv22} & $nd+dL^2R^2$ & $nd^{\omega-1}$\\ \hline
       \cite{mv24} & $nd+dL^2R^2$ & $\nnz(A)+d^2$  \\ \hline
       {\color{red}This work} & {\color{red}$d^2+dL^2R^2$} & {\color{red}$nd^{\omega-1}$} \\ \hline
    \end{tabular}
    \caption{Comparison of algorithms for sampling from a log-concave density $e^{-f}$ over a $d$-dimensional polytope with $n$ constraints, where $f$ is $L$-Lipschitz. We omit $\wt O(\cdot)$ to mixing time and per iteration cost. We assume the evaluation of $f$ can be done in unit time. The first row is the hit-and-run walk. Our algorithm has the fastest mixing time among all Dikin walks, and for $L=O(1)$, $r=O(1)$, our walk mixes faster than hit-and-run.}
    \label{tab:compare_polytope}
\end{table}
Our next result is closely related to semidefinite programming. Given the the set ${\cal K}=\{x\in \R^d: \sum_{i=1}^d x_iA_i\succeq C \}$ for symmetric $n\times n$ matrices $A_1,\ldots,A_d, C$, we consider sampling from a log-concave distribution over ${\cal K}$. We use the popular log-barrier for ${\cal K}$ studied in~\cite{nn94,jkl+20,hjs+22}. To further utilize our robust sampling framework, we develop a novel approach for generating a spectral approximation of the Hessian. Compared to~\cite{jkl+20,hjs+22} in which the Hessian is maintained in a very careful way in conjunction with the interior point method, our spectral approximation does not require to control the changes over iterations and is highly efficient when $n\gg d$, which is a popular regime for semidefinite programs.

\begin{theorem}[Robust sampling for log-concave distribution over spectrahedra]
\label{thm:main_sdp}

Let $\delta\in (0,1)$ and $R\geq 1$. Given a collection of symmetric matrices $A_1,\ldots,A_d\in \R^{n\times n}$ and a target symmetric matrix $C\in \R^{n\times n}$, let $\K:=\{x\in \R^d: \sum_{i=1}^d x_iA_i\succeq C \}$ be the corresponding spectrahedron. Suppose $\K$ is enclosed in a ball of radius $R$ with non-empty interior. Let $f:\K\rightarrow \R$ be an $L$-Lipschitz, convex function with an evaluation oracle. Finally, let $\pi$ be the distribution such that $\pi\propto e^{-f}$.

Suppose we are given an initial point from $\K$ that is sampled from a $w$-warm start distribution with respect to $\pi$ for some $w> 0$, then there is an algorithm (Algorithm~\ref{alg:informal}) that outputs a point from a distribution $\mu$ where $\TV( \mu, \pi)\leq \delta$.

Let $g:{\cal K}\rightarrow \R$ be the logarithmic barrier function with $\nu=n$. There is an algorithm that uses
\begin{align*}
    \wt O( (  nd + d L^2 R^2) \log(w/\delta))
\end{align*}
Markov chain steps. In each step it uses $O(1)$ function evaluations and an extra cost of 
\begin{align*}
    \wt O(n^\omega+n^2d^{3\omega-5}).
\end{align*}
\end{theorem}

Prior work due to~\cite{nr17} provides a walk that mixes in $\wt O(n^2d^3+n^2 d L^2R^2)$ steps and each step could be implemented in $O(dn^\omega+n^2d^{\omega-1})$ time, our algorithm improves upon both mixing and per iteration cost. On the front of per iteration cost, computing the Hessian of the log-barrier of ${\cal K}$ exactly takes time $O(dn^\omega+n^2d^{\omega-1})$, for the regime where $n\gg d$ the dominating term is $dn^\omega$. Note that whenever $n\gg d^{\frac{3\omega-4}{\omega-2}}$, our algorithm is more efficient than that of exact computation. This is a common regime in semidefinite program in which the number of constraints $d$ is small compared to the size of the primal solution. 

\begin{table}[!ht]
    \centering
    \begin{tabular}{|l|l|l|}
    \hline
     {\bf References}    &  {\bf Mixing time} & {\bf Per iteration cost}\\ \hline
       \cite{lv06}  & $d^2R^2/r^2$ & $n^\omega+n^2 d$\\ \hline
       \cite{nr17} & $n^2d^3+n^2dL^2R^2$ & $n^\omega d+n^2 d^{\omega-1}$ \\ \hline
       {\color{red}This work} & {\color{red}$nd+dL^2R^2$} & {\color{red}$n^\omega+n^2d^{3\omega-5}$} \\ \hline
    \end{tabular}
    \caption{Comparison of algorithms for sampling from a log-concave density $e^{-f}$ over a $d$-dimensional spectrahedron with $n\times n$ symmetric matrix constraints, where $f$ is $L$-Lipschitz. We omit $\wt O(\cdot)$ to mixing time and per iteration cost. We assume evaluation of $f$ can be done in unit time. The first row is the hit-and-run walk. Our algorithm mixes faster than~\cite{nr17} in all parameter regimes, and for $L=O(1)$, $r=O(1)$, $d^2R^2\geq n$, our walk mixes faster than hit-and-run.}
    \label{tab:compare_spectrahedron}
\end{table}
\section{Technical Overview}

\begin{algorithm}[!ht]\caption{Sampler for log-concave distribution $\pi \propto \exp(-f)$ over convex body with a barrier function $g$. ${\cal A}$ is the description of the convex set, function $f$ is an $L$-Lipschitz concave function and $g$ is a $\nu$-self-concordant barrier function, $x_0\in \K$ is a $w$-warm start point, and $\delta$ is the total variance distance between our Markov chain and the target distribution. Let $\K\subseteq \R^d$ enclosed by a ball of radius $R$.}\label{alg:informal}
\begin{algorithmic}[1]
\Procedure{SamplingConvexBody}{${\cal A}, f, g, x_0, \delta$}
    \State Initialize $x$ to be a point in $\K$
    \State $T\gets \wt O((\nu d+dL^2R^2 )\log(w/\delta))$
     \For{$t=1 \to T$}
        \State Sample a point $\xi \sim {\cal N}(0,I_d)$
        \State {\color{blue}// $H(x)$ is the exact Hessian matrix at $x$, we never explicitly form it}
        \State Compute $\wt H(x)$ where $(1-1/d) H(x)\preceq \wt H(x)\preceq (1+1/d) H(x)$
        \State $\wt \Phi(x)=d\cdot \wt H(x)+dL^2\cdot I_d $ 
        \State $z \gets x + \wt{\Phi}(x)^{-1/2} \xi$
        \If{$z \in \mathrm{Int}(\K)$}
            \State {\color{blue}// $H(z)$ is the exact Hessian matrix at $z$}
            \State Compute $\wh H(z)$ such that $(1-1/d)H(z)\preceq \wh H(z)\preceq (1+1/d) H(z)$
            \State $\wh \Phi(z)=d\cdot \wh H(z)+dL^2\cdot I_d$
            \State $\tau\gets \frac{ \exp(-f(z)) \cdot ( \det(\wh{\Phi}(z)))^{1/2} \cdot \exp( - 0.5 \| x - z\|_{ \wh{\Phi}(z) }^2 ) }{ \exp(-f(x)) \cdot ( \det( \wt{\Phi}(x)) )^{1/2} \cdot \exp(-0.5 \| x - z \|_{ \wt{\Phi}(x) }^2 ) } $
            \State {\color{blue}// Our walk is lazy, i.e., it only moves to a new point with half probability}
            \State {\bf accept} $x \gets z$ with probability $ 0.5 \cdot \min \{ \tau , 1 \} $ 
        \Else 
            \State {\bf reject} $z$
        \EndIf 
    \EndFor 
\EndProcedure
\end{algorithmic}
\end{algorithm}

Before diving into our techniques, we first illustrate an informal version of our algorithm. The algorithm itself is a variant of the Dikin walk~\citep{kn12} called the soft-threshold Dikin walk~\citep{mv22}. Starting with a $w$-warm start point, the algorithm approximately computes the Hessian of the barrier function, then proposes a new point $z$ from a Gaussian with mean at the current point $x$ and variance $\wt \Phi(x)^{-1}$, where $\wt \Phi$ is a multiple of the approximate Hessian plus a multiple of the identity. The proposal is then accepted via a Metropolis filter. Note that compared to the standard Dikin walk framework, our algorithm crucially allows the Hessian to be approximated with a good precision. This means any improvements on generating spectral approximation leads to runtime improvement of our algorithm in a black-box fashion.

\subsection{Soft-threshold Dikin walk for log-concave sampling via a barrier-based argument}

For uniform sampling from structured convex bodies such as polytopes,~\cite{kn12} introduce a refined ball walk that utilizes a \emph{self-concordant barrier function} associated with the convex body. To extend the Dikin walk framework for log-concave sampling,~\cite{nr17} adds an additional coefficient $\frac{\exp(-f(x))}{\exp(-f(z))}$ to the acceptance probability of Metropolis filter. This achieves a sub-optimal mixing time bound of $\wt O(\nu^2 d)$. By introducing a soft-threshold regularizer to the Hessian,~\cite{mv22} proves that for log-barrier function, it is possible to approach the optimality with an $\wt O(nd+dL^2R^2)$ mixing time. Unfortunately, the method used by~\cite{mv22} is specialized to log-barrier, making it hard to generalize to other barriers. They show that under proper scaling, the true Hessian $H(x)$ under a sequence of polytopes converges to the regularized Hessian under the given polytope in the limit. They define the $j$-th polytope as 
\begin{align*}
   A_j = \begin{bmatrix}
       A \\
       I_d \\
       \vdots \\
       I_d
   \end{bmatrix} , & ~ b_j = \begin{bmatrix}
       b \\
       j\cdot {\bf 1}_d \\
       \vdots \\
       j\cdot {\bf 1}_d
   \end{bmatrix}
\end{align*}
the number of copies of $I_d$ and $j\cdot {\bf 1}_d$ is a function of $j$. While this polytope-based method works well for log-barrier, it functions poorly for more intricate barriers, such as volumetric barrier and Lee-Sidford barrier. Take volumetric barrier as an example, recall that $H_{\rm vol}(x)=A^\top \Sigma(S(x)^{-1}A) S(x)^{-2}A$ where $\Sigma(S(x)^{-1}A)$ is the statistical leverage score~\citep{dmm06,ss11}. Leverage score is a numerical quantity that measures \emph{how important a row of a matrix is, compared to other rows}. This means that, if one duplicates a row infinitely many times, the row will be assigned a score of 0. More concretely, suppose we are given an $n\times d$ matrix with $n$ identical rows, then the leverage score of each row is $\frac{d}{n}$. Taking $n\rightarrow \infty$, it's easy to see that all rows have score 0, and the Hessian of volumetric barrier will zero out the rows contributed to the infinitely many copies of identities. Thus, for volumetric barrier, extra constraints of the polytope sequence vanish. On the other hand, the~\cite{mv22} construction heavily relies on the infinitely many occurrences of $I_d$ for it to converge to the regularized Hessian.

To circumvent this issue and provide a simpler argument, we develop a proof strategy that is \emph{barrier-based}, instead of polytope-based. We example a class of popular barrier functions for polytopes that can be classified as \emph{weighted log-barrier functions}~\citep{v89_cp}. Roughly speaking, these functions take in the form of $g(x)=-\sum_{i=1}^n w_i \log(S(x)_i)$, with $w={\bf 1}_n$ being the log-barrier, $w=\sigma(S(x)^{-1}A)$ being the volumetric barrier~\citep{v89_cp} and if we choose $w$ to be a proper power of the Lewis weights, then the barrier is precisely the Lee-Sidford barrier~\citep{ls19,llv20}. These weights are stable, meaning that if two points are close in certain local norm induced by proper ellipses given by these weights, then their weights are also close~\citep{ls19}. This property proves to be particularly useful when proving the mixing rate of our Dikin walk. In addition, we carve out three sufficient conditions for a weighed log-barrier barrier function to work for log-concave distribution, and has $\wt O(\nu d)$ mixing rate:

\begin{itemize}
    \item {\bf $\ov\nu$-symmetry.} The Hessian $H(x)$ is $\ov\nu$-symmetry, that is, for any $x\in \K$, $E_x(1)\subseteq \K\cap (2x-\K)\subseteq E_x(\sqrt{\ov \nu})$ where $E_x(r)=\{y\in \R^d: (y-x)^\top H(x) (y-x)\leq r^2 \}$. We show that $\ov \nu=\nu$ for log-barrier, volumetric barrier and Lee-Sidford barrier for polytopes, and log-barrier for spectrahedra.
    \item {\bf Bounded local norm.} The variance term $\|H(x)^{-1/2}\cdot \nabla \log\det(H(x))\|_2^2\leq d\poly\log n$. We show this condition holds for all barriers of interest in this paper.
    \item {\bf Convexity of regularized barrier.} The function $F(x)=\log\det(H(x)+I_d)$ is convex at $x$ for any $x\in \K$. We note that for log-barrier, volumetric barrier and Lee-Sidford barrier, it has been shown that $\log\det(H(x))$ is indeed convex at $x$. We further prove that this still holds when a copy of identity is blended in.
\end{itemize}

We want to highlight that our approach is more robust and generic than that of~\cite{mv22}, as it solely depends on the structure of the Hessian matrix. Our argument should be treated as a generalization of~\cite{sv16}, in which they utilize the bounded local norm and convexity of $\log\det(H(x))$ for log-barrier, together with the Gaussianity to conclude that the difference between $\log\det$ are small. We go beyond log-barrier for polytopes and uniform sampling.

\subsection{Approximation preserves the mixing time}

The core premise of our algorithm is we allow the Hessian to be approximated, so that each step can be implemented efficiently. This poses significant challenges in proving mixing, as even if we have a good approximation of one-step in the Markov chain, the approximate chain does not necessarily mixes as fast as the original chain.
For example, if every step is approximated within $\epsilon$-TV distance to the original Markov chain, then in $T$ steps, the TV distance between the resulting distribution under approximate walk and the distribution under original walk could be as large as $T \epsilon$.
This means we have to take $\epsilon$ very small ($O(T_{\text{mix}}^{-1})$) to guarantee same convergence property as original chain, which is unacceptable.
A related issue is that the stationary distribution under approximate walk may not be the same as the target distribution. Therefore we need a way to control the mixing properties of the approximate walk. We resolve this problem by exactly computing the acceptance probability under approximate walk.
That is, after proposing the next step $z$ from $x, \wt{\Phi}(x)$, we sample $\wh{\Phi}(z)$, and accept with probability $\min\{1, \frac{\pi(z) \wh{G}_z(x)}{\pi(x) \wt{G}_x(z)}\}$ (where $\wt G_x(z)$ is the probability of proposing $z$ starting from $x$ under the approximate walk). Recall that in the exact walk, we accept with probability $\min\{1, \frac{\pi(z) G_z(x)}{\pi(x) G_x(z)}\}$ (where $G_x(z)$ is the probability of proposing $z$ stating from $x$ under the exact walk). If we use this acceptability for the approximate walk, then we only have one-step approximation guarantee, and will suffer from the problem mentioned previously.

By using this modified acceptance probability, we can prove reversibility of the approximate walk, and that stationary distribution of the approximate walk is indeed our target distribution.
For the mixing rate, we bound the conductance of the approximate walk, which requires us to prove the following properties:
\begin{enumerate}[label={(\arabic*)}]
    \item The proposed step is accepted with decent probability.
    \item Starting from two points moderately close to each other (i.e., $\|x-z\|_{\Phi(x)} \le \epsilon$), the approximate steps taken $P_x$ and $P_z$ are close in TV distance (i.e., $\TV(P_x, P_z) \le 1-\epsilon'$) for some absolute constants $\epsilon,\epsilon'>0$.
\end{enumerate}
We are able to prove these properties by comparison with the original chain.
That is, assuming the original chain satisfies properties (1)(2), then the approximate chain also satisfies the same properties, as long as the approximation is good enough.
This enables us to prove these key properties in a hassle-free way. To prove these comparison results, we need to prove, for example, $\wt{G}_x(z) \approx G_x(z)$.
Because both $G_x(z)$ and $\wt G_x(z)$ have nice factorizations
\begin{align*}
G_x(z) &\propto \det(\Phi(x))^{1/2} \exp(-\frac 12 \|x-z\|_{\Phi(x)}^2),\\
\wt G_x(z) &\propto \det(\wt \Phi(x))^{1/2} \exp(-\frac 12 \|x-z\|_{\wt \Phi(x)}^2),
\end{align*}
we only need to prove $\det ( \Phi(x) ) \approx \det ( \wt{\Phi}(x) )$ and $\exp(-\frac 12 \|x-z\|_{\Phi(x)}^2) \approx \exp(-\frac 12 \|x-z\|_{\wt{\Phi}(x)}^2)$ separately. Both of these can be handled via $\Phi(x) \approx \wt \Phi(x)$, using our approximation procedure.

\subsection{Approximation of Lewis weights for rapid mixing}

Given our robust and generic Dikin walk framework, we realize it with the Lee-Sidford barrier whose complexity $\nu=d\log^5 n$. This ensures the walk mixes in $\wt O(d^2+dL^2R^2)$ steps, improving upon the log-barrier-based walk of~\cite{mv22,mv24} that mixes in $\wt O(nd+dL^2R^2)$ steps, as the log-barrier has complexity $\nu=n$. To implement the Lee-Sidford barrier, it is imperative to compute the $\ell_p$ (for $p>0$) Lewis weights, which is defined as the following convex program:
\begin{align}\label{eq:lewis_convex}
    \min_{M\succeq 0}~-\log\det M,~\text{subject to}~\sum_{i=1}^n (a_i^\top Ma_i)^{p/2}\leq d,
\end{align}
and the $\ell_p$ Lewis weights is a vector $w_p\in \R^d$ where $(w_p)_i=a_i^\top M_* a_i$ and $M_*$ is the optimum of Program~\eqref{eq:lewis_convex}. Solving the program exactly is not efficient, and a long line of works~\citep{cp15,ls14,ls19,ccly19,flps22,jls22} provide fast algorithms that approximate all weights up to $(1\pm\epsilon)$-factor. The Hessian of Lee-Sidford barrier has the form (up to scaling) $H_{\rm Lewis}(x)=A^\top S(x)^{-1} W_p(S(x)^{-1}A)^{1-2/p} S(x)^{-1}A$ where $W_p(S(x)^{-1}A)$ is the diagonal matrix of $\ell_p$ Lewis weights with respect to $S(x)^{-1}A$. When clear from context, we use $W_p$ as a shorthand for $W_p(S(x)^{-1}A)$. As $W_p$ could be approximated in $\wt O(nd^{\omega-1})$ time, we could then perform subsequent operations exactly to form the approximate Hessian. We instead provide a spectral approximation procedure that runs in $\wt O(\nnz(A)+\Tmat(d,d^3,d))$ given the approximate Lewis weights where $\Tmat(a, b, c)$ is the time of multiplying an $a\times b$ matrix with a $b\times c$ matrix and $\Tmat(d, d^3, d)\approx d^{4.2}$. This is important as even though it doesn't lead to direct runtime improvement, if approximate Lewis weights can be computed in $\wt O(\nnz(A)+\poly(d))$ time, then we obtain a per iteration cost upgrade in a black-box fashion. Our approach is based on spectral sparsification via random sampling: given a matrix $B\in \R^{n\times d}$, one can construct a matrix $\wt B\in \R^{s\times d}$ consists of re-scaled rows of $B$, such that $(1-\epsilon)\cdot B^\top B\preceq \wt B^\top \wt B\preceq (1+\epsilon)\cdot B^\top B$. If we sample according to the leverage scores~\citep{dmm06,ss11}, one can set $s=O(\epsilon^{-2} d\log d)$. In our case, we set $B=W_p^{1/2-1/p}S(x)^{-1}A$, and perform the sampling. However, computing leverage score exactly is as expensive as forming the Hessian exactly, which requires $O(nd^{\omega-1})$ time. To speed up the crucial step, we use randomized sketching technique to reduce the row count from $n$ to $\poly(d)$, then approximately estimate all leverage scores with this small matrix. One can either use the sparse embedding matrix~\citep{nn13} to approximate these scores in time $O(\nnz(A)\log n+\Tmat(d, \epsilon^{-2}d, d))$, or use the subsampled randomized Hadamard transform~\citep{ldfu13} in time $O(nd\log n+\Tmat(d, \epsilon^{-2}d, d))$. We remark that approximating leverage scores for subsampling has many applications in numerical linear algebra, such as low rank approximation~\citep{bwz16,swz17,swz19}. Utilizing this framework, we provide an algorithm that approximates the Hessian of Lee-Sidford barrier in time $\wt O(nd^{\omega-1})+\wt O(\nnz(A)+\Tmat(d, d^3, d))$, as advertised.

\subsection{Approximation of log-barrier Hessian over a spectrahedron}

For a spectrahedron induced by the dual semidefinite program, it is also natural to define its log-barrier with its corresponding Hessian being 
\begin{align*}
    H_{\rm log}(x) = & ~ {\sf A}(S(x)^{-1}\otimes S(x)^{-1}){\sf A}^\top,
\end{align*}
where ${\sf A}\in \R^{d\times n^2}$ with the $i$-th row being $\vect(A_i)$ for $n\times n$ symmetric matrix $A_i$, and $S(x)=\sum_{i=1}^d x_iA_i-C$ where $C$ is also an $n\times n$ symmetric matrix.~\cite{nn94} shows that by smartly arranging the organization of $H_{\rm log}(x)$, it can be computed exactly in time $O(dn^\omega+d^{\omega-1}n^2)$. Beyond exact computation,~\cite{jkl+20,hjs+22} present algorithms that approximately \emph{maintain} the Hessian matrix under low-rank updates. These maintenance approaches are crafted towards solving semidefinite programs in which the trajectory can be carefully controlled. For our soft-threshold Dikin walk, though the proposal generates a point that is relatively close to the starting point, due to its Gaussianity nature, much less structure can be exploited and thus maintained. Instead, we propose a maintenance-free approach that uses randomized sketching to generate a spectral approximation of $H_{\rm log}(x)$.

To illustrate the algorithm, let's first define a matrix 
\begin{align*}
    {\sf B}={\sf A}(S(x)^{-1/2}\otimes S(x)^{-1/2}).
\end{align*}
It is not hard to see that $H_{\rm log}(x)={\sf B}{\sf B}^\top$. Due to the Kronecker product structure of $(S(x)^{-1/2}\otimes S(x)^{-1/2})$, it is natural to consider sketches for Kronecker product of matrices~\citep{dssw18,djssw19,akk+20,swyz21}. Following the standard procedure for using sketch to generate spectral approximation, we can choose a {\sf TensorSRHT} matrix~\citep{akk+20,swyz21} $T$ and compute $R:=(S(x)^{-1/2}\otimes S(x)^{-1/2})T$ then form ${\sf A}R$. This is unfortunately, not efficient enough, as multiplying ${\sf A}$ with $R$ might be too slow. To further optimize the runtime efficiency, we use a more intrinsic approach based on the matrix $T$ and ${\sf B}$. It is well-known that the $i$-th row of ${\sf B}$ can be first computed as $S(x)^{-1/2}A_iS(x)^{-1/2}$ then vectorize. If we can manage to apply the sketch in a row-by-row fashion, then fast matrix multiplication can be utilized for even faster algorithms. We recall that $T=P\cdot (HD_1\otimes HD_2)$ where $H$ is the Hadamard matrix and $P$ is a row sampling matrix. To compute a row, we can first apply $HD_1$ and $HD_2$ individually to $S(x)^{-1/2}$ in $\wt O(n^2)$ time since $H$ is a Hadamard matrix. We can then form two matrices $X$ and $Y$ with rows being the corresponding sampled rows from $HD_1$ and $HD_2$. Finally, we compute $XA_iY^\top$ to form one row. Using fast matrix multiplication, we can form each row in $\Tmat(d^3, n, n)=O(n^2d^{3(\omega-2)})$ time. Applying this procedure to $d$ rows leads to an overall $O(n^2d^{3\omega-5})$ time, which beats the $O(dn^\omega)$ time as long as $n\gg d$. Our runtime is somewhat slow when $n$ is not that larger than $d$, this is due to we have to set $\epsilon=O(1/d)$. For more popular regimes where it suffices to set $\epsilon=O(1)$, our algorithm outperforms exact computation as long as $n\geq d$.

\section{Applications}

There are many applications of our Dikin walk, such as differentially private learning, simulated annealing and regret minimization. For a more comprehensive overview of the applications, we refer readers to~\cite{nr17}.

\paragraph{Differentially private learning.} Let ${\cal X}^m=\{x_1,\ldots,x_m\}$ denote an $m$-point dataset, associate a convex loss function $\ell_i$ to point $x_i$, the learning problem attempts to find an optimal parameter $\theta_*\in \K$ that minimizes $\ell(\theta)=\sum_{i=1}^m \ell_i(\theta)$. We could further enforce \emph{privacy} constraints to learning: we say a randomized algorithm ${\cal M}: {\cal X}^m\rightarrow \R$ is $\epsilon$-DP if for any datasets $X, X'\in {\cal X}^m$ differ by a single point and for any $S\subseteq \R$, $\Pr[{\cal M}(X)\subseteq S]\leq e^\epsilon \Pr[{\cal M}(X')\subseteq S]$. The differentially private learning seeks to solve the learning problem under DP guarantees, and it has been shown that if one allows for \emph{approximate DP}, then it can be achieved via the exponential mechansim and sampling from the distribution $\pi(\theta)\propto e^{-\ell(\theta)}$~\citep{wfs15}. If $\K$ is a polytope or spectrahedron, and in addition $\ell$ is $L$-Lipschitz, one can implement our walk to obtain an $\wt O(d^2+dL^2R^2)$ mixing for polytopes or $\wt O(nd+dL^2R^2)$ mixing for spectrahedra. Via a standard reduction from TV distance bound to infinity distance bound~\citep{mv22_neurips}, our walk can also be adapted for a \emph{pure DP} guarantee~\citep{lmwrb24}.

\paragraph{Simulated annealing for convex optimization.} Let $f: \R^d\rightarrow \R$ be an $L$-Lipschitz and convex function, we consider the problem of minimizing $f$ over $\K$ when we can only access the function value of $f$. This is also referred to as zeroth-order convex optimization. One could solve the problem via the \emph{simulated annealing framework}, in which one needs to sample from the distribution $\pi(x)\propto e^{-f(x)/T}$ where $T$ is the temperature parameter~\citep{kv06}. Our walk can be adapted to the framework for polytopes, it mixes in $\wt O(d^{2.5}+d^{1.5}L^2R^2)$ steps and for spectrahedra, it mixes in $\wt O(nd^{1.5}+d^{1.5}L^2R^2)$ steps, improving upon prior best algorithms that mix in $\min\{d^{5.5}+d^{3.5}L^2R^2, nd^{1.5}+d^{1.5}L^2R^2 \}$ steps for polytopes and $n^2d^{3.5}+n^2d^{1.5}L^2R^2$ steps for spectrahedra~\citep{nr17,mv22}.

 \paragraph{Online convex optimization.} Consider the following online optimization problem: let $\K$ be a convex set that we could choose our actions from and let $\ell_1,\ldots,\ell_T$ be a sequence of unknown convex cost functions with $\ell_t: \K\rightarrow \R$. The goal is to design a good strategy that chooses from $\K$ so that the total cost is small, compared to the offline optimal cost, in which one could choose the strategy after seeing the entire sequence of $\ell_t$'s. The algorithm works as follows: at round $t$, we choose a distribution $\mu_{t-1}$ supported on $\K$ and play the action $Y_t\sim \mu_{t-1}$. Our goal is to minimize the expected regret, defined as ${\rm Reg}_T(U)=\E[\sum_{t=1}^T \ell_t(Y_t)-\sum_{t=1}^T \ell_t(U)]$ with respect to all randomized strategies defined by distribution $p_U$. It turns out if one sets $s_t(x)=\eta \sum_{s=1}^t \ell_t(x)$ where $\eta>0$ is a step size parameter and sets $\mu_t\propto e^{-s_t}$, then this update rule reflects the multiplicative weights update algorithm~\citep{ahk12}. Moreover, one could prove a nearly-optimal regret bound of $O(\sqrt{T})$ if the KL divergence between $\mu_0$ and all possible $p_U$'s are bounded. The algorithmic problem of sampling from $\mu_t$ is equivalent to the log-concave sampling problem we study in this paper, and we could efficiently generate strategies with good regrest, similar to~\cite{nr17}. Let $\ell_t$'s be 1-Lipschitz linear functions,~\cite{nr17} uses their time-varying Dikin walk to achieve a regret of $O(d^{2.5}\sqrt{T})$, while we could significantly improve this bound to $O(d\sqrt{T})$.
\section{Conclusion}\label{sec:conclusion}

In this paper, we design a class of error-robust Dikin walks for sampling from a log-concave and log-Lipschitz density over a convex body. The key features of our walks are that their mixing time depends \emph{linearly} on the complexity of self-concordant barrier function of the convex body, and they allow computationally intensive quantities to be \emph{approximated} rather than computed exactly. For polytopes, our walk mixes in $\wt O(d^2+dL^2R^2)$ steps with a per iteration cost of $\wt O(nd^{\omega-1})$, and for spectrahedra, our walk mixes in $\wt O(nd+dL^2R^2)$ steps with a per iteration cost of $\wt O(n^\omega+n^2d^{3\omega-5})$. For polytopes, our walk is the first successful adaptation of the Lee-Sidford barrier for log-concave sampling under the minimal assumption that $f$ is Lipschitz, improving upon the mixing of prior works~\citep{nr17,mv22,mv24}. For spectrahedra, we improve the mixing of~\cite{nr17} from $n^2d^3+n^2 dL^2R^2$ to $nd+dL^2R^2$, for the term that depends on only $n$ and $d$, we obtain an improvement of $nd^2$, and for the term that depends on $L$ and $R$, we shave off the quadratic dependence on $n$. Moreover, we adapt our error-robust framework and present a sketching algorithm for approximating the Hessian matrix in $\wt O(n^\omega+n^2d^{3\omega-5})$ time. Our results have deep implications, as it could be leveraged for differentially private learning, convex optimization and regret minimization.

While our framework offers the fastest mixing Dikin walk for log-concave sampling, its mixing time has a rather unsatisfactory dependence on the radius of the bounding box, $R$. It would be interesting if one could design a walk that \emph{does not} depend on $R$. We also note that we require the density of be log-Lipschitz rather than Lipschitz.~\cite{kv23} shows that if $f$ in addition satisfies the \emph{relative strongly-convex property}, then there exists a walk whose mixing does not depend on $R$ and one only needs the density of be Lipschitz. Another important direction is to further improve the mixing rate of sampling over a spectrahedron using barrier functions such as volumetric barrier and hybrid barrier~\citep{a00}, while maintaining an small per iteration cost. Finally, extending RHMC to log-concave sampling would be essentially, as it has the potential to give the fastest mixing walk by utilizing Riemannian metrics instead of Euclidean. As our work is theoretical in nature, we don't foresee any potential negative societal impact. It has the potential positive impact to reduce energy consumption and carbon emission when deployed in practice.
\section*{Acknowledgement}

We would like to thank Jonathan Kelner, Yin Tat Lee, Santosh Vempala and Yunbum Kook for valuable discussions and anonymous reviewers for helpful comments.
The research is partially supported by the NSF awards: SCALE MoDL-2134209, CCF-2112665 (TILOS).
It is also supported by the U.S. Department of Energy, the Office of Science, the Facebook Research Award, as well as CDC-RFA-FT-23-0069 from the CDC’s Center for Forecasting and Outbreak Analytics. Lichen Zhang is supported by NSF awards CCF-1955217 and DMS-2022448. For more information related to the paper and adjacent topics, see~\url{https://www.youtube.com/@zhaosong2031} and~\url{https://space.bilibili.com/3546587376650961}.

\ifdefined\isarxiv
\bibliographystyle{alpha}
\bibliography{ref}
\else
\bibliography{ref}

\newcommand{\etalchar}[1]{$^{#1}$}
\begin{thebibliography}{DJRW{\etalchar{+}}16}

\bibitem[AHK12]{ahk12}
Sanjeev Arora, Elad Hazan, and Satyen Kale.
\newblock The multiplicative weights update method: a meta-algorithm and
  applications.
\newblock {\em Theory of Computing}, 8(1):121--164, 2012.

\bibitem[AKK{\etalchar{+}}20]{akk+20}
Thomas~D Ahle, Michael Kapralov, Jakob~BT Knudsen, Rasmus Pagh, Ameya
  Velingker, David~P Woodruff, and Amir Zandieh.
\newblock Oblivious sketching of high-degree polynomial kernels.
\newblock In {\em Proceedings of the Fourteenth Annual ACM-SIAM Symposium on
  Discrete Algorithms (SODA)}, pages 141--160, 2020.

\bibitem[Ans00]{a00}
Kurt~M Anstreicher.
\newblock The volumetric barrier for semidefinite programming.
\newblock {\em Mathematics of Operations Research}, 25(3):365--380, 2000.

\bibitem[AW21]{aw21}
Josh Alman and Virginia~Vassilevska Williams.
\newblock A refined laser method and faster matrix multiplication.
\newblock In {\em Proceedings of the 2021 ACM-SIAM Symposium on Discrete
  Algorithms (SODA)}, pages 522--539. SIAM, 2021.

\bibitem[BDM18]{bdm18}
Nicolas Brosse, Alain Durmus, and Eric Moulines.
\newblock The promises and pitfalls of stochastic gradient langevin dynamics.
\newblock {\em Advances in Neural Information Processing Systems}, 31, 2018.

\bibitem[BWZ16]{bwz16}
Christos Boutsidis, David~P Woodruff, and Peilin Zhong.
\newblock Optimal principal component analysis in distributed and streaming
  models.
\newblock In {\em Proceedings of the forty-eighth annual ACM symposium on
  Theory of Computing (STOC)}, pages 236--249, 2016.

\bibitem[CCLY19]{ccly19}
Michael~B Cohen, Ben Cousins, Yin~Tat Lee, and Xin Yang.
\newblock A near-optimal algorithm for approximating the john ellipsoid.
\newblock In {\em Conference on Learning Theory}, pages 849--873. PMLR, 2019.

\bibitem[CDWY18]{cdwy18}
Yuansi Chen, Raaz Dwivedi, Martin~J. Wainwright, and Bin Yu.
\newblock Fast mcmc sampling algorithms on polytopes.
\newblock {\em J. Mach. Learn. Res.}, 2018.

\bibitem[CE22]{ce22}
Yuansi Chen and Ronen Eldan.
\newblock Hit-and-run mixing via localization schemes, 2022.

\bibitem[CFM{\etalchar{+}}18]{cfmbj18}
Niladri Chatterji, Nicolas Flammarion, Yian Ma, Peter Bartlett, and Michael
  Jordan.
\newblock On the theory of variance reduction for stochastic gradient monte
  carlo.
\newblock In {\em International Conference on Machine Learning}, pages
  764--773. PMLR, 2018.

\bibitem[CP15]{cp15}
Michael~B. Cohen and Richard Peng.
\newblock Lp row sampling by lewis weights.
\newblock In {\em Proceedings of the Forty-Seventh Annual ACM Symposium on
  Theory of Computing}, STOC '15, 2015.

\bibitem[DBLJ14]{dbl14}
Aaron Defazio, Francis Bach, and Simon Lacoste-Julien.
\newblock Saga: A fast incremental gradient method with support for
  non-strongly convex composite objectives.
\newblock {\em Advances in neural information processing systems}, 27, 2014.

\bibitem[DJRW{\etalchar{+}}16]{djw+16}
Kumar~Avinava Dubey, Sashank J~Reddi, Sinead~A Williamson, Barnabas Poczos,
  Alexander~J Smola, and Eric~P Xing.
\newblock Variance reduction in stochastic gradient langevin dynamics.
\newblock {\em Advances in neural information processing systems}, 29, 2016.

\bibitem[DJS{\etalchar{+}}19]{djssw19}
Huaian Diao, Rajesh Jayaram, Zhao Song, Wen Sun, and David Woodruff.
\newblock Optimal sketching for kronecker product regression and low rank
  approximation.
\newblock {\em Advances in neural information processing systems},
  32:4737--4748, 2019.

\bibitem[DK19]{dk19}
Arnak~S Dalalyan and Avetik Karagulyan.
\newblock User-friendly guarantees for the langevin monte carlo with inaccurate
  gradient.
\newblock {\em Stochastic Processes and their Applications},
  129(12):5278--5311, 2019.

\bibitem[DL21]{dl21}
Zhiyan Ding and Qin Li.
\newblock Langevin monte carlo: random coordinate descent and variance
  reduction.
\newblock {\em J. Mach. Learn. Res.}, 22:205--1, 2021.

\bibitem[DMM06]{dmm06}
Petros Drineas, Michael~W. Mahoney, and S.~Muthukrishnan.
\newblock Sampling algorithms for l2 regression and applications.
\newblock In {\em Proceedings of the Seventeenth Annual ACM-SIAM Symposium on
  Discrete Algorithm}, SODA '06, page 1127–1136, USA, 2006. Society for
  Industrial and Applied Mathematics.

\bibitem[DSSW18]{dssw18}
Huaian Diao, Zhao Song, Wen Sun, and David Woodruff.
\newblock Sketching for kronecker product regression and p-splines.
\newblock In {\em International Conference on Artificial Intelligence and
  Statistics}, pages 1299--1308. PMLR, 2018.

\bibitem[DWZ23]{dwz23}
Ran Duan, Hongxun Wu, and Renfei Zhou.
\newblock Faster matrix multiplication via asymmetric hashing.
\newblock In {\em FOCS}, 2023.

\bibitem[FB86]{fb86}
Z~Furedi and I~Barany.
\newblock Computing the volume is difficult.
\newblock In {\em Proceedings of the Eighteenth Annual ACM Symposium on Theory
  of Computing}, STOC '86, page 442–447, New York, NY, USA, 1986. Association
  for Computing Machinery.

\bibitem[FLPS22]{flps22}
Maryam Fazel, Yin~Tat Lee, Swati Padmanabhan, and Aaron Sidford.
\newblock Computing lewis weights to high precision.
\newblock In {\em Proceedings of the 2022 Annual ACM-SIAM Symposium on Discrete
  Algorithms (SODA)}, pages 2723--2742. SIAM, 2022.

\bibitem[Gal24]{lg24}
Francois~Le Gall.
\newblock Faster rectangular matrix multiplication by combination loss
  analysis.
\newblock In {\em Proceedings of the Thirty-Fifth Annual ACM-SIAM Symposium on
  Discrete Algorithms}, SODA'24, 2024.

\bibitem[GKV24]{gkv24}
Khashayar Gatmiry, Jonathan Kelner, and Santosh~S. Vempala.
\newblock Sampling polytopes with riemannian hmc: Faster mixing via the lewis
  weights barrier.
\newblock In Shipra Agrawal and Aaron Roth, editors, {\em Proceedings of Thirty
  Seventh Conference on Learning Theory}, volume 247 of {\em Proceedings of
  Machine Learning Research}, pages 1796--1881. PMLR, 30 Jun--03 Jul 2024.

\bibitem[GU18]{lgu18}
Fran\c{c}ois~Le Gall and Florent Urrutia.
\newblock Improved rectangular matrix multiplication using powers of the
  coppersmith-winograd tensor.
\newblock In {\em Proceedings of the Twenty-Ninth Annual ACM-SIAM Symposium on
  Discrete Algorithms}, SODA '18, page 1029–1046, 2018.

\bibitem[HJ90]{hj91}
Roger~A. Horn and Charles~R. Johnson.
\newblock {\em Matrix Analysis}.
\newblock Cambridge University Press, 1990.

\bibitem[HJS{\etalchar{+}}22]{hjs+22}
Baihe Huang, Shunhua Jiang, Zhao Song, Runzhou Tao, and Ruizhe Zhang.
\newblock Solving sdp faster: A robust ipm framework and efficient
  implementation.
\newblock In {\em FOCS}, 2022.

\bibitem[JKL{\etalchar{+}}20]{jkl+20}
Haotian Jiang, Tarun Kathuria, Yin~Tat Lee, Swati Padmanabhan, and Zhao Song.
\newblock A faster interior point method for semidefinite programming.
\newblock In {\em 2020 IEEE 61st annual symposium on foundations of computer
  science (FOCS)}, pages 910--918. IEEE, 2020.

\bibitem[JL84]{jl84}
William~B Johnson and Joram Lindenstrauss.
\newblock Extensions of lipschitz mappings into a hilbert space.
\newblock {\em Contemporary mathematics}, 26(189-206):1, 1984.

\bibitem[JLS22]{jls22}
Arun Jambulapati, Yang~P. Liu, and Aaron Sidford.
\newblock Improved iteration complexities for overconstrained p-norm
  regression.
\newblock In {\em Proceedings of the 54th Annual ACM SIGACT Symposium on Theory
  of Computing}, STOC 2022, 2022.

\bibitem[JLSW20]{jlsw20}
Haotian Jiang, Yin~Tat Lee, Zhao Song, and Sam Chiu-wai Wong.
\newblock An improved cutting plane method for convex optimization,
  convex-concave games, and its applications.
\newblock In {\em Proceedings of the 52nd Annual ACM SIGACT Symposium on Theory
  of Computing}, pages 944--953, 2020.

\bibitem[JSWZ21]{jswz21}
Shunhua Jiang, Zhao Song, Omri Weinstein, and Hengjie Zhang.
\newblock A faster algorithm for solving general lps.
\newblock In {\em Proceedings of the 53rd Annual ACM SIGACT Symposium on Theory
  of Computing}, pages 823--832, 2021.

\bibitem[JZ13]{jz13}
Rie Johnson and Tong Zhang.
\newblock Accelerating stochastic gradient descent using predictive variance
  reduction.
\newblock {\em Advances in neural information processing systems}, 26, 2013.

\bibitem[KLS95]{kls95}
R.~Kannan, L.~Lov\'{a}sz, and M.~Simonovits.
\newblock Isoperimetric problems for convex bodies and a localization lemma.
\newblock {\em Discrete Comput. Geom.}, 1995.

\bibitem[KN12]{kn12}
Ravindran Kannan and Hariharan Narayanan.
\newblock Random walks on polytopes and an affine interior point method for
  linear programming.
\newblock {\em Mathematics of Operations Research}, 37(1):1--20, 2012.

\bibitem[KV06]{kv06}
Adam~Tauman Kalai and Santosh Vempala.
\newblock Simulated annealing for convex optimization.
\newblock {\em Mathematics of Operations Research}, 31(2):253--266, 2006.

\bibitem[KV24]{kv23}
Yunbum Kook and Santosh~S. Vempala.
\newblock Gaussian cooling and {D}ikin walks: {T}he interior-point method for
  logconcave sampling.
\newblock In Shipra Agrawal and Aaron Roth, editors, {\em Proceedings of Thirty
  Seventh Conference on Learning Theory}, volume 247 of {\em Proceedings of
  Machine Learning Research}, pages 3137--3240. PMLR, 30 Jun--03 Jul 2024.

\bibitem[LDFU13]{ldfu13}
Yichao Lu, Paramveer Dhillon, Dean~P Foster, and Lyle Ungar.
\newblock Faster ridge regression via the subsampled randomized hadamard
  transform.
\newblock In {\em Advances in neural information processing systems}, pages
  369--377, 2013.

\bibitem[LLV20]{llv20}
Aditi Laddha, Yin~Tat Lee, and Santosh Vempala.
\newblock Strong self-concordance and sampling.
\newblock In {\em Proceedings of the 52nd Annual ACM SIGACT Symposium on Theory
  of Computing (STOC)}, pages 1212--1222, 2020.

\bibitem[LM00]{lm00}
Beatrice Laurent and Pascal Massart.
\newblock Adaptive estimation of a quadratic functional by model selection.
\newblock {\em Annals of Statistics}, pages 1302--1338, 2000.

\bibitem[LMW{\etalchar{+}}24]{lmwrb24}
Yingyu Lin, Yian Ma, Yu-Xiang Wang, Rachel~Emily Redberg, and Zhiqi Bu.
\newblock Tractable {MCMC} for private learning with pure and gaussian
  differential privacy.
\newblock In {\em The Twelfth International Conference on Learning
  Representations}, 2024.

\bibitem[LS93]{ls93}
L{\'a}szl{\'o} Lov{\'a}sz and Mikl{\'o}s Simonovits.
\newblock Random walks in a convex body and an improved volume algorithm.
\newblock {\em Random structures \& algorithms}, 4(4):359--412, 1993.

\bibitem[LS14]{ls14}
Yin~Tat Lee and Aaron Sidford.
\newblock Path finding methods for linear programming: Solving linear programs
  in {\~{o}}(sqrt(rank)) iterations and faster algorithms for maximum flow.
\newblock In {\em 55th {IEEE} Annual Symposium on Foundations of Computer
  Science, {FOCS} 2014, Philadelphia, PA, USA, October 18-21, 2014}, pages
  424--433, 2014.

\bibitem[LS19]{ls19}
Yin~Tat Lee and Aaron Sidford.
\newblock Solving linear programs with sqrt (rank) linear system solves.
\newblock {\em arXiv preprint arXiv:1910.08033}, 2019.

\bibitem[LSW15]{lsw15}
Yin~Tat Lee, Aaron Sidford, and Sam Chiu-wai Wong.
\newblock A faster cutting plane method and its implications for combinatorial
  and convex optimization.
\newblock In {\em Foundations of Computer Science (FOCS), 2015 IEEE 56th Annual
  Symposium on}, pages 1049--1065. IEEE, 2015.

\bibitem[LSZ19]{lsz19}
Yin~Tat Lee, Zhao Song, and Qiuyi Zhang.
\newblock Solving empirical risk minimization in the current matrix
  multiplication time.
\newblock In {\em Conference on Learning Theory}, pages 2140--2157. PMLR, 2019.

\bibitem[LV03]{lv03}
L{\'a}szl{\'o} Lov{\'a}sz and Santosh Vempala.
\newblock Hit-and-run is fast and fun.
\newblock {\em preprint, Microsoft Research}, 2003.

\bibitem[LV06]{lv06}
L{\'a}szl{\'o} Lov{\'a}sz and Santosh Vempala.
\newblock Simulated annealing in convex bodies and an ${O}^*(n^4)$ volume
  algorithm.
\newblock {\em Journal of Computer and System Sciences}, 72(2):392--417, 2006.

\bibitem[LV07]{lv07}
L{\'a}szl{\'o} Lov{\'a}sz and Santosh Vempala.
\newblock The geometry of logconcave functions and sampling algorithms.
\newblock {\em Random Structures \& Algorithms}, 30(3):307--358, 2007.

\bibitem[LV17]{lv17}
Yin~Tat Lee and Santosh~S. Vempala.
\newblock Geodesic walks in polytopes.
\newblock In {\em Proceedings of the 49th Annual ACM SIGACT Symposium on Theory
  of Computing}, STOC 2017, 2017.

\bibitem[LV18]{lv18}
Yin~Tat Lee and Santosh~S Vempala.
\newblock Convergence rate of riemannian hamiltonian monte carlo and faster
  polytope volume computation.
\newblock In {\em Proceedings of the 50th Annual ACM SIGACT Symposium on Theory
  of Computing}, pages 1115--1121, 2018.

\bibitem[LY21]{ly18}
Yin~Tat Lee and Man–Chung Yue.
\newblock Universal barrier is $n$-self-concordant.
\newblock {\em Mathematics of Operations Research}, 46(3):1129--1148, 2021.

\bibitem[LZC{\etalchar{+}}19]{lzc+19}
Zhize Li, Tianyi Zhang, Shuyu Cheng, Jun Zhu, and Jian Li.
\newblock Stochastic gradient hamiltonian monte carlo with variance reduction
  for bayesian inference.
\newblock {\em Machine Learning}, 108(8):1701--1727, 2019.

\bibitem[MV22]{mv22_neurips}
Oren Mangoubi and Nisheeth Vishnoi.
\newblock Sampling from log-concave distributions with infinity-distance
  guarantees.
\newblock In S.~Koyejo, S.~Mohamed, A.~Agarwal, D.~Belgrave, K.~Cho, and A.~Oh,
  editors, {\em Advances in Neural Information Processing Systems}, volume~35,
  pages 12633--12646. Curran Associates, Inc., 2022.

\bibitem[MV23]{mv22}
Oren Mangoubi and Nisheeth~K Vishnoi.
\newblock Sampling from structured log-concave distributions via a
  soft-threshold dikin walk.
\newblock In {\em Advances in Neural Information Processing Systems},
  NeurIPS'23, 2023.

\bibitem[MV24]{mv24}
Oren Mangoubi and Nisheeth~K. Vishnoi.
\newblock Faster sampling from log-concave densities over polytopes via
  efficient linear solvers.
\newblock In {\em The Twelfth International Conference on Learning
  Representations}, 2024.

\bibitem[Nar16]{n16}
Hariharan Narayanan.
\newblock Randomized interior point methods for sampling and optimization.
\newblock {\em Ann. Appl. Probab.}, 2016.

\bibitem[NN89]{nn89}
Yurii Nesterov and Arkadii Nemirovskii.
\newblock Self-concordant functions and polynomial-time methods in convex
  programming.
\newblock {\em USSR Academy of Sciences}, 1989.

\bibitem[NN94]{nn94}
Yurii Nesterov and Arkadii Nemirovskii.
\newblock {\em Interior-point polynomial algorithms in convex programming}.
\newblock SIAM, 1994.

\bibitem[NN13]{nn13}
Jelani Nelson and Huy~L Nguy{\^e}n.
\newblock {OSNAP}: Faster numerical linear algebra algorithms via sparser
  subspace embeddings.
\newblock In {\em 54th Annual IEEE Symposium on Foundations of Computer Science
  (FOCS)}, pages 117--126. IEEE, 2013.

\bibitem[NR17]{nr17}
Hariharan Narayanan and Alexander Rakhlin.
\newblock Efficient sampling from time-varying log-concave distributions.
\newblock {\em The Journal of Machine Learning Research}, 18(1):4017--4045,
  2017.

\bibitem[QSZZ23]{qszz23}
Lianke Qin, Zhao Song, Lichen Zhang, and Danyang Zhuo.
\newblock An online and unified algorithm for projection matrix vector
  multiplication with application to empirical risk minimization.
\newblock In {\em AISTATS}, 2023.

\bibitem[Ren88]{r88}
James Renegar.
\newblock A polynomial-time algorithm, based on newton's method, for linear
  programming.
\newblock {\em Math. Program.}, 1988.

\bibitem[RRT17]{rrt17}
Maxim Raginsky, Alexander Rakhlin, and Matus Telgarsky.
\newblock Non-convex learning via stochastic gradient langevin dynamics: a
  nonasymptotic analysis.
\newblock In {\em Conference on Learning Theory}, pages 1674--1703. PMLR, 2017.

\bibitem[SS11]{ss11}
Daniel~A Spielman and Nikhil Srivastava.
\newblock Graph sparsification by effective resistances.
\newblock {\em SIAM Journal on Computing}, 40(6):1913--1926, 2011.

\bibitem[SSZ13]{sz13}
Shai Shalev-Shwartz and Tong Zhang.
\newblock Stochastic dual coordinate ascent methods for regularized loss
  minimization.
\newblock {\em Journal of Machine Learning Research}, 14(1), 2013.

\bibitem[SV16]{sv16}
Sushant Sachdeva and Nisheeth~K Vishnoi.
\newblock The mixing time of the dikin walk in a polytope—a simple proof.
\newblock {\em Operations Research Letters}, 44(5):630--634, 2016.

\bibitem[SWYZ21]{swyz21}
Zhao Song, David Woodruff, Zheng Yu, and Lichen Zhang.
\newblock Fast sketching of polynomial kernels of polynomial degree.
\newblock In {\em International Conference on Machine Learning}, pages
  9812--9823. PMLR, 2021.

\bibitem[SWZ17]{swz17}
Zhao Song, David~P Woodruff, and Peilin Zhong.
\newblock Low rank approximation with entrywise l1-norm error.
\newblock In {\em Proceedings of the 49th Annual ACM SIGACT Symposium on Theory
  of Computing}, pages 688--701, 2017.

\bibitem[SWZ19]{swz19}
Zhao Song, David~P Woodruff, and Peilin Zhong.
\newblock Relative error tensor low rank approximation.
\newblock In {\em Proceedings of the Thirtieth Annual ACM-SIAM Symposium on
  Discrete Algorithms}, pages 2772--2789. SIAM, 2019.

\bibitem[SY21]{sy21}
Zhao Song and Zheng Yu.
\newblock Oblivious sketching-based central path method for solving linear
  programming problems.
\newblock In {\em 38th International Conference on Machine Learning (ICML)},
  2021.

\bibitem[Tro11]{t11}
Joel~A Tropp.
\newblock Improved analysis of the subsampled randomized hadamard transform.
\newblock {\em Advances in Adaptive Data Analysis}, 3(01n02):115--126, 2011.

\bibitem[Vai89]{v89_cp}
Pravin~M Vaidya.
\newblock A new algorithm for minimizing convex functions over convex sets.
\newblock In {\em 30th Annual Symposium on Foundations of Computer Science
  (FOCS)}, pages 338--343. IEEE Computer Society, 1989.

\bibitem[Vem05]{v05}
Santosh Vempala.
\newblock Geometric random walks: a survey.
\newblock {\em Combinatorial and computational geometry}, 52(573-612):2, 2005.

\bibitem[WFS15]{wfs15}
Yu-Xiang Wang, Stephen Fienberg, and Alex Smola.
\newblock Privacy for free: Posterior sampling and stochastic gradient monte
  carlo.
\newblock In Francis Bach and David Blei, editors, {\em Proceedings of the 32nd
  International Conference on Machine Learning}, volume~37 of {\em Proceedings
  of Machine Learning Research}, pages 2493--2502, Lille, France, 07--09 Jul
  2015. PMLR.

\bibitem[Wil12]{w12}
Virginia~Vassilevska Williams.
\newblock Multiplying matrices faster than coppersmith-winograd.
\newblock In {\em Proceedings of the forty-fourth annual ACM symposium on
  Theory of computing (STOC)}, pages 887--898. ACM, 2012.

\bibitem[Woo14]{w14}
David~P. Woodruff.
\newblock Sketching as a tool for numerical linear algebra.
\newblock {\em Foundations and Trends in Theoretical Computer Science},
  10(1--2):1--157, 2014.

\bibitem[WXXZ24]{wxxz23}
Virginia~Vassilevska Williams, Yinzhan Xu, Zixuan Xu, and Renfei Zhou.
\newblock New bounds for matrix multiplication: from alpha to omega.
\newblock In {\em Proceedings of the Thirty-Fifth Annual ACM-SIAM Symposium on
  Discrete Algorithms}, SODA'24, 2024.

\bibitem[ZXG18]{zxg18}
Difan Zou, Pan Xu, and Quanquan Gu.
\newblock Stochastic variance-reduced hamilton monte carlo methods.
\newblock In {\em International Conference on Machine Learning}, pages
  6028--6037. PMLR, 2018.

\end{thebibliography}
\bibliographystyle{plainnat}

\fi


\newpage 
\appendix
\section*{Appendix}
\paragraph{Roadmap.}

Since there are many technical details in the appendix, we provide a roadmap. The appendix can be divided into 5 parts: the first part conveys some preliminary information and states the sufficient conditions we require barrier functions to have, in Section~\ref{sec:preli} and~\ref{sec:assumption}. The second part proves the mixing time of the Dikin walk when barrier functions satisfy the conditions, and the proofs are divided into Section~\ref{sec:key} and~\ref{sec:correct}. The next part focuses on the runtime complexity of sampling from polytopes, including how to generate a spectral approximation of the Hessian, approximate leverage scores and Lewis weights to high precision, in Section~\ref{sec:leverage} and how to incorporate these algorithmic prototypes to implement an efficient sampling algorithm in Section~\ref{sec:time}. We dedicate Section~\ref{sec:sdp} to study the log-barrier for sampling from a spectrahedron, as it is relatively less explored before. Finally, we prove the convexity of the function $\log\det(H(x)+I_d)$ for log-barrier, volumetric barrier and in extension Lee-Sidford barrier in Section~\ref{sec:barrier} and~\ref{sec:vol}.

\section{Preliminaries}\label{sec:preli}

For any positive integer, we use $[n]$ to denote the set $\{1,2,\cdots,n\}$. For a vector $x \in \R^n$, we use $\| x \|_2$ to denote its $\ell_2$ norm, i.e., $ \| x \|_2 : = ( \sum_{i=1}^n x_i^2 )^{1/2}$. We use $\| x \|_1$ to denote its $\ell_1$ norm, $\| x \|_1 := \sum_{i=1}^n |x_i|$. We use $\| x \|_{\infty}$ to denote its $\ell_{\infty}$ norm, i.e., $\| x \|_{\infty} := \max_{i\in [n]} |x_i|$. For a random variable $X$, we use $\E[X]$ to denote its expectation. We use $\Pr[\cdot ]$ to denote probability.

We use ${\bf 0}_d$ to denote a length-$d$ vector where every entry is $0$. We use ${\bf 1}_d$ to denote a length-$d$ vector where every entry is $1$. We use $I_d$ to denote an identity matrix which has size $d \times d$ or simply $I$ when dimension is clear from context.

For a matrix $A$, we use $A^\top$ to denote the transpose of matrix $A$. 
For a square and non-singular matrix $A$, we use $A^{-1}$ to denote the inverse of matrix $A$. 
For a real square matrix $A$, we say it is positive definite (PD, i.e., $A \succ 0$) if for all vectors $x \in \R^n$ (except for ${\bf 0}_n$), we have $x^\top A x > 0$.
For a real square matrix $A$, we say it is positive semi-definite(PSD, i.e., $A \succeq 0$) if for all vectors $x \in \R^n$, we have $x^\top A x \geq 0$. For a square matrix, we use $\det(A)$ to denote the determinant of matrix $A$. For a matrix $A$, we use $\| A \|$ to denote its spectral norm, use $\| A \|_F$ to denote its Frobenius norm, i.e., $\| A \|_F: = ( \sum_{i=1}^n \sum_{j=1}^d A_{i,j}^2 )^{1/2}$ and use $\nnz(A)$ to denote the number of nonzero entries in $A$.

Given a function $f:\K\rightarrow \R$, we say it's convex if for any $x, y\in \K$, $f(x)\geq f(y)+\nabla f(y)^\top (x-y)$. We say it's $L$-Lipschitz if $|f(x)-f(y)|\leq L\cdot \|x-y\|_2$ for a fixed parameter $L > 0$.
 
For two distributions $P_1$ and $P_2$, we use $\TV(P_1,P_2)$ to denote the total variation (TV) distance between $P_1$ and $P_2$.

For a convex body $\K\subseteq \R^d$, we use ${\rm Int}(\K)$ to denote the interior of $\K$.

We use $\otimes$ to denote Kronecker product. We use $A \otimes_S B$ to denote $A \otimes I + I \otimes B$. Given a matrix $A$, we use $\vect(A)$ to denote its vectorization, since we will always apply $\vect(\cdot)$ to a symmetric matrix, whether the vetorization is row-major or column-major doesn't matter. We use $\circ$ to denote Hadamard or element-wise product.

For two vectors $x, y\in \R^d$, we use $\langle x, y\rangle=x^\top y$ to denote the standard Euclidean inner product over $\R^d$, and for two symmetric matrices $A, B\in \R^{d\times d}$, we use $\langle A, B\rangle=\tr[A^\top B]$ to denote the trace inner product.

In Section~\ref{sec:preli:fastmm}, we present several definitions and notations related to fast matrix multiplication. In Section~\ref{sec:preli:geometry}, we provide some backgrounds about convex geometry. In Section~\ref{sec:preli:probability}, we state several basic probability tools. In Section~\ref{sec:preli:fact}, we state several basic facts related to trace, and Kronecker product. In Section~\ref{sec:preli:barrier}, we present the standard notion about self-concordance barrier.

\subsection{Fast matrix multiplication}\label{sec:preli:fastmm}
\begin{definition}[Fast matrix multiplication]
Given three positive integers $a,b,c$, we use $\Tmat(a,b,c)$  to denote the time of multiplying an $a \times b$ matrix with another $b \times c$ matrix.
\end{definition}

For convenience, we also define the $\omega(\cdot,\cdot, \cdot)$ function as follows:
\begin{definition}[Fast matrix multiplication, an alternative notation]
Given $x,y,z$, we use $d^{\omega(x,y,z)}$ to denote the time of multiplying a $d^x \times d^y$ matrix with another $d^y \times d^z$ matrix.
\end{definition}

\begin{lemma}[\cite{lgu18}]
We have the following bounds of $\omega(\cdot,\cdot,\cdot)$:
\begin{itemize}
    \item $\omega(1,1,1) = \omega$,
    \item $\omega(1,3,1) = 4.199712$ (see Table 3 in \cite{lgu18}).
\end{itemize}
\end{lemma}
Here, $\omega$ denotes the exponent of matrix multiplication, currently $\omega \approx 2.373$ \citep{w12,lgu18,aw21,dwz23,wxxz23,lg24}.

\subsection{Convex geometry}\label{sec:preli:geometry}

We define $B(x,R) \subset \R^d$ as 
\begin{align*}
    B(x,R) := \{ y \in \R^d ~:~ \| y - x \|_2 \leq R \}.
\end{align*}

Define polytope $\K$ as 
\begin{align*}
    \K: = \{ x \in \R^d ~:~ A x \leq b \}.
\end{align*}

Define spectrahedron ${\cal K}$ as 
\begin{align*}
    {\cal K} := & ~ \{x\in \R^d~:~\sum_{i=1}^d x_iA_i\succeq C \},
\end{align*}
where $A_1,\ldots,A_d,C\in \R^{n\times n}$ are symmetric matrices.

We will often use the notion of a \emph{Dikin ellipsoid}:
\begin{definition}[Dikin ellipsoid]\label{def:dikin_ellipsoid}
We define the Dikin ellipsoid $D_{\theta} \subset \R^d$ as follows
\begin{align*}
    D_{\theta} := \{ w \in \R^d ~:~ w^\top H^{-1}(\theta) w \leq 1 \}
\end{align*}
where $H(\theta) \in \R^{d \times d}$ is the Hessian of self-concordant barrier at $\theta$/
\end{definition}

We define the standard cross-ratio distance (see Definition E.1 in \cite{mv22} for example).
\begin{definition}[Cross-ratio distance]
\label{def:cross_ratio_dist}

Let $u,v\in \K$. If $u\neq v$, let $p, q$ be two endpoints of the chord in $\K$ which passes through $u$ and $v$ such that the four points lie in the order of $p, u, v, q$, let 
\begin{align*}
    \sigma(u,v) := & ~ \frac{\|u-v\|_2\cdot \|p-q\|_2}{\|p-u\|_2\cdot \|v-q\|_2}.
\end{align*}

We additionally set $\sigma(u,v)=0$ for $u=v$.

For two subsets $S_1,S_2\subseteq \K$, we define
\begin{align*}
    \sigma(S_1, S_2) := & ~ \min\{\sigma(u,v): u\in S_1, v\in S_2 \}.
\end{align*}
\end{definition}

We state the standard isoperimetric inequality for cross-ratio distance.
\begin{lemma}[\cite{lv03}]\label{lem:lv03_hit_and_run}
Let $\pi : \R^d \rightarrow \R$ be a log-concave density, with support on a convex body $\K$. Then for any partition of $\R^d$ into measurable sets $S_1, S_2, S_3$, the induced measure $\pi^*$ satisfies
\begin{align*}
    \pi^*(S_3) \geq \sigma(S_1, S_2) \cdot \pi^*(S_1) \cdot \pi^*(S_2).
\end{align*}
\end{lemma}

\subsection{Probability tools}\label{sec:preli:probability}

We state some useful concentration inequalities.

\begin{lemma}[Lemma 1 on page 1325 of 
\cite{lm00}]\label{lem:chi_square_tail}
Let $X \sim {\cal X}_k^2$ be a chi-squared distributed random variable with $k$ degrees of freedom. Each random variable has zero mean and $\sigma^2$ variance. Then
\begin{align*}
\Pr[ X - k \sigma^2 \geq ( 2 \sqrt{kt} + 2t ) \sigma^2 ] \leq \exp (-t), \\
\Pr[ k \sigma^2 - X \geq 2 \sqrt{k t} \sigma^2 ] \leq \exp(-t).
\end{align*}
\end{lemma}

\begin{lemma}[Matrix Chernoff bound~\citep{t11}]
\label{lem:matrix_chernoff}
Let $X_1,\ldots,X_s$ be independent copies of a symmetric random matrix $X\in \R^{d\times d}$ with $\E[X]=0$, $\|X\|\leq \gamma$ almost surely and $\|\E[X^\top X] \|\leq \sigma^2$. Let $W=\frac{1}{s}\sum_{i\in [s]}X_i$. For any $\epsilon\in (0,1)$, 
\begin{align*}
    \Pr[\|W\|\geq \epsilon]\leq & ~ 2d\cdot \exp\left(-\frac{s\epsilon^2}{\sigma^2+{\gamma\epsilon}/{3}} \right).
\end{align*}
\end{lemma}

\subsection{Basic algebra facts}\label{sec:preli:fact}

\begin{lemma}\label{lem:matrix_trace_psd}
Let $A$ and $B$ be $m \times m$ matrices. Then we have
\begin{itemize}
    \item $\tr[AB] = \tr[BA]$.
    \item If $A$ is symmetric, then $\tr[AB]= \tr[A B^\top]$.
    \item If $A$ and $B$ are PSD, then $\langle A, B \rangle \geq 0$, and $\langle A, B \rangle = 0$ if and only if $A B = {\bf 0}_{m \times m}$.
    \item If $A \succeq 0$ and $B \succeq C$, then $\langle A, B \rangle \geq \langle A , C \rangle$.
\end{itemize}
\end{lemma}

\begin{lemma}\label{lem:matrix_otimes_oplus}
Let $A,B,C,D$ be conforming matrices. Then
\begin{itemize}
    \item $(A \otimes B) (C \otimes D) = AC \otimes BD$.
    \item $(A \otimes_S B) (C \otimes_S D) = \frac{1}{2} ( AC \otimes_S BD + AD \otimes_S BC )$.
    \item $(A \otimes B)^\top = A^\top \otimes B^\top$.
    \item If $A$ and $B$ are non-singular, then $A \otimes B$ is non-singular, and $(A \otimes B)^{-1} = A^{-1} \otimes B^{-1}$.
    \item $\vect(ABC) = (C^\top \otimes A) \cdot \vect(B)$.
\end{itemize}
\end{lemma}

\subsection{Self-concordant barrier}\label{sec:preli:barrier}

We provide a standard definition about self-concordance barrier,
\begin{definition}[Self-concordant barrier]
\label{def:self_concordant_barrier}
A real-valued function $F: \mathrm{Int}(\K) \rightarrow \R$, is a
regular self-concordant barrier if it satisfies the conditions stated below. For convenience, if $x \not \in \mathrm{Int}(\K)$, we define $F(x) = \infty$.
\begin{enumerate} \item (Convex, Smooth) $F$ is a convex thrice continuously differentiable function on $\mathrm{Int}(\K)$. \item (Barrier) For every
sequence of points $\{x_i\} \in \mathrm{Int}(\K)$ converging to a point $x \not \in \mathrm{Int}(\K)$,  $\lim_{i \rightarrow \infty} f(x_i) = \infty$. \item (Differential
Inequalities)    For all $h \in \R^d$ and all $x \in \mathrm{Int}(\K)$, the following inequalities hold.

\begin{enumerate}
\item $D^2 F(x)[h, h]$ is $2$-Lipschitz continuous with respect to the local norm, which is equivalent to
$$D^3 F(x)[h, h, h] \leq 2 (D^2 F(x)[h, h])^{\frac{3}{2}}.$$ 
\item $F(x)$ is $\nu$-Lipschitz continuous with respect to the local norm defined by $F$,
$$\|\nabla_h F(x)\|_2^2 \leq \nu \cdot \|h\|^2_{H(x)}.$$ We call the smallest positive integer $\nu$ for which this holds, \emph{the self-concordance  parameter} of the barrier. 
\end{enumerate} 
\end{enumerate}
\end{definition}

\section{Sufficient Conditions for Log-concave Sampling}\label{sec:assumption}

To prove that our algorithm works for log-concave distribution, we state three key assumptions (see Section~\ref{sec:assumption:convexity_variance}) which are the sufficient conditions to prove the mixing rate of log-concave distribution, for different barrier functions.  

\subsection{Conditions: \texorpdfstring{$\ov\nu$}{}-symmetry, convexity and bounded local norm}\label{sec:assumption:convexity_variance}

We state three key conditions in order to lower bound the conductance for log-concave sampling.

\begin{assumption} \label{ass:assumptions}
Given a convex body $\K$, let $H:\K \rightarrow \R^{d\times d}$ be a self-concordant matrix function, we assume that
\begin{enumerate}[label=(\roman*)]
    \item {\bf $\ov\nu$-symmetry.} For any $x\in \K$, we have $E_x(1)\subseteq \K\cap (2x-\K)\subseteq E_x(\sqrt{\ov\nu})$ where $E_x(r)=\{y\in \R^d: (y-x)^\top H(x) (y-x)\leq r^2 \}$. \label{ass:nu_symmetry}
   
    \item {\bf Convexity.} Let $F:\K\rightarrow \R$ be defined as $F(x)=\log(\det(H(x)+I_d))$, then $F$ is convex in $x$. \label{ass:convex}
    \item {\bf Bounded local norm.} Let $\nabla \log(\det(H(x)))$ denote the gradient of $\log(\det(H(x)))$ in $x$. For any $x\in \K$, we have
    \begin{align*}
        \|(H(x))^{-1/2} \cdot \nabla \log(\det(H(x)))\|_2^2 
        \leq & ~  \wt O(d).
\end{align*} \label{ass:bound_variance}
\end{enumerate}
\end{assumption}
For the $\ov\nu$-symmetry assumption, we will prove that for barriers of the concern (log-barrier, Lee-Sidford barrier for polytopes and log-barrier for spectrahedra), $\ov\nu=\nu$. This characteristic is proved in~\cite{llv20} for polytopes, and we prove that for spectrahedra in Section~\ref{sec:sdp}. For convexity, due to the large amount of calculations, we defer to Section~\ref{sec:barrier} and~\ref{sec:vol}. For bounded local norm, the results for polytopes are similarly shown in~\cite{llv20}, and we will prove for log-barrier over spectrahedra.

\subsection{Local PSD approximation for any self-concordant matrix function}

\label{sec:correct:local_psd}

In this section, we prove a generalization of the matrix function self-concordance to regularized matrix function. When $H$ is Hessian of the log barrier, this fact was proved in~\cite[Lemma E.3]{mv22}. We generalize this fact to general barriers.
\begin{lemma}
\label{lem:related_local_norm}
Let $\alpha \in (0,1)$.
Let $H: \K\to \R$ be a self-concordant matrix function and $\Phi(u) := \alpha^{-1} H(u) + \eta^{-1} I_d$.
For any $u,v \in \K$ such that $\| u - v \|_{\Phi(u)} \leq \frac{1}{ \alpha^{1/2} }$ we have
\begin{align*}
    (1-\alpha^{1/2} \cdot \| u - v \|_{\Phi(u)} )^2 \cdot \Phi(v) \preceq \Phi(u) \preceq (1+\alpha^{1/2}  \cdot \| u - v \|_{\Phi(u)} )^2 \cdot \Phi(v).
\end{align*}
\end{lemma}
\begin{proof}
    By \cite[Lemma 1.1]{llv20}, we have
    \begin{align*}
    (1- \| u - v \|_{H(u)} )^2 \cdot H(v) \preceq H(u) \preceq (1 +  \| u - v \|_{H(u)} )^2 \cdot H(v).
    \end{align*}
    Because $H(u) \preceq \alpha \cdot \Phi(u)$, we get 
    \begin{align*}
    (1- \alpha^{1/2} \cdot \| u - v \|_{\Phi(u)} )^2 \cdot H(v) \preceq H(u) \preceq (1 + \alpha^{1/2} \cdot \| u - v \|_{\Phi(u)} )^2 \cdot H(v).
    \end{align*}
    Using $\Phi(u) = \alpha^{-1} H(u) + \eta^{-1} I_d$ again we get 
    \begin{align*}
    (1- \alpha^{1/2} \cdot \| u - v \|_{\Phi(u)} )^2 \cdot \Phi(v) \preceq \Phi(u) \preceq (1 + \alpha^{1/2} \cdot \| u - v \|_{\Phi(u)} )^2 \cdot \Phi(v). 
    \end{align*}
    This completes the proof.
\end{proof}

\subsection{Bounded local norm for regularized Hessian function}
\label{sec:correct:variance}

We note that the bounded local norm condition generally holds for standard barrier functions, but in our core argument, we will instead rely on the bounded local norm of the \emph{regularized Hessian function}. We prove that the bounded local norm condition implies the bounded local norm on the regularized Hessian function.

\begin{lemma}
Let $H:\K\rightarrow \R^{d\times d}$ be a matrix function, define $F(x)=\log\det(H(x)+L^2I_d)$ for $L\in \R$, then we have the following inequality:
\begin{align*}
    \|(H(x)+L^2I_d)^{-1/2}\nabla F(x)\|_2^2 \leq & ~ \|H(x)^{-1/2}\nabla \log\det(H(x))\|_2^2.
\end{align*}
\end{lemma}

\begin{proof}
Let $H(x)=U\Lambda U^\top$ be the eigendecomposition of the matrix $H(x)$, let $y=U^\top \nabla H(x)$, then
\begin{align*}
    \|(H(x)+L^2I_d)^{-1/2}\nabla F(x)\|_2^2= & ~ \|(H(x)+L^2I_d)^{-3/2}\nabla H(x)\|_2^2 \\
    = & ~ \|U(\Lambda+L^2I_d)^{-3/2} U^\top \nabla H(x)\|_2^2 \\
    = & ~ \|(\Lambda+L^2I_d)^{-3/2}y\|_2^2,
\end{align*}
the given condition implies
\begin{align*}
    \|H(x)^{-1/2}\nabla \log\det(H(x))\|_2^2 = & ~ \|H(x)^{-3/2} \nabla H(x)\|_2^2 \\
    = & ~ \|U\Lambda^{-3/2}U^\top \nabla H(x)\|_2^2 \\
    = & ~ \|\Lambda^{-3/2} y\|_2^2.
\end{align*}
As $L^2\geq 0$, we can expand the regularized squared $\ell_2$ norm as 
\begin{align*}
    \|(\Lambda+L^2I_d)^{-3/2}y\|_2^2 = & ~ \sum_{i=1}^d \frac{1}{(\lambda_i+L^2)^3} y_i^2 \\
    \leq & ~ \sum_{i=1}^d \frac{1}{\lambda_i^3} y_i^2 \\
    = & ~ \|\Lambda^{-3/2}y\|_2^2.
\end{align*}
This completes the proof.
\end{proof}
\section{Key Tools for Robust Sampling}\label{sec:key}

In this section, we list several tools which can be shared in the core proofs of all the main theorems. This section is organized as follows. In Section~\ref{sec:key:gaussian_concentration}, we provide a tool for Gaussian concentration.  In Section~\ref{sec:key:psd_implies_determinant}, we show that PSD approximation implies the determinant approximation (by losing a factor of $d$). One of the major idea in this work is instead of using exact accept probability, we will only need to use approximate accept probability. In Section~\ref{sec:key:approximate_accept_probability}, we prove several useful robust properties for approximate accept probability. In Section~\ref{sec:key:tv_between_exact_and_approximate_general}, we present a general lemma for bounding the TV distance between exact process and approximate process. In Section~\ref{sec:key:tv_between_exact_and_approximate_special}, we provide some specific choices for parameters and then bound the TV distance. In Section~\ref{sec:key:lower_bound_on_conductance_definition}, we provide a definition which will be used in proof of lower bound on conductance. In Section~\ref{sec:key:lower_bound_on_conductance_lemma}, we show how to lower bound the conductance by cross ratio.

\subsection{Gaussian concentration lemma}\label{sec:key:gaussian_concentration}

\begin{lemma}\label{lem:xi_bound}
Let $\xi\sim \N(0, I_d)$, then for any $t>\sqrt{2d}$, we have
\begin{align*}
    \Pr[\|\xi\|_2\geq t] \leq & ~ \exp(-(t^2-d)/8).
\end{align*}
\end{lemma}

\begin{proof}
We consider the squared $\ell_2$ norm of $\xi$: $\|\xi\|_2^2=\sum_{i=1}^d \xi_i^2$, which is a $\chi^2$ distribution with degree of freedom $d$. By Lemma~\ref{lem:chi_square_tail}, we know that

\begin{align*}
    \Pr[\|\xi\|_2^2\geq 2\sqrt{d}k+2k^2+d] \leq & ~ \exp(-k^2),
\end{align*}
set $k^2=(t^2-d)/8$, we obtain the desired bound.
\end{proof}

\subsection{Spectral approximation to determinant approximation}\label{sec:key:psd_implies_determinant}

\begin{lemma}\label{lem:det_approximation}
Let $\epsilon_{\Phi} \in (0,1)$. Given four psd matrices satify the following conditions 
\begin{align*}
    (1-\epsilon_{ \Phi}) \Phi(x) \preceq \wt \Phi(x) \preceq (1+\epsilon_{\Phi}) \Phi(x), \\
    (1-\epsilon_{\Phi})\Phi(z) \preceq \wh \Phi(z) \preceq (1+\epsilon_{\Phi}) \Phi(z).
\end{align*}
Then we have
\begin{align*}
    ( 1 - 10 \cdot \epsilon_{\Phi} \cdot d ) \cdot \frac{\det(\Phi(x))}{\det( \Phi(z))}\leq \frac{\det(\wt \Phi(x))}{\det(\wh{\Phi}(z))} \leq ( 1+ 10 \cdot \epsilon_{\Phi} \cdot d ) \cdot \frac{\det(\Phi(x))}{\det(\Phi(z))}
\end{align*}
\end{lemma}

\begin{proof}
Since $\wt \Phi(x) \in (1\pm \epsilon_{\Phi}) \Phi(x)$, we have the following spectral bound:
\begin{align*}
    (1-\epsilon_\Phi) I_d \preceq \Phi(x)^{-1/2} \wt \Phi(x) \Phi(x)^{-1/2} \preceq (1+\epsilon_\Phi) I_d,
\end{align*}
we can take the determinant of the middle term:
\begin{align*}
    \det(\Phi(x)^{-1/2} \wt \Phi(x) \Phi(x)^{-1/2}) = & ~ \det(\Phi(x)^{-1/2}) \det(\wt \Phi(x))  \det(\Phi(x)^{-1/2}) \\
    = & ~ \frac{\det(\wt \Phi(x))}{\det(\Phi(x))}  \\
    \in & ~ \det((1\pm\epsilon_\Phi) I_d) \\
    \in & ~ (1\pm\epsilon_\Phi)^d \\
    \in & ~ 1+10\cdot \epsilon_\Phi\cdot d.
\end{align*}
Thus, we complete the proof.
\end{proof}

\subsection{Approximate accept probability}\label{sec:key:approximate_accept_probability}

We start with some definitions:
\begin{definition}
 We define
 \begin{align*}
     \wh{G}_u(x) := & ~ \det( \wh{\Phi}(u) )^{1/2} \cdot \exp(-0.5 \| u - x \|_{\wh{\Phi}(u)}^2 ) \\
     G_x(u) := & ~ \det( \Phi(x) )^{1/2} \cdot \exp(-0.5 \| u - x \|_{\Phi(x)}^2 ) \\
     \wt{G}_x(u) := & ~ \det( \wt{\Phi}(x) )^{1/2} \cdot \exp(-0.5 \| u - x \|_{\wt{\Phi}(x)}^2 )
 \end{align*}
 \end{definition}
 
 \begin{definition}
We define
\begin{align*}
   P_{x,\Phi(x)}^{\mathrm{accept}} (u) := \E_{\wh{\Phi}(u)} \big[ \min\{1, \frac{\hat G_u(x) \exp(-f(u))}{ G_x(u) \exp(-f(x))}\} \big] \\
   P_{x,\wt{\Phi}(x)}^{\mathrm{accept}} (u) := \E_{\wh{\Phi}(u)} \big[ \min\{1, \frac{\hat G_u(x) \exp(-f(u))}{ \wt{G}_x(u) \exp(-f(x))}\} \big]
\end{align*}
We will use $P_x$, $\wt P_x$ as a shorthand for the above notations, and we call $P_x$ the \emph{exact process} and $\wt P_x$ the \emph{approximate process}.
\end{definition}

\begin{lemma}\label{lem:pro_accept_diff}
Fix $x$ and $\wt{\Phi}(x)$.
Suppose $\wt{\Phi}(x)$ is $(\epsilon_H,\delta_H)$-good approximation to $\Phi(x)$.
Then
\begin{align*}
\epsilon_p := \E_{u\sim \N(x, \Phi^{-1}(x))}[ |P_{x, \Phi(x)}^{\mathrm{accept}}(u) - P_{x, \wt{\Phi}(x)}^{\mathrm{accept}}(u)|]
\le 0.001.
\end{align*}
\end{lemma}

\begin{proof}
Note that
\begin{align*}
\frac{G_x(u)}{\wt{G}_x(u)} = \frac{(\det \Phi(x))^{1/2} \exp(-\frac 12 \|x-u\|^2_{\Phi(x)})}{(\det \wt{\Phi}(x))^{1/2} \exp(-\frac 12 \|x-u\|^2_{\wt{\Phi}(x)})}.
\end{align*}

Because $\wt{\Phi}(x)$ is a good approximation to $\Phi(x)$, by Lemma~\ref{lem:det_approximation} we have
\begin{align*}
(1-\epsilon_H)^{d/2} \le \frac{(\det \Phi(x))^{1/2}}{(\det \wt{\Phi}(x))^{1/2}} \le (1+\epsilon_H)^{d/2}
\end{align*}
and
\begin{align*}
(1-\epsilon_H)^2 \le \frac{\|x-u\|_{\Phi(x)}^2}{\|x-u\|_{\wt{\Phi}(x)}^2} \le (1+\epsilon_H)^2.
\end{align*}

By choosing step size sufficiently small, we have $\|u-x\|_{\Phi(x)}^2 < d $ 
with probability $0.9999$.
Therefore
\begin{align*}
  \Pr \Big[ | \|x-u\|^2_{\Phi(x)} - \|x-u\|^2_{\wt{\Phi}(x)}  | \le 3 \epsilon_H d \Big] \ge 0.9999.
\end{align*}
Then
\begin{align*}
\Pr \Big[ 1 - 3 \epsilon_H d \le \frac{\exp(- 0.5 \|x-u\|^2_{\Phi(x)})}{\exp(- 0.5 \|x-u\|^2_{\wt{\Phi}(x)})} \le 1 + 3 \epsilon_H d \Big] \ge 0.9999.
\end{align*}

For $\epsilon_H d$ small enough, we have
\begin{align*}
\Pr \Big[ 0.9999 \le \frac{P_{x, \Phi(x)}^{\mathrm{accept}}(u)}{P_{x, \wt{\Phi}(x)}^{\mathrm{accept}}(u)} \le 1.0001 \Big] \ge 0.9998,
\end{align*}
and therefore 
\begin{align*}
\Pr[| {P_{x, \Phi(x)}^{\mathrm{accept}}(u)}- {P_{x, \wt{\Phi}(x)}^{\mathrm{accept}}(u)}| \le 0.0001] \ge 0.9998.
\end{align*}

So
\begin{align*}
\E_{u\sim \N(x, \Phi^{-1}(x))}[ |P_{x, \Phi(x)}^{\mathrm{accept}}(u) - P_{x, \wt{\Phi}(x)}^{\mathrm{accept}}(u)|] 
\le&~ 0.9998 \cdot 0.0001 +  0.0002 \cdot 1\\
\le&~ 0.001. 
\end{align*}
This completes the proof.
\end{proof}

\subsection{TV distance between exact process and approximate process}\label{sec:key:tv_between_exact_and_approximate_general}

\begin{lemma}[TV distance between exact process and approximate process] \label{lem:general_exact_step_vs_approx_step}
 
Let $\wt{\Phi}$ denote the $(1\pm  \epsilon_H)$ approximation of $\Phi$. Let $\delta_H$ denote the failure probability.  Let $\epsilon_p$ be defined as Lemma~\ref{lem:pro_accept_diff}.
Let $P_x$ denote the exact process. Let $\wt{P}_x$ denote the approximate process.  If $x\in {\cal K}$, then
\begin{align*}
    \TV (P_x, \wt{P}_x) \le \delta_H + \epsilon_p + \sqrt{ d \epsilon_H}.
\end{align*}
\end{lemma}
\begin{proof}
We say $\wt \Phi(x)$ is good if it is an $(\epsilon_H, \delta_H)$-approximation to $\Phi(x)$, and $\wt \Phi(x)$ is bad otherwise. We can upper bound $\TV (P_x, \wt{P}_x) $ as follows:
\begin{align*}
    \TV (P_x, \wt{P}_x) 
    \le & ~
    \E[\TV(P_{x, \Phi(x)}, P_{x, \wt{\Phi}(x)})] \\
    = & ~  \E[\TV(P_{x, \Phi(x)}, P_{x, \wt{\Phi}(x)}) ~|~    \wt{\Phi}(x) \text{~is~bad} ] \\
    + & ~ \E[\TV(P_{x, \Phi(x)}, P_{x, \wt{\Phi}(x)}) ~|~  \wt{\Phi}(x) \text{~is~good} ] \\
    \le & ~ \Pr[\wt{\Phi}(x) \text{~is~bad}] + 
    \E[\TV(P_{x, \Phi(x)}, P_{x, \wt{\Phi}(x)}) ~|~ \wt{\Phi}(x) \text{~is~good}]
    \\
    \le & ~ \delta_H + \E[\TV(P_{x, \Phi(x)}, P_{x, \wt{\Phi}(x)}) ~|~ \wt{\Phi}(x) \text{~is~good}]. 
\end{align*}
where the third step follows from $\TV(\cdot, \cdot )\leq 1$, the last step follows from Lemma~\ref{lem:tilde_H}.

It remains to compute the second term in the above equation. By Pinsker's inequality, when $\wt{\Phi}(x)$ is good, we have
\begin{align}\label{eq:tv_gaussian_diff}
    & ~ \TV (\N(x, \wt{\Phi}^{-1}(x)), \N(x, \Phi^{-1}(x)))^2 \notag \\
    \le & ~ 0.5 \cdot D(\N(x, \wt{\Phi}^{-1}(x)) || \N(x, \Phi^{-1}(x))) \notag \\
    = & ~ 0.25 \cdot (\tr[ \wt{\Phi}(x)^{-1/2} \Phi(x) \wt{\Phi}(x)^{-1/2} ] -d + \log \frac{\det \wt \Phi(x) }{\det {\Phi(x)}}) \notag \\
    \le & ~ 0.25 \cdot ((1\pm \epsilon_H) d - d + \log (1\pm \epsilon_H)^d) \notag \\
    \le & ~ 0.5 \cdot d \epsilon_H. 
\end{align}
where the first step follows from Pinsker inequality.

Now, we can upper bound $\TV(P_x,\wt{P}_x)$ in the following sense,
\begin{align*}
    \TV (P_x, \wt{P}_x)
    \le &~ \delta_H + \E[\TV(P_{x, \Phi(x)}, P_{x, \wt{\Phi}(x)}) ~|~ \wt{\Phi}(x) \text{~is~good}] \\
    \le &~ \delta_H + \E_{\substack{u\sim \N(x, \Phi^{-1}(x)) \\ u'\sim \N(x, \wt{\Phi}^{-1}(x))\\ \text{any coupling}}}  [|P^{\mathrm{accept}}_{x, \Phi(x)} (u) - P^{\mathrm{accept}}_{x, \wt{\Phi}(x)} (u)| \cdot \mathbbm{1}\{u=u'\} + \mathbbm{1}\{u \ne u'\}]\\
    \leq & ~ \delta_H + \E_{u \sim {\cal N}(x, \Phi^{-1}(x))} [ | P_{x, \Phi(x)}^{\mathrm{accept}}(u) - P_{x, \wt{\Phi}(x)}^{\mathrm{accept}}(u) | ] + \TV( {\cal N}(x, \wt{\Phi}^{-1}(x)) , {\cal N}(x, \Phi^{-1}(x) ) ) \\
    = &~ \delta_H + \epsilon_p + \TV( {\cal N}(x, \wt{\Phi}^{-1}(x)) , {\cal N}(x, \Phi^{-1}(x) ) ) \\\
    \leq & ~ \delta_H + \epsilon_p + \sqrt{ d \epsilon_H}
\end{align*}
The forth step follows from definition of $\epsilon_p$, the fifth step follows from Eq.~\eqref{eq:tv_gaussian_diff}.  
\end{proof}

\subsection{TV distance between exact and approximate process: instantiation}\label{sec:key:tv_between_exact_and_approximate_special}

\begin{lemma}\label{lem:exact_step_vs_approx_step}
If $d\epsilon_H < 0.001$ and $\delta_H < 0.001$. Then, we have 
\begin{align*}
    \TV(P_x, \wt{P}_x) \leq 0.01
\end{align*}
\end{lemma}
\begin{proof}

We can upper bound it as follows
\begin{align*}
    \TV(P_x, \wt{P}_x) \leq & ~ \delta_H + \epsilon_p + \sqrt{ d \epsilon_H} \\
    \leq & ~ 0.001 + \epsilon_p + 0.001 \\
    \leq & ~ 0.01
\end{align*}
where the first step follows from Lemma~\ref{lem:general_exact_step_vs_approx_step}, the second step follows from choice $\epsilon_H, \delta_H$, and the last step follows from Lemma~\ref{lem:pro_accept_diff}.
\end{proof}

\subsection{Lower bound on conductance, definition}\label{sec:key:lower_bound_on_conductance_definition}

\begin{definition}\label{def:S_1'_S_2'}
Let $\beta \in (0,0.1)$ denote some fixed constant. Let $S_1 \subset \K$ and $S_2 := \K \backslash S_1$ and $\pi(S_1) \leq 1/2$.
We define $S_1'$ and $S_2'$ as follows 
\begin{align*} 
    S_1' := & ~ \{x \in S_1 : \wt{P}_x(S_2) \leq \beta \},\\
    S_2' := & ~ \{z \in S_2 : \wt{P}_z(S_1) \leq \beta \}.
\end{align*}
\end{definition}

\subsection{Lower bound on conductance, lemma}\label{sec:key:lower_bound_on_conductance_lemma}

\begin{lemma}[Lower bound the conductance by cross ratio]\label{lem:bound_conductance_by_sigma}
Let $S_1'$ and $S_2'$ be defined as Definition~\ref{def:S_1'_S_2'}. 
The conductance $\phi$ satisfies
\begin{align*}
    \phi \geq \frac{1}{16} \beta \cdot \sigma(S_1',S_2').
\end{align*}
\end{lemma}
\begin{proof} 
The proof follows the general format for conductance proofs for geometric Markov chains (see e.g. Section 5 of \cite{v05}).

Let $S_1 \subseteq \K$ and let $S_2 = \K \backslash S_1$. Let $S_1'$ and $S_2'$ be defined as Definition~\ref{def:S_1'_S_2'}. We define $S_3'$
\begin{align*}
    S_3' := (\K \backslash S_1')\backslash S_2'.
\end{align*}

By Lemma~\ref{lem:robust_lemma_5.12_in_mv22}, 
we have that

\begin{equation}\label{eq_h6}
   \int_{x \in S_1} \pi(x) \wt{P}_x(S_2) \mathrm{d}  x =    \int_{x \in S_2} \pi(x) \wt{P}_x(S_1) \mathrm{d}  x.
\end{equation}
Thus, by Lemma \ref{lem:lv03_hit_and_run}, we have 
\begin{equation}\label{eq_h7}
    \pi^\star(S_3') \geq \sigma(S_1', S_2') \pi^\star(S_1') \pi^\star(S_2').
\end{equation}

{\bf Case 1.} First, we assume that both $\pi^\star(S_1') \geq \frac{1}{4} \pi^\star(S_1)$ and $\pi^\star(S_2') \geq \frac{1}{4} \pi^\star(S_2)$.
In this case we have
\begin{align}\label{eq_h8}
    \int_{S_1} \wt{P}_x(S_2) \pi(x) \mathrm{d} x 
    = & ~ \frac{1}{2}\int_{S_1} \wt{P}_x(S_2) \pi(x) \mathrm{d} x + \frac{1}{2}\int_{S_2} \wt{P}_x(S_1) \pi(x) \mathrm{d} x \notag \\
    \geq & ~ \frac{\beta}{2} \cdot \pi^\star(S_3')\notag \\
    \geq & ~ \frac{\beta}{2} \cdot \sigma(S_1', S_2') \cdot \pi^\star(S_1') \pi^\star(S_2)\notag \\
    \geq & ~ \frac{\beta}{4} \cdot \sigma(S_1',S_2') \cdot \min(\pi^\star(S_1'),  \pi^\star(S_2'))\notag \\
    \geq & ~ \frac{\beta}{16} \cdot \sigma(S_1',S_2') \cdot \min(\pi^\star(S_1),   \pi^\star(S_2)).
\end{align}
where the first step follows from Eq.~\eqref{eq_h6}, the second step follows from Definition~\ref{def:S_1'_S_2'},  
the third step follows from Eq.~\eqref{eq_h7}, the fifth step follows from one of $\pi^\star(S_1')$ and $\pi^{\star}(S_2')$ is at least $1/2$.

{\bf Case 2.} 
Now suppose that instead either $\pi^\star(S_1') < \frac{1}{4} \pi^\star(S_1)$ or $\pi^\star(S_2') < \frac{1}{4} \pi^\star(S_2)$.

{\bf Case 2a.}
If  $\pi^\star(S_1') < \frac{1}{4} \pi^\star(S_1)$  then we have
\begin{align}\label{eq_h9}
\int_{S_1} \wt{P}_x(S_2) \pi(x) \mathrm{d} x 
{=} & ~ \frac{1}{2}\int_{S_1} \wt{P}_x(S_2) \pi(x) \mathrm{d} x + \frac{1}{2}\int_{S_2} \wt{P}_x(S_1) \pi(x) \mathrm{d} x \notag \\
\geq & ~ \frac{1}{2}\int_{S_1\backslash S_1'} \wt{P}_x(S_2) \pi(x) \mathrm{d}  x \notag \\
{\geq} & ~ \frac{1}{2}\cdot \frac{3}{4} \cdot \beta\pi^\star(S_1) \notag \\
\geq & ~ \frac{3}{8}\beta \min(\pi^\star(S_1),  \pi^\star(S_2)).
\end{align}
where the first step follows from Eq.~\eqref{eq_h6}, and the third step follows from Definition~\ref{def:S_1'_S_2'}.

{\bf Case 2b.}
Similarly, if  $\pi^\star(S_2') < \frac{1}{4} \pi^\star(S_2)$  we have
\begin{align}\label{eq_h10}
\int_{S_1} \wt{P}_x(S_2) \pi(x) \mathrm{d} x  = & ~  \frac{1}{2}\int_{S_1} \wt{P}_x(S_2) \pi(x) \mathrm{d} x + \frac{1}{2}\int_{S_2} \wt{P}_x(S_1) \pi(x) \mathrm{d} x\notag \\
\geq & ~ \frac{1}{2}\int_{S_2\backslash S_2'} \wt{P}_x(S_1) \pi(x) \mathrm{d} x\notag \\
  {\geq} & ~  \frac{1}{2}\cdot \frac{3}{4} \cdot \beta \pi^\star(S_2)\notag \\
   \geq & ~ \frac 38\beta \min(\pi^\star(S_1),  \pi^\star(S_2)).
\end{align}
where the first step follows from Eq.~\eqref{eq_h6}, and the third step follows from Definition~\ref{def:S_1'_S_2'}.

Therefore, Eq.~\eqref{eq_h8}, \eqref{eq_h9}, and \eqref{eq_h10} together imply that
\begin{equation} \label{eq_h11}
\frac{1}{\min(\pi^\star(S_1),  \pi^\star(S_2))}   \int_{S_1} \wt{P}_x(S_2) \pi(x) \mathrm{d} x \geq  \frac {\beta}{16} \cdot \sigma(S_1',S_2') 
\end{equation}
for every partition $S_1\cup S_2 = \K$.

Hence, Eq.~\eqref{eq_h11} implies that
\begin{align*}
    \phi = & ~ \inf_{S \subseteq \K : \pi^\star(S)\leq \frac{1}{2}} \frac{1}{\pi^\star(S)} \int_S \wt{P}_x (\K \backslash S) \pi(x) \mathrm{d} x  \geq \frac{\beta}{16} \sigma(S_1',S_2'). \qedhere
\end{align*}
\end{proof}

\section{Correctness for General Barrier Functions with Regularization}\label{sec:correct}
In Section~\ref{sec:correct:bound_function}, we show how to bound the density ratios. In Section~\ref{sec:correct:bound_determinant}, we show how to bound the determinant. In Section~\ref{sec:correct:bound_local_norm}, we show how to bound the local norm. In Section~\ref{sec:correct:tv_between_exact_process}, we explain how to bound the TV distance between two exact distributions. In Section~\ref{sec:correct:tv_between_approximate_process}, we show how to bound the TV distance between two approximate processes. In Section~\ref{sec:correct:tv_between_gaussians}, we show how to bound the TV distance between two Gaussians. In Section~\ref{sec:correct:reversibility_and_stationary_distribution}, we prove reversibility and stationary distribution. In Section~\ref{sec:correct:lower_bound_conductance}, we prove a lower bound on cross ratio distance. In Section~\ref{sec:conductance}, we relate the cross ratio distance to local norm by utilizing $\ov\nu$-symmetry.

\begin{algorithm}[!ht]\caption{
Our algorithm for sampling from polytope with log-barrier (formal version of Algorithm~\ref{alg:informal}). As the only differences between log-barrier for polytopes and others are the choice of $\nu$ and how to generate the approximate the Hessian, we only present this algorithm. 
}
\label{alg:main_logconcave}
\begin{algorithmic}[1]
\Procedure{Main}{$A\in \R^{n\times d}, b\in \R^n, \delta\in (0,1), x_0\in \R^d,\alpha>0, \eta>0$} \Comment{Theorem~\ref{thm:main_polytope}}
\State \Comment{$\K:=\{x\in \R^d: Ax\leq b \}$}
\State Let $g$ be the log-barrier for $\K$ with $\nu=n$
\State $\alpha \gets \Theta(1/d)$
\State $\eta \gets \Theta(1/(dL^2))$
\State $T \gets (\nu \alpha^{-1} + \eta^{-1} R^2) \cdot \log(w/\delta)$
\State $x\gets x_0$ \Comment{We are given $x_0 \in \mathrm{Int}(\K)$}
\State $\epsilon \gets \Theta(1)$
\State $\epsilon_H \gets \Theta(\epsilon/d )$
\For{$t=1\to T$}
\State Sample a point $\xi\sim \N(0, I_d)$
\State \Comment{Let $H$ denote the Hessian of barrier function $g$}
\State $\wt{H}(x) \gets \textsc{SubSample} (A,b,x,n,d,\epsilon_H)$ \Comment{Algorithm~\ref{alg:subsample}}
\State 
\Comment{$\wt{H}(x)$ can be written $\sum_{i \in S} \wt{\sigma}_i a_i a_i^\top$ where $|S| = \epsilon_H^{-2} d \log d$}
\State \Comment{The above step takes $\nnz(A) \log n + \Tmat(d, \epsilon_H^{-2} d, d)$ time}
\State $\wt{\Phi}(x) \gets \alpha^{-1} \wt{H}(x)+\eta^{-1} I_d$
\State $z\gets x+ \wt{\Phi}(x)^{-1/2}\xi$
\If{$z\in {\rm Int}(\K)$}
\State $\wh{H}(z) \gets \textsc{SubSample}(A,b,z,n,d,\epsilon_H)  $ \Comment{Algorithm~\ref{alg:subsample}}
\State $\wh{\Phi}(z)\gets \alpha^{-1} \wh{H}(z)+\eta^{-1}I_d$
\State {\bf accept} $x\gets z$ with probability 
\begin{align*} 
    \frac{1}{2}\cdot \min \Big\{ \frac{ \exp(-f(z)) \cdot ( \det(\wh{\Phi}(z)))^{1/2} \cdot \exp( - 0.5 \| x - z\|_{ \wh{\Phi}(z) }^2 ) }{ \exp(-f(x)) \cdot ( \det( \wt{\Phi}(x)) )^{1/2} \cdot \exp(-0.5 \| x - z \|_{ \wt{\Phi}(x) }^2 ) },1 \Big\}
\end{align*}
\Else
\State {\bf reject} $z$
\EndIf
\EndFor
\State \Return $x$
\EndProcedure

\end{algorithmic}
\end{algorithm}

\subsection{Bounding the density ratio}\label{sec:correct:bound_function}
\begin{lemma}[Bounding the density ratio]
\label{lem:bound_function}
Let $x \in \mathrm{Int}(\K)$. Suppose $f$ is $L$-Lipschitz and $\eta \leq 1/(10 d L^2)$, then 
\begin{align*}
    \Pr_{z \sim {\cal N}(x, {\Phi}^{-1} (x) ) } \Big[ \frac{ \exp(-f(z)) }{ \exp(-f(x)) } \geq \frac{1}{2} \Big] \geq 0.999.
\end{align*}
If in addition, $f$ is $\beta$-smooth and $\eta \leq 1/(10 d \beta)$, then we further have
\begin{align*}
    \Pr_{z \sim {\cal N}(x, {\Phi}^{-1} (x) ) } \Big[ \frac{ \exp(-f(x)) }{ \exp(-f(z)) } \geq \frac{1}{2} \Big] \geq 0.499.
\end{align*}
\end{lemma}
\begin{proof}

{\bf Proof of Part 1.} Since $z \sim {\cal N}(x, {\Phi}(x)^{-1} )$, we have that
\begin{align*}
    z = & ~ x + {\Phi}(x)^{-1/2} \xi \\
    = & ~ x + ( \alpha^{-1} \cdot {H}(x) + \eta^{-1} I_d )^{-1/2}
\end{align*}
for some $\xi \sim {\cal N}(0, I_d)$.

Since $\alpha^{-1} {H}(x) + \eta^{-1} I_d \succeq \eta^{-1} I_d$, and ${H}( x )$ and $I_d$ are both positive definite, we have that
\begin{align}\label{eq:7}
    \eta \cdot I_d \succeq ( \alpha^{-1} {H}(x) + \eta^{-1} I_d )^{-1}
\end{align}

Thus, we can upper bound $\| z - x \|_2$ as follows:
\begin{align*}
    \| z - x \|_2 
    = & ~ \| ( \alpha^{-1} {H}(x) + \eta^{-1} I_d )^{-1/2} \xi \|_2 \\
    = & ~ \sqrt{ \xi^{\top} ( \alpha^{-1} {H}(x) + \eta^{-1} I_d )^{-1} \xi  } \\
    \leq & ~ \sqrt{ \xi^\top \eta I_d \xi } \\
    = & ~ \sqrt{\eta} \cdot \| \xi \|_2
\end{align*}
where the third step follows from Eq.~\eqref{eq:7},

Recall that $\xi$ is sampled from ${\cal N}(0,I_d)$, using Lemma~\ref{lem:xi_bound}, we can show
\begin{align*}
    \Pr[ \| \xi \|_2 > t ] \leq \exp( -(t^2 -d)/8 ), ~~~\forall t > \sqrt{2d}. 
\end{align*}
Combining the above two equations, we have
\begin{align}\label{eq:bound_z_minus_theta}
    \Pr[ \| z - x \|_2 > \sqrt{\eta} \sqrt{20 d} ] \leq \exp(-\frac{19}{8} d) < 0.001.
\end{align}
Using $\eta \leq 1/(80 d L^2)$, we have
\begin{align}\label{eq:bound_z_minus_theta_L}
    \Pr[ \| z - x \|_2 > 1/(2L) ]  < 0.001.
\end{align}
Since $f$ is $L$-Lipschitz, then we have
\begin{align*}
    \frac{ \exp(-f(z)) }{ \exp(-f(x)) } = \exp( - (f(z) - f(x))) \geq \exp(- L \| z - x \|_2).
\end{align*}
Therefore,

\begin{align*}
\Pr_{z \sim N(x, {\Phi}^{-1}(x))} \Big[  \frac{\exp(-f(z))}{\exp(-f(x))} \geq {1}/{2} \Big] 
\geq & ~\Pr_{z \sim N(x,  \Phi^{-1}(x))} \Big[ \exp(-L\|z-x\|_2) \geq {1}/{2} \Big] \\
 = & ~\Pr_{z \sim N(x,  \Phi^{-1}(x))} \Big[ \|z-x\|_2 \leq {\log(2)}/{L} \Big] \\
 \geq & ~ \Pr_{z \sim N(x,  \Phi^{-1}(x))} \Big[ \|z-x\|_2 \leq 1/(2L) \Big] \\
\geq & ~ 0.999
\end{align*}
where the last inequality holds by Eq.~\eqref{eq:bound_z_minus_theta_L}. 

{\bf Proof of Part 2.}
Moreover, in the setting where $f$ is differentiable and  $\beta$-smooth, we have that, since $z-x$ is a multivariate Gaussian random variable, 
\begin{align*}
\Pr[(z-x)^\top \nabla f(x) \leq 0] = \frac{1}{2}. 
\end{align*}
If $(z-x)^\top \nabla f(x) \leq 0$, we have that
\begin{align*}
    f(z) - f(x) &\leq (z-x)^\top \nabla f(x) + \beta \|z- x\|_2^2\\
    &\leq \beta \|z- x\|_2^2.
\end{align*}
Therefore,
\begin{align*}
& ~\Pr_{z \sim N(x, \wt \Phi^{-1}(x))} [ \frac{\pi(z)}{\pi(x)} \geq \frac{1}{2} ]\\
\geq & ~\Pr_{z \sim N(x, \wt \Phi^{-1}(x))} [  \frac{\pi(z)}{\pi(x)} \geq \frac{1}{2} \mathrm{~~~and~~~}  (z-x)^\top \nabla f(x) \leq 0  ] -  \Pr_{z \sim N(x, \wt \Phi^{-1}(x))} [ (z-x)^\top \nabla f(x) > 0 ]\\
\geq & ~ \Pr_{z \sim N(x, \wt \Phi^{-1}(x))} [ e^{-\beta \|z- x\|_2^2} \geq \frac{1}{2} ] - 0.5\\
= & ~ \Pr_{z \sim N(x, \wt \Phi^{-1}(x))} [ \|z-x\|_2 \leq \frac{\sqrt{\log(2)}}{\sqrt{\beta}} ]-0.5\\
\geq & ~ 0.999-0.5\\
= & ~ 0.499,
\end{align*}
where the last Inequality holds by Eq.~\eqref{eq:bound_z_minus_theta} 
since $\eta \leq \frac{1}{20 d \beta}$.
\end{proof}

\begin{lemma}\label{lem:bound_function_shifted}
Let $\| x - z \|_{\Phi(x)} < 0.001$. Let $\eta \leq 0.01/L^2$. Then we have
\begin{align*}
    | f(x) - f(z) | \leq 0.1
\end{align*}
\end{lemma}
\begin{proof}
We have
\begin{align*}
    | f(x) - f(z) | \leq & ~ L \cdot \| x - z \|_2 \\
     \leq & ~ L \cdot \sqrt{\eta} \cdot \| x -z \|_{\Phi(x)} \\
     \leq & ~ 0.1
\end{align*}
where the second step follows from $\Phi(x) \succeq \eta^{-1} I_d$, the last step follows from $\eta \leq 0.01/L^2$.
\end{proof}

\subsection{Bounding the determinant}\label{sec:correct:bound_determinant}

\begin{lemma}[Bounding the determinant]\label{lem:bound_determinant}

Consider any $x \in \mathrm{Int}(\K)$, and $\xi \sim {\cal N}(0, I_d)$. Let $z = x + ( \Phi(x) )^{-1/2} \xi$.  Then
\begin{align*}
    \Pr_{\xi} \Big[  \log(\det(\Phi(z))) - \log( \det(\Phi(x)))  \ge -0.1 \Big] \geq 0.999.
\end{align*}

\end{lemma}

\begin{proof}
First, note that 
\begin{align*}
    \log(\det(\Phi(z)))-\log(\det(\Phi(x))) = & ~ \log(\frac{\det(\Phi(z))}{\det(\Phi(x))}) \\
    = & ~ \log(\frac{\det(d\cdot H(z)+dL^2\cdot I_d)}{\det(d\cdot H(x)+dL^2\cdot I_d)})) \\
    = & ~ \log(\frac{\det(H(z)+L^2\cdot I_d)}{\det(H(x)+L^2\cdot I_d)}),
\end{align*}

thus, it suffices to consider the function $F(x)=\log(\det(H(x)+L^2\cdot I_d))$ and bound $F(z)-F(x)$. By Assumption~\ref{ass:convex}, 
we know that $F(x)$ is convex.

Thus, 
\begin{align*}
    F(z) - F(x) \geq (z-x)^\top \nabla F(x).
\end{align*}

We know that
\begin{align*}
    z = x + ( \Phi(x) )^{-1/2} \xi
\end{align*}
where $\xi \sim {\cal N}(0,I_d)$.

Thus,
\begin{align*}
    F(z) - F(x) \geq \xi^\top ( \Phi(x) )^{-1/2} \nabla F(x)
\end{align*}

By property of Gaussian distribution, we know that
\begin{align*}
    \xi^\top ( \Phi(x) )^{-1/2} \nabla F(x)
\end{align*}
is a Gaussian with mean $0$ and variance $\| (\Phi(x))^{-1/2} \nabla F(x) \|_2^2$.

By definition of $\Phi$ and Assumption~\ref{ass:bound_variance} 
, we have 
\begin{align*}
    \| (H(x)+L^2\cdot I_d)^{-1/2} \nabla F(x) \|_2^2 = O(d)
\end{align*}
and therefore
\begin{align*}
    \| \Phi(x)^{-1/2} \nabla F(x) \|_2^2 = O(1).
\end{align*}

Thus, rescaling the function and applying the standard concentration inequality, we complete the proof. \qedhere
\end{proof}

\begin{lemma}[Bounding the determinant, shifted]\label{lem:bound_determinant_shift}
Let $\alpha < 0.001/d$.
For fixed $x, z \in \K$ such that $\| z - x \|_{\Phi(x)} < 0.001$. 
Then, we have
\begin{align*}
    \log(\det(\Phi(z))) - \log( \det(\Phi(x))) \ge -0.01.
\end{align*}
\end{lemma}
\begin{proof}
Let $F(x) = \log  \det (H(x) + L^2 \cdot I_d)$.
Then
\begin{align*}
\log(\det(\Phi(z))) - \log( \det(\Phi(x))) = F(z) - F(x) \ge (z-x)^\top \nabla F(x)
\end{align*}
where the second step is because $F(x)$ is convex (by Assumption~\ref{ass:convex}).

We have
\begin{align*}
((z-x)^\top \nabla F(x))^2 &= ((z-x)^\top\Phi(x)^{1/2} \cdot \Phi(x)^{-1/2} \nabla F(x))^2 \\
&\le \|(z-x)^\top\Phi(x)^{1/2}\|_2 \cdot \|\Phi(x)^{-1/2} \nabla F(x)\|_2 \\
&\le 0.001 \cdot \|\Phi(x)^{-1/2} \nabla F(x)\|_2 \\
&\le 0.001 \cdot 0.1.
\end{align*}
where the second step is by Cauchy-Schwarz, the third step is by assumption $\|z-x\|_{\Phi(x)} \le 0.001$,
the fourth step is by $\alpha < 0.001/d$ and Assumption~\ref{ass:bound_variance}.
\end{proof}

\subsection{Bounding the local norm}\label{sec:correct:bound_local_norm}

\begin{lemma}[Bounding the local norm]\label{lem:bound_local_norm}
Consider any $x \in \mathrm{Int}(\K)$, and $\xi \sim {\cal N}(0, I_d)$. Let $\eta \leq 0.001/d$. Let $z = x + ( \Phi(x) )^{-1/2} \xi$. Then 
\begin{align*}
\Pr_{\xi} \Big[ | \|z-x\|_{{\Phi}(z)}^2 - \|z-x\|_{{\Phi}(x)}^2 | \leq 0.01 \Big]\ge 0.999
\end{align*}
\end{lemma}
\begin{proof}
Our proof is a generalization of~\cite[Proposition 7]{sv16}, where they prove the result for non-regularized log-barrier. Note that their proof also works for regularized barriers because two points are close in the $\Phi(x)$ norm indicates that they are close in the $H(x)$ norm. The only difference is that when computing the Gaussian polynomial as in~\cite{sv16}, we need to handle non-uniform weights instead of the uniform weights of log-barrier. This could also be handled by observing that if two points are close in $H(x)$ norm, then their weights are close in the sense that $\|w_p(x)^{-1}(w_p(x)-w_p(z)) \|_\infty \leq c_p$ where $c_p<1$ is some small constant that depends on $p$ (see~\cite[Lemma 34]{ls19}). One could then check that the argument of~\cite[Proposition 7]{sv16} is robust under small perturbation to the weights, and conclude similar conclusions.
\end{proof}

\begin{lemma}[Bounding the local norm, shifted]\label{lem:bound_local_norm_shifted_1}
Consider any $x \in \mathrm{Int}(\K)$, and $\xi \sim {\cal N}(0, I_d)$. Let $\eta \leq 0.001/d$. Let $\| z - x \|_{\Phi(x)} < 0.001$. Let $u = x + ( \Phi(x) )^{-1/2} \xi$. Then 
\begin{align*}
\Pr_{\xi} \Big[ | \|u-z\|_{{\Phi}(u)}^2 - \|u-x \|_{{\Phi}(u)}^2 | \le  0.1 \Big]\ge 0.999
\end{align*}
\end{lemma}
\begin{proof}
Consider the following inequalities:
\begin{align*}
   & ~ | \|u-z\|_{{\Phi}(u)}^2 - \|u-x \|_{{\Phi}(u)}^2 | \\
 \leq & ~ \| u -  x \|_{\Phi(u)}^2  + 2 | \langle u - x , x-z  \rangle_{\Phi(u)} | \\
 \leq & ~ 2 \| u -x \|_{\Phi(x)}^2 + 4 | \langle u - x , x-z  \rangle_{\Phi(x)} | \\
 = & ~ 2 \| u -x \|_{\Phi(x)}^2 + 4 |w|
 \end{align*}
 where the second step follows from Lemma~\ref{lem:related_local_norm}, in the last step $w \sim {\cal N}(0, \| \Phi(x)^{1/2} (x-z) \|_2^2  )$.
 
 First, using $\eta$ is small, we can show $\| u-x \|_{\Phi(x)}^2 < 0.01$ with probability $0.9999$.
 
 Second, using $\| z - x \|_{\Phi(x)} < 0.001$ and Gaussian concentration, we can show that
 \begin{align*}
     \Pr[ |w| < 0.01] \geq & ~  0.9999.
 \end{align*} 
Put them together and union bound, we obtain the desired result.
\end{proof}

\begin{lemma}[Bounding the local norm, shifted]\label{lem:bound_local_norm_shifted_2}
Consider any $x \in \mathrm{Int}(\K)$, and $\xi \sim {\cal N}(0, I_d)$. Let $\eta \leq 0.001/d$. Let $\| z - x \|_{\Phi(x)} < 0.001$. Let $u = x + ( \Phi(x) )^{-1/2} \xi$.
We have
\begin{align*}
   \Pr[ |  \|u-z \|^2_{  \Phi(z) } -  \| u -x \|^2_{\Phi(u)} | < 0.1 ] \geq 0.998
\end{align*}
\end{lemma}
\begin{proof}
Let $u' \sim z + \Phi(z)^{-1/2} \xi'$ where $\xi' \sim \N(0, I_d)$.
We have
\begin{align*}
    & ~| \|u-z \|^2_{  \Phi(z) } -  \| u -x \|^2_{\Phi(u)} | \\
    \leq &~ 
    { | \|u-z \|^2_{  \Phi(z) } -  \| u' -z \|^2_{\Phi(z)} | }
    + { | \|u'-z \|^2_{  \Phi(z) } -  \| u' -z \|^2_{\Phi(u')} | }
    \\ 
    & ~+ { | \|u'-z \|^2_{  \Phi(u') } -  \| u -z \|^2_{\Phi(u)} | }
    + { | \|u-z \|^2_{  \Phi(u) } -  \| u -x \|^2_{\Phi(u)} | }
\end{align*}
So
\begin{align*}
&~ \Pr[ |  \|u-z \|^2_{  \Phi(z) } -  \| u -x \|^2_{\Phi(u)} | > 0.1 ] \\ 
\le &~ \underbrace{\Pr[{ | \|u'-z \|^2_{  \Phi(z) } -  \| u' -z \|^2_{\Phi(u')} | } > 0.05]}_{\text{Lemma~\ref{lem:bound_local_norm}}} \\
& ~+ \underbrace{\Pr[{ | \|u-z \|^2_{  \Phi(u) } -  \| u -x \|^2_{\Phi(u)} | } > 0.05]}_{\text{Lemma~\ref{lem:bound_local_norm_shifted_1}}}
+ \underbrace{\Pr[u \ne u']}_{\text{Lemma~\ref{lem:TV_Gaussian_x_z}}}\\
\le & ~ 0.001 + 0.001 + 0.001 = 0.003.
\end{align*}
where first step is by union bound, second step is by Lemma~\ref{lem:bound_local_norm}, Lemma~\ref{lem:bound_local_norm_shifted_1}, and by $\TV(u,u') \le 0.001$ when $\|x-z\|_{\Phi(x)} \le 0.001$ (Lemma~\ref{lem:TV_Gaussian_x_z}).
\end{proof}

\subsection{Bounding TV distance between exact processes}\label{sec:correct:tv_between_exact_process}
Recall that we define $P_x(S) = \Pr[x\in S]$. We prove a bound between exact distributions of two points $x$ and $z$. 

\begin{lemma}[TV distance between the exact distribution]\label{lem:lemma_5.10_in_mv22}
For any $x, z \in \K$ such that $\| x - z \|_{ \Phi(x) } \leq 0.001$, we have that
\begin{align*}
    \TV(P_{x}, P_z) \leq & ~ 0.99. 
\end{align*}
\end{lemma}

\begin{proof}
We define 
\begin{align*} 
{G}_{x,\Phi(x)}(u) := ( \det( \Phi(x)) )^{1/2} \cdot \exp( -0.5 \|u-x \|^2_{  \Phi(x) } ).
\end{align*}
Then
\begin{align*}
    \TV( P_{x,\Phi(x)}, P_{z,\Phi(z)}) 
    =  1 - p_{\min},
\end{align*}
where
\begin{align*}
    p_{\min} := \E_{u\sim \N(x, {\Phi}^{-1}(x))} [\min\{1, \frac{{G}_z(u)}{{G}_x(u)},
    \frac{\exp(-f(u)) {G}_u(x)}{\exp(-f(x)) {G}_x(u)}, \frac{\exp(-f(u)) {G}_u(z)}{\exp(-f(z)) {G}_z(u)} \cdot \frac{{G}_z(u)}{{G}_x(u)}\}].
\end{align*}
For convenience, we define
\begin{align*}
    a_1 := & ~ \frac{{G}_z(u)}{{G}_x(u)} \\
    a_2 := & ~ \frac{\exp(-f(u)) {G}_u(x)}{\exp(-f(x)) {G}_x(u)}  \\
    a_3 := & ~ \frac{\exp(-f(u)) {G}_u(z)}{\exp(-f(z)) {G}_z(u)} \cdot \frac{{G}_z(u)}{{G}_x(u)}
\end{align*}

Let $\theta \in (0,1)$. We have the following:
\begin{align*}
    \TV(P_x, P_z) 
    = & ~ 1- p_{\min} \\
    = & ~ 1 - \E[ \min \{1,a_1,a_2,a_3 \} ] \\
    = & ~ 1 - \min\{1,\theta\} \cdot \Pr[ \min\{a_1,a_2,a_3 \} \geq \theta ] - \min\{a_1,a_2,a_3\} \cdot \Pr[  \min\{a_1,a_2,a_3 \} < \theta ] \\
    \leq & ~ 1 - \min\{1,\theta\}  \cdot \Pr[ \min\{a_1,a_2,a_3 \} \geq \theta ]  \\
    \leq & ~ 1 -  \theta \cdot \Pr[ \min\{a_1,a_2,a_3 \} \geq \theta ] \\
    \leq & ~ 1- \theta \cdot \theta_0 \\
    \leq & ~ 1 - 0.01 \cdot 0.95 \\
    \leq & ~ 0.99
\end{align*}
where the sixth step follows from Eq.~\eqref{eq:theta_bound}, and the last step follows from choice of $\theta$ and $\theta_0$.

Let $a_i = e^{-b_i}$. It suffices to prove that for some $\theta \in (0,1)$, we have
\begin{align}\label{eq:theta_bound}
    \Pr[ \min \{a_1, a_2, a_3 \} \geq \theta ] \geq \theta_0.
\end{align}

Suppose for some $\tau \in [1, 5]$ we can show that
\begin{align*}
    \Pr[ b_1 < \tau ] \geq \theta_1 \\
    \Pr[ b_2 < \tau ] \geq \theta_2 \\
    \Pr[ b_3 < \tau ] \geq \theta_3 
\end{align*}
Then by union bound, we can know that
\begin{align*}
     & ~ \Pr[ \max \{b_1, b_2, b_3 \} < \tau ] \\
    = & ~\Pr[ \min \{a_1, a_2, a_3 \} > e^{-\tau} ]\\
    = & ~\Pr[ \min \{a_1, a_2, a_3 \} > \theta ]\\
    \geq & ~ 1- (1-\theta_1) - (1-\theta_2) - (1- \theta_3) \\ 
    \geq & ~ 0.95
\end{align*}
where the second step follows from $e^{-\tau} = \theta \geq 0.01$, and the last step follows from $\theta_1=\theta_2= \theta_3 = 0.99$. In the following, we will establish the three bounds of interest.

{\bf Part 1.} By definition, we have
\begin{align*}
    a_1 
    = & ~ \frac{{G}_z(u)}{{G}_x(u)} \\
    = & ~ \frac{( \det( \Phi(z)) )^{1/2}}{ ( \det( \Phi(x)) )^{1/2}} \cdot \exp( - 0.5 \|u-z \|^2_{  \Phi(z) } + 0.5 \|u-x \|^2_{  \Phi(x) } ) \\
    = & ~ \exp ( - 0.5 \|u-z \|^2_{  \Phi(z) } + 0.5 \|u-x \|^2_{  \Phi(x) } + 0.5 \log \det(\Phi(z)) - 0.5 \log(\det(\Phi(x))) ) 
\end{align*}

Using Lemma~\ref{lem:bound_local_norm_shifted_2} (on the term $0.5 \|u-z \|^2_{  \Phi(z) } - 0.5 \| u -x \|^2_{\Phi(u)}$) and Lemma~\ref{lem:bound_local_norm} (on the term $0.5 \| u -x \|^2_{\Phi(u)} -  0.5 \|u-x \|^2_{  \Phi(x) } $), we have
\begin{align*}
    \Pr_{u}[ \underbrace{ 0.5 \|u-z \|^2_{  \Phi(z) } - 0.5 \| u -x \|^2_{\Phi(u)} }_{\rm Lemma~\ref{lem:bound_local_norm_shifted_2}} + \underbrace{ 0.5 \| u -x \|^2_{\Phi(u)} -  0.5 \|u-x \|^2_{  \Phi(x) } }_{\rm Lemma~\ref{lem:bound_local_norm} } \leq 0.1 \tau ]  \geq  0.999,
\end{align*}

From Lemma~\ref{lem:bound_determinant_shift}, 
we know
\begin{align*}
    0.5 \log (\det(\Phi(x))) - 0.5 \log(\det(\Phi(z))) \leq 0.1 \tau.
\end{align*}
Thus, 
\begin{align*}
    \Pr[b_1 \leq \tau ] \geq 0.99
\end{align*}

{\bf Part 2.} By definition, for $a_2$ we have
\begin{align*}
    a_2 = & ~ \frac{\exp(-f(u)) {G}_u(x)}{\exp(-f(x)) {G}_x(u)} \\
    = & ~ \frac{\exp(-f(u))}{\exp(-f(x))} \cdot \frac{( \det( \Phi(u)) )^{1/2}}{ ( \det( \Phi(x)) )^{1/2}} \cdot \exp( - 0.5 \|x-u \|^2_{  \Phi(u) } + 0.5 \|u-x \|^2_{  \Phi(x) } ) \\
    = & ~ \exp( -f(u) + f(x) + 0.5 \log(\det(\Phi(u))) - 0.5 \log(\det(\Phi(x))) - 0.5 \| x- u \|_{\Phi(u)}^2 + 0.5 \| u -x \|_{\Phi(x)}^2 )
\end{align*}

By Lemma~\ref{lem:bound_function}, we have
\begin{align*}
    \Pr_{u}[ f(u) - f(x) \leq 0.1 \tau ] \geq 0.999
\end{align*}

By Lemma~\ref{lem:bound_determinant}, we have
\begin{align*}
    \Pr_u[ 0.5 \log(\det(\Phi(x))) - 0.5 \log(\det (\Phi(u))) \leq 0.1\tau  ] \geq 0.999
\end{align*}

By Lemma~\ref{lem:bound_local_norm}, we have
\begin{align*}
    \Pr_u[ 0.5 \| x - u \|_{\Phi(u)}^2 - 0.5 \| u - x \|_{\Phi(x)}^2 \leq 0.1 \tau ] \geq 0.999
\end{align*}

Then, combining the above three equations, we can get the following result:
\begin{align*}
    \Pr[ b_2 \leq \tau] \geq 0.99
\end{align*}

{\bf Part 3. }
\begin{align*}
    a_3 = & ~ \frac{\exp(-f(u)) {G}_u(z)}{\exp(-f(z)) {G}_z(u)} \cdot \frac{{G}_z(u)}{{G}_x(u)}  \\
    = & ~ \frac{\exp(-f(u))}{\exp(-f(z))} \cdot \frac{( \det( \Phi(u)) )^{1/2}}{ ( \det( \Phi(x)) )^{1/2}} \cdot \exp( - 0.5 \|z-u \|^2_{  \Phi(u) } + 0.5 \|u-x \|^2_{  \Phi(x) } ) \\
    = &~ \exp(-f(u) + f(z) + 0.5 \log(\det(\Phi(u))) -0.5 \log(\det(\Phi(x))) -  0.5 \|z-u \|^2_{  \Phi(u) } + 0.5 \|u-x \|^2_{  \Phi(x) }  )
\end{align*}

By Lemma~\ref{lem:bound_function} (on the term $f(u) - f(x)$) and Lemma~\ref{lem:bound_function_shifted} (on the term $f(x) - f(z)$), we have
\begin{align*}
    \Pr_u[ \underbrace{ f(u) -f(x) }_{ \rm Lemma~\ref{lem:bound_function} } + \underbrace{ f(x) - f(z) }_{ \rm Lemma~\ref{lem:bound_function_shifted} } \leq 0.1 \tau ] \geq 0.999 .
\end{align*}

By Lemma~\ref{lem:bound_determinant}, we have
\begin{align*}
    \Pr_u[ 0.5 \log(\det(\Phi(x))) - 0.5 \log(\det (\Phi(u))) \leq 0.1 \tau  ] \geq 0.999.
\end{align*}

By Lemma~\ref{lem:bound_local_norm} (on the term $0.5 \| u - x\|_{\Phi(u)}^2 - 0.5 \| u- x \|_{\Phi(x)}^2$) and Lemma~\ref{lem:bound_local_norm_shifted_1} (on the term $0.5 \| z - u \|_{\Phi(u)}^2 - 0.5 \| u - x \|_{\Phi(u)}^2$), we have
\begin{align*}
    \Pr_u[  \underbrace{ 0.5 \| z - u \|_{\Phi(u)}^2 -0.5 \| u - x \|_{\Phi(u)}^2 }_{\rm Lemma~\ref{lem:bound_local_norm_shifted_1}} + \underbrace{ 0.5 \| u - x \|_{\Phi(u)}^2 - 0.5 \| u - x \|_{\Phi(x)}^2 }_{\rm Lemma~\ref{lem:bound_local_norm} } \leq 0.2 \tau ] \geq 0.999
\end{align*}
Thus, combining the above three equations, we obtain the following:
\begin{align*}
    \Pr[ b_3 \leq \tau] \geq & ~ 0.99. \qedhere
\end{align*}
\end{proof}

\subsection{Bounding TV distance between approximate processes}\label{sec:correct:tv_between_approximate_process}
\begin{lemma}[Robust version of Lemma E.10 in \cite{mv22}]\label{lem:robust_lemma_5.10_in_mv22}
 
If $\delta_H < 0.001$, $d \alpha < 0.001$, $d \epsilon_H <0.001$ and for any $x, z \in \K$, $\| x - z \|_{{\Phi}(x)} < 0.001$, then we have
\begin{align*}
    \TV( \wt{P}_x, \wt{P}_z ) \leq 0.99.
\end{align*}

\end{lemma}
\begin{proof}
We can upper bound $\TV( \wt{P}_x, \wt{P}_y )$ as follows:
    \begin{align*}
         \TV(\wt{P}_x, \wt{P}_y)  
        \le & ~
        \TV(\wt{P}_x, P_x) + \TV(P_x, P_y) + \TV(P_y, \wt{P}_y) \\
        \le & ~ \TV(\wt{P}_x, P_x) + 0.9 + \TV(P_y, \wt{P}_y) \\
        \le & ~ 0.01 + 0.9 + 0.01 \\
        \leq & ~ 0.99
    \end{align*}
    First step is by triangle inequality. Second step is by  Lemma~\ref{lem:lemma_5.10_in_mv22}. The third step is by Lemma~\ref{lem:exact_step_vs_approx_step}.  
\end{proof}

\subsection{Bounding TV distance between Gaussians}\label{sec:correct:tv_between_gaussians}

\begin{lemma}[Lemma E.6 in \cite{mv22}] \label{lem:TV_Gaussian_x_z}
For any $x, z \in \K$ such that $\| x - z\|_{ \Phi( x ) } \leq 0.001$, we have
\begin{align*}
    \TV({\cal N} (x, \Phi^{-1} (x) ) , {\cal N} (z , \Phi^{-1} (z) ) ) \leq \sqrt{3 d \alpha+1/2}\cdot \| x - z \|_{ \Phi(x) }.
\end{align*}
Further, if $\alpha d < 0.001 $, then we have
\begin{align*}
    \TV({\cal N} (x, \Phi^{-1} (x) ) , {\cal N} (z , \Phi^{-1} (z) ) ) \leq 0.001.
\end{align*}
\end{lemma}

\begin{lemma}[Robust version of Lemma~\ref{lem:TV_Gaussian_x_z}] \label{lem:robust_lemma_5.6_in_mv22}
For any $x, z \in \K$ such that $\| x - z\|_{ \Phi( x ) } \leq \frac{1}{4 \alpha^{1/2}}$,
with probability at least $1-1/1000$, we have
\begin{align*}
    \TV( {\cal N} (x, \wt{\Phi}^{-1} (x) ) , {\cal N} (z , \wh{\Phi}^{-1} (z) ) ) \leq \sqrt{2 d \epsilon_H} + \sqrt{3 d \alpha+1/2} \cdot \| x - z \|_{ \Phi(x) }
\end{align*}
\end{lemma}

\begin{proof}
By triangle inequality,
\begin{align*}
& ~ \TV( {\cal N} (x, \wt{\Phi}^{-1} (x) ) , {\cal N} (z , \wh{\Phi}^{-1} (z) ))\\
\le & ~
\TV({\cal N} (x, \wt{\Phi}^{-1} (x) ) , {\cal N} (x , \Phi^{-1} (x) )  ) \\
+ & ~ 
\TV({\cal N} (x, \Phi^{-1} (x) ) , {\cal N} (z , \Phi^{-1} (z) ) )\\
+ & ~
\TV({\cal N} (z, \Phi^{-1} (z) ) , {\cal N} (z , \wh{\Phi}^{-1} (z) ) ).
\end{align*}

Second term is bounded using Lemma \ref{lem:TV_Gaussian_x_z}.
First and third term are bounded using similar ways.
We have
\begin{align*}
& ~ \TV( {\cal N} (x, \wt{\Phi}^{-1} (x) ) , {\cal N} (x , \Phi^{-1} (x) ))^2 \\
\le & ~ \frac 12 D({\cal N} (x, \wt{\Phi}^{-1} (x) ) \| {\cal N} (x , \Phi^{-1} (x) )  ) \\
= & ~ \frac 14 (\tr[\wt{\Phi}(x)^{-1/2} \Phi(x) \wt{\Phi}(x)^{-1/2}] -d + \log \frac{\det \wt \Phi(x) }{\det {\Phi(x)}}) \\
\le & ~ \frac 14 ((1\pm \epsilon_H) d - d + \log (1\pm \epsilon_H)^d) \\
\le & ~ \frac 12 d \epsilon_H.
\end{align*}
Thus, we complete the proof.
\end{proof}

\subsection{Reversibility and stationary distribution}\label{sec:correct:reversibility_and_stationary_distribution}

For any $x \in \K$, we define the random variable $Z_{x}$ to be the step taken by the Markov chain in Algorithm~\ref{alg:main_logconcave} from the point $x$, that is, set $z = x + \wt{\Phi}(x)^{-1/2} \xi$ where $\xi \sim {\cal N}(0,I_d)$. If $z \in \K$, set $Z_x = z$ with probability 
\begin{align*} 
\frac{1}{2}\cdot \min \Big\{ \frac{ \exp(-f(z)) \cdot ( \det(\wh{\Phi}(z)))^{1/2} \cdot \exp( - 0.5 \| x - z\|_{ \wh{\Phi}(z) }^2 ) }{ \exp(-f(x)) \cdot ( \det( \wt{\Phi}(x)) )^{1/2} \cdot \exp(-0.5 \| x - z \|_{ \wt{\Phi}(x) }^2 ) },1 \Big\}.
\end{align*}
Else, we set $z = x$.

We provide a modified version of Proposition E.12 in \cite{mv22} where our Markov chain differs from theirs.
\begin{lemma}[Reversibility and stationary distribution]\label{lem:robust_lemma_5.12_in_mv22}
For any $S_1, S_2 \subseteq \K$ we have that
\begin{align*}
    \int_{x \in S_1} \pi(x) \wt{P}_{x} (S_2) \mathrm{d} x = \int_{y \in S_2} \pi(y) \wt{P}_y(S_1) \mathrm{d} y.
\end{align*}
\end{lemma}

\begin{proof}
Let $a,b$ be two i.i.d. random variables (one can view $a$ and $b$ as random coins to generate the corresponding sparsifers) such that $\tilde \Phi_{a,x}$ is a function of $a,x$ and $\hat \Phi_{b,y}$ is a function of $b,y$.

Let $p_{a,x}(\cdot)$ be pdf of $\N(0, \tilde \Phi_{a,x}^{-1})$ and $p_{b,y}(\cdot)$ be the pdf of $\N(0, \hat \Phi_{b,y}^{-1})$.
Because $a$ and $b$ are i.i.d, they are interchangeable, i.e., $(a,b)$ has the same distribution as $(b,a)$.

\begin{align*}
    & \int_{x\in S_1} \pi(x) \wt{P}_x(S_2) \mathrm{d} x \\
    =& \int_{x\in S_1} \pi(x) \int_a q(a) \int_{y\in S_2} p_{a,x}(y) \int_b q(b) \cdot \min\{\frac{\pi(y) p_{b,y}(x)}{\pi(x) p_{a,x}(y)}, 1\} \mathrm{d} b~\mathrm{d} y~\mathrm{d} a~\mathrm{d} x \\
    =&  \int_a q(a) \int_b q(b) \int_{x\in S_1} \int_{y\in S_2}  \min\{\pi(y) p_{b,y}(x), \pi(x) p_{a,x}(y)\} \mathrm{d} y~\mathrm{d} x~\mathrm{d} b~\mathrm{d} a\\
    =&  \int_a q(a) \int_b q(b) \int_{x\in S_1} \int_{y\in S_2}  \min\{\pi(y) p_{a,y}(x), \pi(x) p_{b,x}(y)\} \mathrm{d} y~\mathrm{d} x~\mathrm{d} b~\mathrm{d} a \\
    =& \int_{y\in S_2} \int_a q(a) \int_{x\in S_1} \int_b q(b) \min\{\pi(y) p_{a,y}(x), \pi(x) p_{b,x}(y)\} \mathrm{d} b~ \mathrm{d} x~\mathrm{d} a~\mathrm{d} y \\
    =& \int_{y\in S_2} \pi(y) \wt{P}_y(S_1) \mathrm{d} y.
\end{align*}
The first and fifth steps are by definition of $\wt{P}_x$. The second and fourth steps are by changing order of integration. The third step is by interchangeability of $a$ and $b$.

This proves reversibility and stationary distribution.
\end{proof}

\subsection{Lower bound on cross ratio distance}\label{sec:correct:lower_bound_conductance}

\begin{definition}\label{def:F}
We define
\begin{align*}
    \mathsf{F}(n,\alpha,\eta,R):= 0.1/ \sqrt{n\alpha^{-1} + \eta^{-1} R^2}.
\end{align*}
For simplicity, we use $\mathsf{F}$ to denote $\mathsf{F}(n,\alpha,\beta,R)$. 
\end{definition}

\begin{lemma}
Let $S_1'$ and $S_2'$ be defined as Definition~\ref{def:S_1'_S_2'}. Le $\mathsf{F}$ be defined as Definition~\ref{def:F}. 
We have
\begin{align*}
    \sigma(S_1',S_2') > 1000 \mathsf{F}.
\end{align*}
\end{lemma}
\begin{proof}

Using Lemma~\ref{lem:lemma_5.2_in_mv22},
we have that for any $u,v \in \K$,
\begin{align*}
    \sigma(u,v) \geq \mathsf{F} \cdot \| u - v \|_{{\Phi}(u)}.
\end{align*}
We will try to prove $\sigma(u,v) > 1000 {\sf F}$ in the following. To do that, we will prove it by making a contradiction.

Suppose $\sigma(u,v) \leq 1000 \mathsf{F}$, then we have to require $\| u - v \|_{ {\Phi}(u) } \leq 0.001$.

Once we had that $\| u - v \|_{{\Phi}(u)} \leq 0.001$, using Lemma~\ref{lem:robust_lemma_5.10_in_mv22}, for any $u,v \in \K$ we have that
\begin{align}\label{eq_h1}
\TV( \wt{P}_{u}, \wt{P}_v)
\leq & ~ 0.99
\end{align}

On the other hand, Definition~\ref{def:S_1'_S_2'} implies that, for any $u \in S_1'$, $v \in S_2'$ we have that  
\begin{equation}\label{eq_h4}
    \TV( \wt{P}_u , \wt{P}_v ) \geq  1 - \wt{P}_u(S_2) - \wt{P}_v(S_1) \geq 1- 2\beta > 0.99.
\end{equation}
where the last step follows from $1-2\beta > 0.99$.

Thus, we obtain a contradiction, which means that
\begin{align*} 
& ~ \sigma(S_1', S_2') \geq \sigma(u,v) > 1000 \mathsf{F}. \qedhere
\end{align*}
\end{proof}

\subsection{Bounding cross ratio distance through \texorpdfstring{$\ov\nu$}{}-symmetry}\label{sec:conductance}

In Lemma E.2 of \cite{mv22}, they only prove the cross ratio distance bound for log-barrier function. Here, we generalize it to arbitrary barrier function. The key concept our proof relies on is the notion of $\ov\nu$-symmetry (Assumption~\ref{ass:nu_symmetry}).

\begin{lemma}
\label{lem:p_u_close}
Let $H\in \K\rightarrow \R^{d\times d}$ be a $\ov\nu$-symmetric function, and $u, v\in \K$. Consider the chord that passes through $u, v$ and intersects with $\K$. Let $p, q$ be the two endpoints of the chord, with the order $p, u, v, q$. Then, we have
\begin{align*}
    \|p-u\|_{H(u)} \leq & ~ \sqrt{\ov\nu}.
\end{align*}
\end{lemma}

\begin{proof}
Suppose we can show that $p\in \K\cap (2u- \K)$, then it naturally follows that $p\in E_u(\sqrt{\ov\nu})$ and $\|p-u\|_{H(u)}\leq \sqrt{\ov\nu}$. 

$p\in \K$ trivially follows by the construction of the chord. To see $p\in 2u-\K$, it is enough to show that for some $y\in \K$, we have $p=2u-y$. We claim $y=2u-p$ is such a choice. To see $2u-p\in \K$, we need to show that $2Au-Ap\leq b$. We partition the constraints into two sets.

{\bf Case 1.} Suppose for $a_i$, we have $a_i^\top u\leq a_i^\top p$, then $2a_i^\top u \leq 2a_i^\top p\leq a_i^\top p+b_i$, therefore, $2a_i^\top u-a_i^\top p\leq b_i$.

{\bf Case 2.} Suppose otherwise, $a_i^\top u>a_i^\top p$, then consider
\begin{align*}
    2a_i^\top u-a_i^\top p-b_i = & ~ (a_i^\top u-a_i^\top p)+(a_i^\top u-b_i) \\
    = & ~ \underbrace{(a_i^\top u-b_i)-(a_i^\top p-b_i)}_{d_1}-\underbrace{(b_i-a_i^\top u)}_{d_2}.
\end{align*}

We shall show that $d_1\leq d_2$ via a geometric argument. Consider the projection of both $p$ and $u$ onto the hyperplane $a_ix=b_i$, and the two chords that pass through $p$ and $u$ respectively, orthogonal to the hyperplane. Let us denote them with $l_p, l_u$ respectively. Observe that $l_p$ and $l_u$ are parallel, and $\|l_p-l_u\|_2=(a_i^\top u-b_i)-(a_i^\top p-b_i)=d_1$. It should then be obvious that $d_1\leq d_2$: let ${\rm proj}(u)$ denote the orthogonal projection of $u$ onto the hyperplane, then
\begin{align*}
    \underbrace{\|l_u-{\rm proj}(u)\|_2}_{d_2} = & ~ \underbrace{\|l_u-l_p\|_2}_{d_1}+\|l_p-{\rm proj}(p)\|_2
\end{align*}
since $\|l_p-{\rm proj}(p)\|_2\geq 0$, we have that $d_1\leq d_2$, and consequently, 
\begin{align*}
    2a_i^\top u-a_i^\top p \leq & ~ b_i.
\end{align*}

This proves that $2u-p\in \K$, and we can conclude that $p\in 2u-\K$, which yields our desired result.
\end{proof}

\begin{remark}
The above argument can be extended to spectrahedra and more general convex sets.
\end{remark}

Let $u, v\in \K$ be two arbitrary points, and consider the chord that passes through $u, v$ with two endpoints being $p, q$. Note that $p, q\in \partial \K$.

\begin{lemma}
\label{lem:lemma_5.2_in_mv22}
For any $u,v \in \K$, let $H$ be any matrix function satisfying Assumption~\ref{ass:nu_symmetry},~\ref{ass:convex} and~\ref{ass:bound_variance}.  We assume that $\K$ is contained in ball of radius $R$ and has nonempty interior.

For any parameter $\alpha \in (0,1), \eta\in (0,1)$, we define matrix function  $\Phi(u) := \alpha^{-1} \cdot H (u) +\eta^{-1}I_d$.

Then, we have
\begin{align*}
    \sigma(u,v) \geq \frac{1}{ \sqrt{2 \nu \alpha^{-1} + \eta^{-1} R^2} } \cdot \| u - v \|_{ \Phi(u) }
\end{align*}
\end{lemma}

\begin{proof} 

Without loss of generality, we can assume that
\begin{align}\label{eq:upper_bound_p_minus_u}
    \| p - u \|_2 \leq \| u - q \|_2.
\end{align}

We can lower bound $\sigma^2(u,v)$ as follows:

\begin{align*}
    \sigma^2(u,v) = & ~ (\frac{\|u-v\|_2^2\cdot \|p-q\|_2^2}{\|p-u\|_2^2\cdot \|v-q\|_2^2}) \\
    \geq & ~ \max\{\frac{\|u-v\|_2^2}{\|p-u\|_2^2},\frac{\|u-v\|_2^2}{\|v-q\|_2^2 } \} \\
    \geq & ~ \max\{\frac{\|u-v\|_2^2}{\|p-u\|_2^2},\frac{\|u-v\|_2^2}{\|u-q\|_2^2 } \} \\
    \geq & ~ \max\{\frac{\|u-v\|_2^2}{\|p-u\|_2^2}, \frac{\|u-v\|_2^2}{\|u-q\|_2^2}, \frac{\|u-v\|_2^2}{\|p-q\|_2^2} \} \\
    = & ~ \max\{\frac{\|u-v\|_2^2}{\|p-u\|_2^2},\frac{\|u-v\|_2^2}{\|p-q\|_2^2} \} \\
    \geq & ~ \frac{1}{2}\frac{\|u-v\|_2^2}{\|p-u\|_2^2}+\frac{1}{2} \frac{\|u-v\|_2^2}{\|p-q\|_2^2} \\
    \geq & ~ \frac{1}{2} \frac{\|u-v\|_H^2}{\|p-u\|_H^2}+\frac{1}{2}\frac{\|u-v\|_2^2}{R^2} \\
    \geq & ~ \frac{1}{2} \frac{\|u-v\|_H^2}{\nu}+\frac{1}{2}\frac{\|u-v\|_2^2}{R^2} \\
    = & ~ (u-v)^\top (\frac{1}{2\nu\alpha^{-1}}\times \alpha^{-1}H+\frac{1}{2R^2\eta^{-1}}\times \eta^{-1}I_d)(u-v) \\
    \geq & ~ \frac{1}{2\nu\alpha^{-1}+2\eta^{-1}R^2} \|u-v\|_{\Phi(u)}^2
\end{align*}
where the second step is by $\|p-q\|_2\geq \max\{\|p-u\|_2, \|v-q\|_2 \}$, the third step is by $\|v-q\|_2\leq \|u-q\|_2$, the fourth step is by $\|p-q\|_2\geq \|p-u\|_2$, the fifth step follows from $\|p-u\|_2\leq \|u-q\|_2$ (see Eq.~\eqref{eq:upper_bound_p_minus_u}), sixth step is by $\max\{A,B\} \geq 0.5 \cdot (A+B)$, where the seventh step follows from $\|p-q\|_2\leq R$ and Eq.~\eqref{eq:upper_bound_p_minus_u_H}. The eighth step is due to Lemma~\ref{lem:p_u_close}.

It remains to show Eq.~\eqref{eq:upper_bound_p_minus_u_H}. Due to the fact that $p, u, v$ are on the same chord, vector $u-v$ and $p-u$ are on the same direction. This means we can write $u-v=c\cdot (p-u)$ for some $c\in \R$, and 
\begin{align}\label{eq:upper_bound_p_minus_u_H}
    \frac{\|u-v\|_H}{\|p-u\|_H} = & ~ \frac{\|c\cdot (p-u)\|_H}{\|p-u\|_H} \notag \\
    = & ~ c \notag \\
    = & ~ \frac{\|u-v\|_2}{\|p-u\|_2}.
\end{align}
This completes the proof.
\end{proof}

\section{Computing Approximate Lewis Weights and Leverage Scores}\label{sec:leverage}
In Section~\ref{sec:leverage:subsampling}, we present an algorithm to approximate the Hessian. In Section~\ref{sec:leverage:leverage}, we show how to compute leverage score to low precision. In Section~\ref{sec:leverage:lewis}, we show how to compute Lewis weights to high precision.

\subsection{Subsampling to approximate the Hessian}\label{sec:leverage:subsampling}

\begin{algorithm}\caption{Subsampling Algorithm}\label{alg:subsample}
\begin{algorithmic}[1]
\Procedure{Subsample}{$A \in \R^{n \times d}, b \in \R^n, x, n, d, \epsilon_H \in (0,1)$}
    \State $\epsilon_{\sigma} \gets 0.001$
    \State $m \gets \epsilon_H^{-2} d \log d$
    \State Form $B \in \R^{n \times d}$ where $i$-th row of $B$ is $a_i \cdot ( \langle a_i, x \rangle - b_i )^{-1}$
    \State Compute the $O(1\pm \epsilon_{\sigma})$-approximation to the leverage score of $B$ 
    \State Sample a matrix $\wt{B} \in \R^{m \times d}$ such that $(1-\epsilon_H) B^\top B \preceq \wt{B}^\top \wt{B} \preceq (1+\epsilon_H) B^\top B$
    \State \Return $\wt{B}^\top \wt{B}$
\EndProcedure
\end{algorithmic}
\end{algorithm}

In this section, we provide a sparsification result for $H(w)$ using leverage score sampling.

To better monitor the whole process, it is useful to write $H(w)$ as $A^\top S(w)^{-2} A$, where $A\in \R^{n\times d}$ is the constraint matrix and $S(w)$ is a diagonal matrix with $S(w)_i=a_i^\top w-b_i$. The sparsification process is then sample the rows from the matrix $S(w)^{-1}A$.

\begin{definition}
Let $B\in \R^{n\times d}$ be a full rank matrix. We define the leverage score of the $i$-th row of $B$ as 
\begin{align*}
    \sigma_i(B):= & ~ b_i^\top (B^\top B)^{-1} b_i,
\end{align*}
where $b_i$ is the $i$-th row of $B$.
\end{definition}

\begin{definition}[Sampling process]
\label{def:sample_process}
For any $w\in K$, let $H(w)=A^\top S(w)^{-2}A$. Let $p_i\geq \frac{\beta\cdot\sigma_i(S^{-1}(w)A)}{d}$, suppose we sample with replacement independently for $s$ rows of matrix $S(w)^{-1}A$, with probability $p_i$ of sampling row $i$ for some $\beta\geq 1$. Let $i(j)$ denote the index of the row sampled in the $j$-th trial. Define the generated sampling matrix as
\begin{align*}
    \wt H(w) := & ~ \frac{1}{s} \sum_{j=1}^s \frac{1}{p_{i(j)}} \frac{a_{i(j)}a_{i(j)}^\top}{(a_{i(j)}^\top w-b_{i(j)})^2}.
\end{align*}
\end{definition}

\begin{lemma}[Sample using Matrix Chernoff]
\label{lem:tilde_H}
Let $\epsilon, \delta\in (0,1)$ be precision and failure probability parameters, respectively. Suppose $\wt H(w)$ is generated as in Definition~\ref{def:sample_process}, then with probability at least $1-\delta$, we have
\begin{align*}
    (1-\epsilon)\cdot H(w)\preceq \wt H(w) \preceq (1+\epsilon)\cdot H(w).
\end{align*}
Moreover, the number of rows $s=\Theta(\beta\cdot \epsilon^{-2}d\log(d/\delta))$.
\end{lemma}

\begin{proof}
The proof will be designing a family of random matrices $X$. Let $y_i=(A^\top S(w)^{-2}A)^{-1/2}S(w)^{-1}_{i,i}\cdot a_i$ be the $i$-th sampled row and set $Y_i=\frac{1}{p_i} y_iy_i^\top$. Let $X_i=Y_i-I_d$. Note that
\begin{align}\label{eq:sum_y_i}
 \sum_{i=1}^n y_iy_i^\top = & ~ \sum_{i=1}^n (A^\top S(w)^{-2}A)^{-1/2}S(w)^{-2}_{i,i}\cdot a_ia_i^\top (A^\top S(w)^{-2}A)^{-1/2} \notag\\
 = & ~ (A^\top S(w)^{-2}A)^{-1/2} (\sum_{i=1}^n S(w)^{-2}_{i,i}a_ia_i^\top) (A^\top S(w)^{-2}A)^{-1/2} \notag\\
 = & ~ (A^\top S(w)^{-2}A)^{-1/2} (A^\top S(w)^{-2}A) (A^\top S(w)^{-2}A)^{-1/2} \notag\\
 = & ~ I_d.
\end{align}

Also, the norm of $y_i$ connects directly to the leverage score:
\begin{align}\label{eq:y_i_leverage}
    \|y_i\|_2^2 = & ~ S(w)^{-1}_{i,i}a_i^\top (A^\top S(w)^{-2}A)^{-1} S(w)^{-1}_{i,i}a_i\notag \\
    = & ~ \sigma_i(S^{-1}A).
\end{align}
We use $i(j)$ to denote the index of row that has been sampled during $j$-th trial. 

{\bf Unbiased Estimator.}
Note that
\begin{align*}
    \E[X] = & ~ \E[Y]-I_d \\
    = & ~ (\sum_{i=1}^n p_i\cdot \frac{1}{p_i}y_iy_i^\top)-I_d \\
    = & ~ 0.
\end{align*}

{\bf Bound on $\|X\|$.}
To bound $\|X\|$, we provide a bound for any $\|X_i\|$ as follows:
\begin{align*}
    \|X_i\| = & ~ \|Y_i-I_d\| \\
    \leq & ~ 1+\|Y_i\| \\
    = & ~ 1+\frac{\|y_iy_i^\top\|}{p_i} \\
    \leq & ~ 1+\frac{\|y_i\|_2^2}{\beta\cdot \sigma_i(S^{-1}(w)A)} \\
    = & ~ \frac{d}{\beta\cdot \sigma_i(S^{-1}(w)A)} \|(S(w)A)_i\|_2^2+1. \\
    = & ~ 1+\frac{d}{\beta}.
\end{align*}

{\bf Bound on $\|\E[X^\top X]\|$.}
We compute the spectral norm of the covariance matrix:
\begin{align*}
    \E[X_{i(j)}^\top X_{i(j)}] = & ~ I_d+\E[\frac{y_{i(j)}y_{i(j)}^\top y_{i(j)}y_{i(j)}^\top}{p_i^2}]-2\E[\frac{y_{i(j)}y_{i(j)}^\top}{p_i}] \\
    = & ~ I_d+(\sum_{i=1}^n \frac{\sigma_i(S(w)^{-1}A)}{p_i} y_iy_i^\top)-2I_d \\
    \leq & ~ \sum_{i=1}^n \frac{d}{\beta} y_iy_i^\top - I_d \\
    = & ~ (\frac{d}{\beta}-1)I_d,
\end{align*}
the spectral norm is then
\begin{align*}
    \|\E[X_{i(j)}^\top X_{i(j)}]\| \leq & ~ \frac{d}{\beta}-1.
\end{align*}

{\bf Put things together.} Set $\gamma=1+\frac{d}{\beta}$ and $\sigma^2=\frac{d}{\beta}-1$, we apply Matrix Chernoff Bound as in Lemma~\ref{lem:matrix_chernoff}: 
\begin{align*}
    \Pr[\|W\|\geq \epsilon] \leq & ~ 2d\cdot \exp\left(-\frac{s\epsilon^2}{d/\beta-1+(1+d/\beta)\epsilon/3}\right) \\
    = & ~ 2d\cdot \exp(-s\epsilon^2 \cdot \Theta(\beta/d)) \\
    \leq & ~ \delta
\end{align*}

where we choose $s=\Theta(\beta\cdot \epsilon^{-2}d\log(d/\delta))$. Finally, we notice that
\begin{align*}
    W = & ~ \frac{1}{s}(\sum_{j=1}^s \frac{1}{p_{i(j)}}y_{i(j)}y_{i(j)}^\top -I_d) \\
    = & ~  (A^\top S(w)^{-2}A)^{-1/2}(\frac{1}{s}\sum_{j=1}^s \frac{1}{p_{i(j)}} \frac{a_{i(j)}a_{i(j)}^\top}{(a_{i(j)}^\top w-b_{i(j)})^2}  ) (A^\top S(w)^{-2}A)^{-1/2}-I_d \\
    = & ~ H(w)^{-1/2} \wt H(w) H(w)^{-1/2}-I_d.
\end{align*}
Therefore, we can conclude the desired result via $\|W\|\geq \epsilon$.
\end{proof}

\subsection{Computing leverage score}\label{sec:leverage:leverage}

In this section, we provide an algorithm that approximates leverage scores in near-input sparsity time. Our algorithm makes use of the sparse embedding matrix of~\cite{nn13} that has $O(\log (n/\epsilon))$ nonzero entries per column. 

\begin{lemma}[Approximate leverage scores]
Let $A\in \R^{n\times d}$, there exists an algorithm that runs in time $\wt O(\epsilon^{-1}(\nnz(A)+\epsilon^{-1}d^\omega))$ that outputs a vector $\wt w_2(A)\in \R^n$, such that $\wt w_2(A)\approx_\epsilon w_2(A)$.
\end{lemma}

\begin{proof}
We follow an approach of~\cite{w14}, but instead we use a high precision sparse embedding matrix~\cite{nn13} and prove a two-sided bound on leverage score.

Let $S$ be a sparse embedding matrix with $r= O(\epsilon^{-2}d\poly\log(d/(\epsilon\delta)))$ rows and each column has $O(\epsilon^{-1}\log(nd/\epsilon))$ nonzero entries. We first compute $SA$ in time $\wt O(\epsilon^{-1}\nnz(A))$, then compute the QR decomposition $SA=QR$ in time $\wt O(\epsilon^{-2}d^\omega)$. Note that $R\in \R^{d\times d}$ hence $R^{-1}$ can be computed in $O(d^\omega)$ time.

Now, let $G\in \R^{d\times t}$ matrix with $t=O(\epsilon^{-2}\log (n/\delta))$, each entry of $G$ is i.i.d. $\N(0,1/t)$ random variables. Set $q_i=\|e_i^\top AR^{-1}G \|_2^2$ for all $i\in [n]$. We argue $q_i$ is a good approximation to $(w_2(A))_i$.

First, with failure probability at most $\delta/n$, we have that $q_i\approx_\epsilon \|e_i^\top AR^{-1}\|_2^2$ via Johnson-Lindenstrauss lemma~\citep{jl84}. Now, it suffices to argue that $\|e_i^\top AR^{-1}\|_2^2$ approximates $\|e_i^\top U\|_2^2$ well, where $U\in \R^{n\times d}$ is the left singular vectors of $A$. To see this, first observe that for any $x\in \R^d$,
\begin{align*}
    \|AR^{-1} x\|_2^2 = & ~ (1 \pm \epsilon) \cdot \|S A R^{-1} x \|_2^2 \\
    = & ~ (1 \pm \epsilon) \cdot \|Qx\|_2^2 \\
    = & ~ (1 \pm \epsilon) \cdot \|x\|_2^2,
\end{align*}
where the last step is due to $Q$ has orthonormal columns. This means that all singular values of $AR^{-1}$ are in the range $[1-\epsilon, 1+\epsilon]$. Now, since $U$ is an orthonormal basis for the column space of $A$, $AR^{-1}$ and $U$ has the same column space (since $R$ is full rank). This means that there exists a change of basis matrix $T\in \R^{d\times d}$ with $AR^{-1}T=U$. Our goal is to provide a bound on all singular values of $T$. For the upper bound, we claim the largest singular value is at most $1+2\epsilon$, to see this, suppose for the contradiction that the largest singular is larger than $1+2\epsilon$ and let $v$ be its corresponding (unit) singular vector. Since the smallest singular value of $AR^{-1}$ is at least $1-\epsilon$, we have
\begin{align*}
    \|AR^{-1}Tv\|_2^2 \geq & ~ (1-\epsilon) \|T v\|_2^2 \\
    > & ~ (1-\epsilon)(1+2\epsilon) \\
    > & ~ 1,
\end{align*}
however, recall $AR^{-1}T=U$, therefore $\|AR^{-1}Tv\|_2^2=\|Uv\|_2^2=\|v\|_2^2=1$, a contradiction. One can similarly establish a lower bound of $1-2\epsilon$. Hence, the singular values of $T$ are in the range of $[1-2\epsilon, 1+2\epsilon]$. This means that
\begin{align*}
    \|e_i^\top AR^{-1}\|_2^2 = & ~ \|e_i^\top UT^{-1}\|_2^2 \\
    = & ~ (1\pm 2\epsilon)\|e_i^\top U\|_2^2 \\
    = & ~ (1\pm 2\epsilon)(w_2(A))_i,
\end{align*}
as desired. Scaling $\epsilon$ to $\epsilon/2$ yields the approximation result. 

Now, regarding the running time of computing $q_i$, note that we can first multiply $R^{-1}$ with $G$ in time $\wt O(\epsilon^{-2}d^2)$, this gives a matrix of size $d\times t$. Multiplying this matrix with $A$ takes $\wt O(\epsilon^{-1}\nnz(A))$ time. Hence, the overall time for computing $\wt w_2(A)=q\in \R^n$ is $\wt O(\epsilon^{-1} \nnz(A)+\epsilon^{-2} d^\omega)$.
\end{proof}

\begin{remark}
It suffices to choose $\epsilon=\frac{1}{10}$ for generating approximate leverage scores and subsampling the target matrix $H(x)$. 
\end{remark}

\subsection{Computing Lewis weights}\label{sec:leverage:lewis}

Before stating the main technical tool for this section, we want to make several remarks regarding the Lewis weights. 

Since its introduction by Cohen and Peng~\citep{cp15} as an algorithmic tool for $\ell_p$ row sampling, Lewis weights, as its natural formulation follows from a convex program, is \emph{not known to be solved exactly in polynomial time}. All known algorithms~\citep{cp15,ls19,flps22,ccly19,jls22} can \emph{only compute $\epsilon$-approximate Lewis weights}. 

\begin{definition}[$\ell_p$ Lewis weights]
Let $A\in \R^{n\times d}$. The $\ell_p$ Lewis weights of $A$ is a vector $w_p(A)\in \R^n$ satisfying
\begin{align*}
    w_p(A) = & ~ \sigma(W_p(A)^{\frac{1}{2}-\frac{1}{p}}A),
\end{align*}
where $W_p(A)\in \R^{n\times n}$ is the diagonal matrix that puts $w_p(A)$ on the diagonal, $\sigma: \R^{n\times d}\rightarrow \R^n$ is the operation that outputs all leverage scores of input matrix.
\end{definition}

We are now posed to state a main tool from~\cite{ls19}.

\begin{theorem}[Theorem 46 in \cite{ls19}]
\label{thm:lewis_weights}
Let $P = \{ x : A x > b \}$ denote the interior of non-empty polytope for non-degenerate $A \in \R^{n \times d}$. There is an $O(d \log^5 n)$-self concordant barrier $\psi$ defined by $\ell_p$ Lewis weight with $p= \Theta(\log n)$ satisfying
\begin{align*}
A_x^\top W_x A_x \preceq \nabla^2 \psi(x) \preceq (q+1) A_x^\top W_x A_x
\end{align*}
where $A_x = \mathrm{diag}(Ax - b)$ and $w_x$ is the $\ell_p$ Lewis weight of the matrix $A_x$. Furthermore, we can compute or update the $w_x$, $\nabla \psi(x)$ and $\nabla^2 \psi(x)$ as follows:
\begin{itemize}
    \item Initial Weight: For any $x \in \R^d$, one can compute a vector $\wt{w}_x$ such that $(1-\epsilon) w_x \leq \wt{w}_x \leq (1+\epsilon) w_x$ in $O(n d^{\omega-1/2} \cdot \log^3 n \log(n/\epsilon) )$ time.
    \item Update Weight and Compute Gradient/Hessian: Given a vector $\wt{w}_x$ such that $\wt{w}_x = (1\pm \frac{1}{100}) w_x$ for any $y$ with $\| x - y \|_{A_x^\top W_x A_x} \leq \frac{c}{\log^2 n}$ for some small constant $c>0$, we can compute $\wt{w}_y$, $v$ and $H$ such that $\wt{w}_y = (1\pm\epsilon) w_y$,
    \begin{align*}
        \| v- \nabla \psi(x) \|_{ \nabla^2 \psi(x)^{-1} } \leq \epsilon, \mathrm{~~~and~~~} (1-\epsilon) \nabla^2 \psi(x) \preceq H \preceq (1+\epsilon) \nabla^2 \psi(x)
    \end{align*}
    in $O(n d^{\omega-1} \cdot \log n \cdot \log(n/\epsilon))$ time.
\end{itemize}
\end{theorem}

\section{Fast Approximate Sampling from Polytopes: Complexity}\label{sec:time}

In this section, we provide the runtime analysis for sampling from polytopes via log-barrier and Lee-Sidford barrier. The result for log-barrier is in Section~\ref{sec:time:log_barrier} and for Lee-Sidford barrier is in Section~\ref{sec:time:lee_sidford_barrier}.

\subsection{Log-barrier in nearly-linear time}\label{sec:time:log_barrier}

Throughout this section, we let $\alpha \in (0, 1/(10^5 d) )$ and $\eta \in (0, 1/(20 d L^2) )$.

\begin{lemma} \label{lem:overall_time}
If $g$ is log-barrier function with $\nu = n$, then each iteration of Algorithm \ref{alg:main_logconcave} can be implemented in $ O(\nnz(A) + \Tmat(d,d^3,d))$ time plus $O(1)$ calls to the oracle of $f$. 
\end{lemma}
\begin{proof}
We go through each step of Algorithm \ref{alg:main_logconcave} and add up the time and oracle calls for each step:

We first need to sample a $d$-dimensional Gaussian random vector $\xi \sim N(0, I_d)$, which can be performed in $O(d)$ time.

At each iteration, by the definition of log-barrier function, we can write the Hessian as follows:
\begin{align*}
    H(x) := \sum_{j=1}^{n} \frac{a_j a_j^\top}{(a_j^\top x - b_j)^2}.
\end{align*}

We define define matrix $B \in \R^{n \times d}$ as follows
\begin{align*}
    B := S^{-1} A
\end{align*}
where $s_i = ( \langle a_i, x \rangle -b_i )$, $\forall i \in [n]$. Then we can write $H(x) = B^\top B \in \R^{d \times d}$.

For matrix $B$, we can compute a constant approximation  of its leverage score, then we can sample according to leverage, and generate a diagonal matrix $D$ such that
\begin{align*}
    (1-\epsilon) B^\top B \preceq B^\top D D B \preceq (1+\epsilon) B^\top B.
\end{align*}
This step takes $\wt{O}( \nnz(A) + d^{\omega(1,3,1)} )$ time.

Computing $\wt{\Phi}(x)$ can be done in $d^2$ time and computing the vector $z$
\begin{align*}
    z = & ~ x + \wt{\Phi}(x)^{-\frac{1}{2}} \xi
\end{align*}
requires to invert the matrix $\Phi(x)$, taking the square root and multiplying with a vector $\xi$. Inversion and square root can be performed in time $O(d^\omega)$ by computing its spectral decomposition, and the matrix-vector product can be done in time $O(d^2)$. Hence, the overall time of this step is $O(d^\omega)$.
    
  To determine whether $z \in K$, one can simplify verify $Az\leq b$, in time $O(\nnz(A))$.

  We also need to compute $\wh{H}(z)$ and $\wh{\Phi}(z)$. This is similar to computing $\wt{H}(x)$ and $\wt{\Phi}(x)$.
   
 We then need to compute the determinant $\det(\wt{\Phi}(x))$ and $\det(\wh{\Phi}(z))$  which can be done in $O(d^\omega)$ time via a spectral decomposition. 

 We then need $O(1)$ oracle queries to the function values of $f$.

Therefore, adding up the number of arithmetic operations and oracle calls from all the different steps of Algorithm \ref{alg:main_logconcave}, we get that each iteration of Algorithm \ref{alg:main_logconcave} can be computed in $ \wt{O}(\nnz(A) + \Tmat(d,d^3, d))$ arithmetic operations plus $O(1)$ calls to the oracle of $f$.
\end{proof}

\begin{remark}
We note a recent work~\cite{mv24} has provided an algorithm that runs in $\nnz(A)+d^2$ time per iteration, by adapting techniques from~\cite{llv20}. Their per iteration cost improvement only works for log-barrier, and they could only manage to obtain an $\wt O(nd+dL^2R^2)$ mixing time. Our framework on the other hand generalizes to other barriers such as Lee-Sidford barrier, which provides a nearly-optimal mixing rate. We also adapt the framework to sampling from spectrahedra.
\end{remark}

\subsection{Lee-Sidford barrier via approximation scheme}\label{sec:time:lee_sidford_barrier}

\begin{lemma}
If $g$ is the Lee-Sidford barrierfunction with $\nu=d\log^5 n$, then, each iteration of Algorithm~\ref{alg:main_logconcave} can be implemented in $\wt O(nd^{\omega-1})$ time plus $O(1)$ calls to the oracle of $f$.
\end{lemma}
\begin{proof}
For Lee-Sidford barrier, it suffices to use $H(x)=A_x^\top W_x A_x$ as where $W_x$ is the $\ell_p$ Lewis weights for $p=\Theta(\log m)$. However, as Lewis weights cannot be computed exactly, we shall use the algorithm of~\cite{ls19} to compute an $\epsilon$-approximation. 

To invoke Theorem~\ref{thm:lewis_weights}, we need to make sure that $z$ and $x$ satisfy $\|z-x\|_{A_x^\top W_x A_x}\leq \frac{c}{\log^2 n}$. We can achieve so by scaling down $\alpha$ by a factor of $\log^2 n$. This only blows up the convergence by $\wt O(1)$ factor, so it is acceptable. 

The runtime analysis is then similar to Lemma~\ref{lem:overall_time}. Computing the initial Lewis weights takes $\wt O(nd^{\omega-1/2}\log(1/\epsilon))$ time owing to Theorem~\ref{thm:lewis_weights}. As the algorithm requires $\wt O(d^2)$ iterations, this time can be amortized. For each iteration, it needs to query the Lewis weights data structure that outputs an $\epsilon$-Lewis weights in $\wt O(nd^{\omega-1}\log(1/\epsilon))$ time. As we will choose $\epsilon$ as $O(1/d)$, this time is $\wt O(nd^{\omega-1})$. We can then use this approximate Hessian to progress the algorithm. All subsequent operations take $O(d^\omega)$ time. Thus, the cost per iteration is $\wt O(nd^{\omega-1})$.
\end{proof}

\section{Approximate Sampling from a Spectrahedron}\label{sec:sdp}

We present an algorithm that efficiently and approximately samples from a spectrahedron, which is a popular convex body utilized by semidefinite programs.
\subsection{Definitions}

\begin{definition}
Let $A_1,\ldots,A_d\in \R^{n\times n}$ be a collection of symmetric matrices and $C\in \R^{n\times n}$ be symmetric. We define the corresponding spectrahedron as
\begin{align*}
    {\cal K} = & ~ \{x\in \R^d: \sum_{i=1}^d x_iA_i\succeq C \}.
\end{align*}
\end{definition}

We define the log-barrier for spectrahedron and its corresponding Hessian.

\begin{definition}[\cite{nn94}]\label{def:sdp_log_barrier}
Let ${\cal K}$ be a spectrahedron described by symmetric matrices $\{A_1,\ldots,A_d \}\subseteq \R^{n\times n}$ and $C\in \R^{n\times n}$. The log barrier $\phi_{\rm log}(x)$ is defined as
\begin{align*}
    \phi_{\rm log}(x) = & ~ -\log\det(S(x))
\end{align*}
and its corresponding Hessian is defined as
\begin{align*}
    H_{\rm log}(x) = & ~ {\sf A}(S(x)^{-1}\otimes S(x)^{-1}){\sf A}^\top,
\end{align*}
where $S(x):=\sum_{i=1}^d x_iA_i-C\in \R^{n\times n}$ and ${\sf A}\in \R^{d\times n^2}$ and the $i$-th row of ${\sf A}$ is the vectorization of $A_i \in \R^{n \times n}$.
\end{definition}

To compute $H_{\rm log}(x)$, it is handy to define a matrix ${\sf B} \in \R^{d \times n^2}$.

\begin{definition}
\label{def:sdp_B}
We define a matrix ${\sf B}\in \R^{d\times n^2}$ as ${\sf B}={\sf A}(S(x)^{-1/2}\otimes S(x)^{-1/2})$. Consequently, $H_{\rm log}(x)={\sf B}{\sf B}^\top$.
\end{definition}

We state a useful lemma for computing the matrix ${\sf B}$.

\begin{lemma}
Let matrix ${\sf B}\in \R^{d\times n^2}$ be defined as in Def.~\ref{def:sdp_B}. Then, the $i$-th row of ${\sf B}$ can be computed as $\vect(S(x)^{-1/2}A_iS(x)^{-1/2})$.
\end{lemma}
\begin{proof}
The proof relies on a simple fact of Kronecker product and vectorization:
\begin{align*}
\vect(S(x)^{-1/2}A_iS(x)^{-1/2}) = & ~ (S(x)^{-1/2}\otimes S(x)^{-1/2})\vect(A_i),
\end{align*}
which is the definition of the $i$-th row of ${\sf B}$.
\end{proof}

\subsection{\texorpdfstring{$n$}{}-symmetry of \texorpdfstring{$H_{\rm log}$}{} for spectrahedra}

In this section, we prove that $H_{\rm log}$ is $n$-symmetry for spectrahedra.

\begin{lemma}
$H_{\rm log}(x)$ is $n$-symmetry, that is, for any $x\in \K$, 
\begin{align*}
    E_x(1) \subseteq \K \cap (2x-\K) \subseteq E_x(\sqrt n).
\end{align*}
\end{lemma}

\begin{proof}
Pick a point $y\in E_x(1)$, then we know that $(y-x)^\top {\sf A}(S(x)^{-1}\otimes S(x)^{-1}){\sf A}^\top (y-x)\leq 1$. We note that the set $\K\cap (2x-\K)$ can be characterized as the set of all points $u\in \R^d$ such that
\begin{align*}
    \sum_{i=1}^d u_i A_i \succeq & ~C, \\
    \sum_{i=1}^d (2x-u)_i A_i \succeq & ~ C.
\end{align*}
These two conditions imply
\begin{align*}
    -S(x) \preceq \sum_{i=1}^d (u_i-x_i)A_i \preceq S(x), \\
    -I_n \preceq \sum_{i=1}^d (u_i-x_i)S(x)^{-1/2}A_iS(x)^{-1/2}\preceq I_n.
\end{align*}
We will prove $y$ satisfies these conditions. Note that $y\in E_x(1)$ implies that
\begin{align*}
    \|\sum_{i=1}^d (y_i-x_i) \vect(S(x)^{-1/2}A_i S(x)^{-1/2})\|_2^2 \leq & ~ 1, \\
    \|\sum_{i=1}^d (y_i-x_i) S(x)^{-1/2}A_i S(x)^{-1/2}\|_F^2 \leq & ~ 1
\end{align*}
if we let $M$ denote $\sum_{i=1}^d (y_i-x_i) S(x)^{-1/2}A_i S(x)^{-1/2}$, then the above condition implies that $M^\top M\preceq I_n$, meaning all the eigenvalues of $M$ lie between $[-1, 1]$, thus we have shown that $E_x(1)\subseteq \K\cap (2x-\K)$.

For the other direction, we know that
\begin{align*}
    -I_n \preceq \sum_{i=1}^d (y_i-x_i)S(x)^{-1/2}A_iS(x)^{-1/2}\preceq I_n,
\end{align*}
and henceforth
\begin{align*}
    & ~ (y-x)^\top {\sf A}(S(x)^{-1}\otimes S(x)^{-1}){\sf A}^\top (y-x) \\
    = & ~ \|\sum_{i=1}^d (y_i-x_i)S(x)^{-1/2}A_iS(x)^{-1/2}\|_F^2 \\
    \leq & ~ \|I_n\|_F^2 \\
    = & ~ n,
\end{align*}
where we use the fact that $-I_n\preceq M\preceq I_n$ hence all eigenvalues of $M$ lie between $[-1, 1]$, and the squared eigenvalues of $M$ have its magnitude at most 1. We then use the squared Frobenius norm is equivalent to squared $\ell_2$ norm of singular values of $M$, as desired.
\end{proof}

\subsection{Bounded local norm of \texorpdfstring{$H_{\rm log}(x)$}{}}

We prove that $H_{\rm log}(x)$ has the bounded local norm property.
\begin{lemma}
\label{lem:sdp_log_convex}
Let $H_{\rm log}(x)\in \R^{d\times d}$ be defined as in Def.~\ref{def:sdp_log_barrier}. Let $F(x)=\log\det (H_{\rm log}(x))$. Then, $F(x)$ is convex in $x$.
\end{lemma}

\begin{proof}
The proof is by observe that the function $F(x)$ is actually the volumetric barrier function for the SDP spectrahedron. Thus, the function $F$ is convex, for more details, see~\cite{nn94}.
\end{proof}

\begin{lemma}
Let $F(x)$ be defined as in Lemma~\ref{lem:sdp_log_convex}. Then, we have
\begin{align*}
    \nabla F(x) = & ~ \begin{bmatrix}
        -\tr[{\sf P}((S(x)^{-1/2}A_1S(x)^{-1/2})\otimes_S (S(x)^{-1/2}A_1S(x)^{-1/2}))] \\
        -\tr[{\sf P}((S(x)^{-1/2}A_2S(x)^{-1/2})\otimes_S (S(x)^{-1/2}A_2S(x)^{-1/2}))] \\
        \vdots \\
        -\tr[{\sf P}((S(x)^{-1/2}A_mS(x)^{-1/2})\otimes_S (S(x)^{-1/2}A_mS(x)^{-1/2}))]
    \end{bmatrix}
\end{align*}
where ${\sf P}={\sf B}^\top H(x)^{-1}{\sf B}$.
\end{lemma}

\begin{proof}
For simplicity, we let $H$ to denote $H_{\rm log}$. For each coordinate $j\in [d]$, we have
\begin{align*}
    \frac{\d F}{\d x_j} = & ~ \tr[H(x)^{-1} \frac{\d H(x)}{\d x_j}] \\
    = & ~ \tr[H(x)^{-1}{\sf A}\frac{\d (S(x)^{-1}\otimes S(x)^{-1})}{\d x_j} {\sf A}^\top] \\
    = & ~ \tr[H(x)^{-1}{\sf A}(\frac{\d S(x)^{-1}}{\d x_j}\otimes S(x)^{-1}+S(x)^{-1}\otimes \frac{\d S(x)^{-1}}{\d x_j}) {\sf A}^\top] \\
    = & ~ -\tr[H(x)^{-1}{\sf A}((S(x)^{-1}\frac{\d S(x)}{\d x_j}S(x)^{-1})\otimes S(x)^{-1}+S(x)^{-1}\otimes (S(x)^{-1}\frac{\d S(x)}{\d x_j}S(x)^{-1})) {\sf A}^\top] \\
    = & ~ -\tr[H(x)^{-1}{\sf A}((S(x)^{-1}A_jS(x)^{-1})\otimes S(x)^{-1}+S(x)^{-1}\otimes (S(x)^{-1}A_jS(x)^{-1})) {\sf A}^\top],
\end{align*}
examine the following term:
\begin{align*}
    & ~ \tr[H(x)^{-1}{\sf A}((S(x)^{-1}A_jS(x)^{-1})\otimes S(x)^{-1}){\sf A^\top}] \\
    = & ~ \tr[{\sf A}^\top H(x)^{-1}{\sf A}((S(x)^{-1}A_j)\otimes I)(S(x)^{-1}\otimes S(x)^{-1})] \\
    = & ~ \tr[(S(x)^{-1/2}\otimes S(x)^{-1/2}){\sf A}^\top H(x)^{-1}{\sf A}(S(x)^{-1/2}\otimes S(x)^{-1/2})\\
    & ~ \cdot ((S(x)^{-1/2}A_j)\otimes S(x)^{1/2})(S(x)^{-1/2}\otimes S(x)^{-1/2})] \\
    = & ~ \tr[{\sf P}((S(x)^{-1/2}A_j)\otimes S(x)^{1/2})(S(x)^{-1/2}\otimes S(x)^{-1/2})] \\
    = & ~ \tr[{\sf P}((S(x)^{-1/2}A_jS(x)^{-1/2})\otimes I_n)],
\end{align*}
where ${\sf P}\in \R^{n^2\times n^2}$ is the projection of ${\sf B}^\top$, i.e., ${\sf P}={\sf B}^\top H(x)^{-1}{\sf B}$. Thus,
\begin{align*}
    \nabla F(x) = & ~ \begin{bmatrix}
        -\tr[{\sf P}((S(x)^{-1/2}A_1S(x)^{-1/2})\otimes_S (S(x)^{-1/2}A_1S(x)^{-1/2}))] \\
        -\tr[{\sf P}((S(x)^{-1/2}A_2S(x)^{-1/2})\otimes_S (S(x)^{-1/2}A_2S(x)^{-1/2}))] \\
        \vdots \\
        -\tr[{\sf P}((S(x)^{-1/2}A_mS(x)^{-1/2})\otimes_S (S(x)^{-1/2}A_mS(x)^{-1/2}))]
    \end{bmatrix}. 
\end{align*}
This completes the proof.
\end{proof}

\begin{lemma}
\label{lem:sdp_log_barrier_is_strong_sc}
The matrix function $H(x) = {\sf A} (S^{-1} \otimes S^{-1}) {\sf A}^\top \in \R^{d \times d}$ has the following bounded norm property: for any direction $h\in \R^d$,
\begin{align*}
    \|H(x)^{-1/2} DH(x)[h] H(x)^{-1/2} \|_F^2 \leq & ~ 4d \|h\|_{H(x)}^2.
\end{align*}
\end{lemma}
\begin{proof}
Let $x_t = x + t \cdot h$ for some fixed vector $h \in \R^{d}$, $S_t = \sum_{i=1}^m x_{t,i} A_i - C \in \R^{n \times n}$, ${\sf A}_t = {\sf A}  (S_t^{-1/2} \otimes S_t^{-1/2} ) \in \R^{d \times n^2}$, $P_t = {\sf A}_t^\top ({\sf A}_t {\sf A}_t^\top)^{-1} {\sf A}_t \in \R^{n^2 \times n^2}$ and $\Sigma_t = \sum_{i=1}^n (e_i^\top\otimes I_n) P_t (e_i\otimes I_n)=\sum_{i=1}^n (I_n\otimes e_i^\top)P_t (I_n\otimes e_i)$. Note that for any matrix $M\in \R^{n\times n}$, $\tr[P_t\cdot (M\otimes I_n)]=\tr[P_t\cdot (I_n\otimes M)]=\tr[\Sigma_t\cdot M]$. Also note that $\tr[\Sigma_t]=\tr[P_t]=d$.



Let $H_t = {\sf A}^\top ( S_t^{-1} \otimes S_t^{-1} ) {\sf A} \in \R^{d \times d}$.
We have
\begin{align*}
& ~ \| H_t^{-1/2} ( \frac{\d}{\d t} H_t ) H_t^{-1/2} \|_F^2 \\
= & ~ \tr[ H_t^{-1} \cdot ( \frac{\d }{\d t} H_t ) \cdot H_t^{-1} \cdot (\frac{\d}{\d t} H_t) ] \\
= & ~ \tr[ H_t^{-1} \cdot {\sf A}^\top \frac{\d S_t^{-1} \otimes S_t^{-1}}{ \d t} {\sf A} \cdot H_t^{-1} {\sf A}^\top \frac{ \d ( S_t^{-1} \otimes S_t^{-1} ) }{\d t} {\sf A} ] \\
= & ~ \tr[ {\sf A} H_t^{-1} {\sf A}^\top \cdot \frac{\d S_t^{-1} \otimes S_t^{-1}}{ \d t} \cdot {\sf A}  H_t^{-1} {\sf A}^\top \cdot \frac{ \d ( S_t^{-1} \otimes S_t^{-1} ) }{\d t}  ] \\
= & ~ \tr[ P_t (S_t^{-1/2} \otimes S_t^{-1/2})^{-1} \frac{\d S_t^{-1} \otimes S_t^{-1}}{ \d t} (S_t^{-1/2} \otimes S_t^{-1/2})^{-1} \\
& ~ \cdot P_t (S_t^{-1/2} \otimes S_t^{-1/2})^{-1} \frac{\d S_t^{-1} \otimes S_t^{-1}}{ \d t} (S_t^{-1/2} \otimes S_t^{-1/2})^{-1} ] \\
\leq & ~ \tr[ P_t \cdot (S_t^{-1/2}\otimes S_t^{-1/2})^{-1}\frac{\d S_t^{-1} \otimes S_t^{-1}}{ \d t}(S_t^{-1} \otimes S_t^{-1} )^{-1} \frac{\d S_t^{-1} \otimes S_t^{-1}}{ \d t} (S_t^{-1/2}\otimes S_t^{-1/2})^{-1}] \\
= & ~ \tr[P_t\cdot (S_t^{-1/2}(\sum_{i=1}^d h_iA_i)S_t^{-1/2}\otimes_S I_n)^2] \\
= & ~ \tr[P_t\cdot ((S_t^{-1/2}(\sum_{i=1}^d h_iA_i)S_t^{-1/2})^2\otimes_S I_n+(S_t^{-1/2}(\sum_{i=1}^d h_iA_i)S_t^{-1/2}\otimes S_t^{-1/2}(\sum_{i=1}^d h_iA_i)S_t^{-1/2}))] \\
= & ~ 2\tr[S_t^{-1/2}(\sum_{i=1}^d h_iA_i)S_t^{-1/2}\cdot \Sigma_t \cdot S_t^{-1/2}(\sum_{i=1}^d h_iA_i)S_t^{-1/2}]\\
+ & ~ \tr[P_t\cdot (S_t^{-1/2}(\sum_{i=1}^d h_iA_i)S_t^{-1/2}\otimes S_t^{-1/2}(\sum_{i=1}^d h_iA_i)S_t^{-1/2})]\\ 
\leq & ~ 4\tr[S_t^{-1/2}(\sum_{i=1}^d h_iA_i)S_t^{-1/2}\cdot \Sigma_t \cdot S_t^{-1/2}(\sum_{i=1}^d h_iA_i)S_t^{-1/2}] \\
\leq & ~ 4\|\Sigma_t\| \tr[(S_t^{-1/2}(\sum_{i=1}^d h_iA_i)S_t^{-1/2})^2 ] \\
\leq & ~ 4\tr[\Sigma_t] \tr[(S_t^{-1/2}(\sum_{i=1}^d h_iA_i)S_t^{-1/2})^2] \\
= & ~ 4d \|h\|_{H_t}^2,
\end{align*}
the fifth step is by for PSD matrix $M$, we have $M P M \preceq M^2$ where $P$ is a projection matrix, the sixth step is by the below derivation, the seventh step is by $\tr[P_t\cdot (M^2\otimes_S I_n)]=2\tr[M \Sigma_t M]$, the eighth step is by for PSD matrix $M$, $2M\otimes_S I_n\succeq M\otimes M$, the ninth step is by H{\"o}lder's inequality where $\langle A, B\rangle\leq \|A\|\cdot \|B\|_{s_1}$ for $\|B\|_{s_1}$ be its Schatten-1 norm. If $B$ is PSD, $\|B\|_1=\tr[B]$, the tenth step is by for PSD matrix $A$, $\|A\|\leq \tr[A]$.

Note that
\begin{align*}
\frac{\d S_t^{-1}}{ \d t} = & ~ -S_t^{-1} \frac{\d S_t}{ \d t} S_t^{-1} \\
= & ~ -S_t^{-1} ( \sum_{i=1}^d h_i A_i ) S_t^{-1}
\end{align*}
Thus, 
\begin{align*}
\frac{\d S_t^{-1} \otimes S_t^{-1}}{ \d t}
= & ~ -  S_t^{-1} ( \sum_{i=1}^d h_i A_i ) S_t^{-1} \otimes S_t^{-1}\\
& ~ -   S_t^{-1} \otimes S_t^{-1} ( \sum_{i=1}^d h_i A_i ) S_t^{-1}
\end{align*}
and we have
\begin{align*}
    & ~ (S_t^{-1/2}\otimes S_t^{-1/2})^{-1}\frac{\d S_t^{-1} \otimes S_t^{-1}}{ \d t}(S_t^{-1} \otimes S_t^{-1} )^{-1} \frac{\d S_t^{-1} \otimes S_t^{-1}}{ \d t} (S_t^{-1/2}\otimes S_t^{-1/2})^{-1} \\
    = & ~ (S_t^{-1/2}\otimes S_t^{-1/2})^{-1}(S_t^{-1} ( \sum_{i=1}^d h_i A_i ) S_t^{-1} \otimes S_t^{-1}+ S_t^{-1} \otimes S_t^{-1} ( \sum_{i=1}^d h_i A_i ) S_t^{-1})(S_t^{-1/2} \otimes S_t^{-1/2} )^{-1} \\
    \cdot & ~ (S_t^{-1/2} \otimes S_t^{-1/2} )^{-1}(S_t^{-1} ( \sum_{i=1}^d h_i A_i ) S_t^{-1} \otimes S_t^{-1}+ S_t^{-1} \otimes S_t^{-1} ( \sum_{i=1}^d h_i A_i ) S_t^{-1}) (S_t^{-1/2}\otimes S_t^{-1/2})^{-1} \\
    = & ~ (S_t^{-1/2}(\sum_{i=1}^d h_iA_i)S_t^{-1/2}\otimes_S I_n)^2,
\end{align*}
plug in the above derivation we obtain the desired result. Finally, we send $t\rightarrow 0$ and conclude the proof.
\end{proof}

\begin{lemma}
Let $H(x)$ be the matrix function ${\sf A}(S^{-1}\otimes S^{-1}) {\sf A}^\top$, then we have
\begin{align*}
    \|H(x)^{-1/2}\nabla F(x)\|_2 \leq & ~ \wt O(d).
\end{align*}
\end{lemma}
\begin{proof}
We have
\begin{align*}
\| H(x)^{-1/2} \nabla F(x) \|_2
= & ~ \max_{v: \| v \|_2 =1} ( H(x)^{-1/2} \nabla F(x) )^\top v \\
= & ~ \max_{v: \| v \|_2 =1} \tr[ H(x)^{-1} D H(x) [ H(x)^{-1/2} v ]  ] \\
= & ~ \max_{u: \| u \|_{H(x)} =1} \tr[ H(x)^{-1/2} D H(x)[u] H(x)^{-1/2} ] \\
\leq & ~ \max_{u: \| u \|_{H(x)} =1} \sqrt{d} \| H(x)^{-1/2} D H(x)[u] H(x)^{-1/2} \|_F \\
\leq & ~ 2 d\cdot\| u \|_{H(x)} \\
= & ~ 2d,
\end{align*}
where the fourth step is by $\tr[M]\leq \sqrt{d}\cdot \|M\|_F$ for $d\times d$ PSD matrix $M$, the fifth step follows from Lemma~\ref{lem:sdp_log_barrier_is_strong_sc}, the last step follows from $\| u \|_{H(x)}=1$.
\end{proof}

\subsection{Fast Hessian approximation}

We need a particular type of sketch for Kronecker product of matrices.

\begin{definition}[\textsf{TensorSRHT} \citep{swyz21}]\label{def:tensor_srht}
The \textsf{TensorSRHT} $\Pi: \R^n \times \R^n \to \R^s$ is defined as $S := \frac{1}{\sqrt{s}} P \cdot (HD_1 \otimes HD_2)$, where each row of $P \in \{0, 1\}^{s \times n^2}$ contains only one $1$ at a random coordinate and one can view $P$ as a sampling matrix. $H$ is a $n \times n$ Hadamard matrix, and $D_1$, $D_2$ are two $n \times n$ independent diagonal matrices with diagonals that are each independently set to be a Rademacher random variable (uniform in $\{-1, 1\}$).  
\end{definition}

\begin{lemma}[Lemma 2.12 in~\cite{swyz21}]\label{lem:tensor_srht_implies_ose}
    Let $\Pi$ be a {\sf TensorSRHT} matrix defined in Definition~\ref{def:tensor_srht}. If $s=O(\epsilon^{-2}d \log^3(nd / (\epsilon\delta) ) )$, then for any orthonormal basis $U\in \R^{n^2\times d}$, we have that with probability at least $1-\delta$, the singular values of $\Pi U$ lie in the range of $[1-\epsilon, 1+\epsilon]$. 
\end{lemma}

The following result provides an efficient embedding for ${\sf B}$.

\begin{lemma}
\label{lem:sketch_B}
Let ${\sf B}\in \R^{d\times n^2}$ be defined as in Def.~\ref{def:sdp_B}, $\epsilon \in (0,1/10)$ denote an accuracy parameter and $\delta \in (0,1/10)$ denote a failure probability.

Let $\Pi\in \R^{s\times n^2}$ be a {\sf TensorSRHT} matrix with $s=\Theta(\epsilon^{-2}d\log^3(nd/ (\epsilon\delta) ))$, then we have 
\begin{align*}
   \Pr[  \|\Pi{\sf B}^\top x\|_2 = & ~ (1\pm\epsilon) \|{\sf B}^\top x\|_2, \forall x \in \R^d ] \geq 1-\delta.
\end{align*}

Moreover, $\Pi{\sf B}^\top$ can be computed in time 
\begin{align*}
    O(d\cdot \Tmat(n,n,s)).
\end{align*}
\end{lemma}

\begin{proof}
The correctness part follows directly from Lemma~\ref{lem:tensor_srht_implies_ose}. It remains to argue for the running time. We need to unravel the construction of both $S$ and ${\sf B}$ (see Definition~\ref{def:sdp_log_barrier}). Recall that $\Pi=\frac{1}{\sqrt s}P\cdot (HD_1\otimes HD_2)$ and 
\begin{align*}
    & ~ \Pi(S(x)^{-1/2}\otimes S(x)^{-1/2})\vect(A_i) \\
    = & ~ \frac{1}{\sqrt s} P\cdot (HD_1\otimes HD_2)\cdot (S(x)^{-1/2}\otimes S(x)^{-1/2})\vect(A_i) \\
    = & ~ \frac{1}{\sqrt s} P\cdot (HD_1S(x)^{-1/2}\otimes HD_2S(x)^{-1/2})\vect(A_i) \\
    = & ~ \frac{1}{\sqrt s}P\cdot \vect(HD_2S(x)^{-1/2}A_iS(x)^{-1/2}D_1H^\top)
\end{align*}
since $P$ is a row sampling matrix, the product can be computed as follows:
\begin{itemize}
    \item First compute $HD_1S(x)^{-1/2}$ and $HD_2S(x)^{-1/2}$. Since $H$ is a Hadamard matrix, this step can be carried out in $O(n^2\log n)$ time.
    \item Applying $P$ to the vector can be interpreted as sampling $s$ coordinates from the matrix $HD_2S(x)^{-1/2}A_iS(x)^{-1/2}D_1H^\top$. Let $(i_1,j_1),\ldots,(i_s,j_s)$ denote the coordinates sampled by $P$. We form two matrices $X, Y\in \R^{s\times n}$ where the $k$-th row of $X$ is $(HD_2S(x)^{-1/2})_{i_k,*}$ and the $k$-th row of $Y$ is $(HD_1S(x)^{-1/2})_{j_k,*}$. It is easy to verify that the $(i_k,j_k)$-th entry of $XA_iY^\top \in \R^{s \times s}$ is the corresponding entry of $HD_2S(x)^{-1/2}A_iS(x)^{-1/2}D_1H^\top$. This step therefore takes $\Tmat(n, n, s)$ time.
\end{itemize}
As we need to apply the second step to $d$ rows, the total runtime is 
\begin{align*}
   & ~  O(d\cdot \Tmat(n,n,s)). \qedhere
\end{align*}
\end{proof}

\begin{remark}
For comparison, compute ${\sf B}$ straightforwardly using the Kronecker product identity takes $O(dn^\omega)$ time. As long as the row count $s$ is independent of $n$, Lemma~\ref{lem:sketch_B} approximately computes ${\sf B}$ in time $\wt O(n^2\cdot \poly(d))$. For the regime where $d\ll n$, our algorithm is much more efficient.
\end{remark}

\section{Convexity of Regularized Log-Barrier Function}
\label{sec:barrier}

In this section, we prove the convexity of $\log\det(H_{\rm log}(x)+I_d)$ for both polytopes and spectrahedra.

\subsection{Convexity of \texorpdfstring{$\log\det(H_{\rm log}(x)+I_d)$}{}: polytopes}\label{sec:assumption:convexity_holds}

We prove that for $H(x)=\sum_{i=1}^n \frac{a_ia_i^\top}{(a_i^\top x-b_i)^2}$, the Hessian of the log barrier, we have the convexity.

\begin{definition}[Ridge leverage score]

Let $B\in \R^{n\times d}$, we define the $\lambda$-ridge leverage score of $B$ as 
\begin{align*}
    \wt \sigma_{\lambda, i}(B) := & ~ b_i^\top (B^\top B+\lambda I_d)^{-1} b_i.
\end{align*}
If $\lambda=1$, we abbrieviate as $\wt \sigma_i(B)$.

For convenient of the proof, we also define
\begin{align*}
    \wt \sigma_{\lambda, i,j}(B) := & ~ b_i^\top (B^\top B+\lambda I_d)^{-1} b_j.
\end{align*}
We also define $\wh{\sigma}_{\lambda,i}(B)$ as follows:
\begin{align*}
    \wh{\sigma}_{\lambda, i}(B) := & ~ b_i^\top (B^\top B+\lambda I_d)^{-2} b_i.
\end{align*}
\end{definition}

\begin{fact}\label{fac:d_X_inverse}
Let $X$ be $\R^{d \times d}$ be invertible and symmetric. Then we have
\begin{align*}
    \frac{\d X^{-1}}{\d t} = - X^{-1} \frac{\d X}{ \d t} X^{-1}.
\end{align*}
\end{fact}

\begin{lemma}
Let $X\in \R^{d\times d}$ be invertible and symmetric. Then,
\begin{align*}
    \d \log \det X = & ~ X^{-1} \d X.
\end{align*}
\end{lemma}

\begin{proof}
The proof is by chain rule:

\begin{align*}
    \d \log \det X = & ~ \frac{1}{\det X} \d \det X,
\end{align*}
to compute $\d \det X$, recall the adjugate of $X$, denoted by ${\rm adj}(X)$ where $X^{-1}=\frac{1}{\det X}~{\rm adj}(X)$, and it remains to show that $\d \det X = {\rm adj}(X)$, this can be derived using cofactor expansion of a matrix, which states that fix a row $i$, we have
\begin{align*}
    \det X = & ~ \sum_{k=1}^d X_{i, k}\cdot  {\rm adj}(X)_{i, k},
\end{align*}
by product rule,
\begin{align*}
    \frac{\d \det X}{\d X_{i, j}} = & ~ \sum_{k=1}^d \frac{\d X_{i,k}}{\d X_{i,j}} \cdot {\rm adj}(X)_{i,k}+X_{i,k} \cdot \frac{\d~{\rm adj}(X)_{i,k}}{\d X_{i,j}} \\
    = & ~ {\rm adj}(X)_{i,j},
\end{align*}
where $\frac{\d~{\rm adj}_{i,k}}{\d X_{i,j}}=0$ for all $k$ since the cofactor of $i, k$ is the principal minor by removing row $i$ and column $k$. Therefore,
\begin{align*}
    \d \det X = & ~ {\rm adj}(X)
\end{align*}
and consequently,
\begin{align*}
    \d \log \det X = & ~ \frac{1}{\det X} \d \det X \\
    = & ~ \frac{1}{\det X}~{\rm adj}(X) \\
    = & ~ X^{-1}. \qedhere
\end{align*}
\end{proof}

The goal of this section is to prove the $d \times d$ matrix $\nabla^2 F(x) \succ 0$. We start with computing $\nabla F(x) \in \R^d$.
\begin{lemma}
We define $s_{x,i} = a_i^\top x - b_i$ for each $i \in [n]$. Let $S_x$ denote a diagonal matrix such that $S_x = \mathrm{diag}(s_x)$.  
Let $H(x) = A^\top S_x^{-2} A \in \R^{d \times d}$.  
Let $F(x) = \log \det (H(x)+I_d)$, then 
\begin{align*}
    \nabla F(x) = & ~ -2\sum_{i=1}^n \wt \sigma_i(S_x^{-1}A) \frac{a_i}{ s_{x,i} } .
\end{align*}
\end{lemma}

\begin{proof}

First, we know that
\begin{align}\label{eq:d_s_x_i_-2}
    \frac{ \d s_{x,i}^{-2} }{ \d x_j } = \frac{-2a_{i,j}}{s_{x,i}^3}.
\end{align}

For each $j \in [d]$, we can write $\frac{\d F}{\d x_j}$ as follows:
\begin{align*}
    \frac{\d F}{\d x_j} = & ~ \tr[(H(x)+I_d)^{-1} \frac{\d (H(x)+I_d)}{\d x_j}] \\
    = & ~ \tr[(H(x)+I_d)^{-1} \frac{\d H(x)}{x_j}] \\
    = & ~ \tr[(H(x)+I_d)^{-1} \sum_{i=1}^n a_i a_i^\top \frac{\d s_{x,i}^{-2}}{\d x_j}] \\ 
    = & ~ \tr[(H(x)+I_d)^{-1} \sum_{i=1}^n -2a_i a_i^\top \frac{a_{i,j}}{ s_{x,i}^3 }] \\
    = & ~ -2 \sum_{i=1}^n\tr[\frac{a_{i,j}}{s_{x,i}} \frac{a_i^\top (H(x)+I_d)^{-1}a_i}{s_{x,i}^2}] \\
    = & ~ -2 \sum_{i=1}^n \wt \sigma_i(S_x^{-1}A) \frac{a_{i,j}}{ s_{x,i} }.
\end{align*}
where the forth step follows from Eq.~\eqref{eq:d_s_x_i_-2}, the fifth step follows from $\tr[AB] = \tr[BA]$, and the last step follows from $\wt{\sigma}$.

We use chain rule:
\begin{align*}
    \frac{\d F}{\d x} 
    = & ~ \begin{bmatrix} \frac{\d F }{\d x_1} & \frac{\d F }{\d x_2} & \cdots & \frac{\d F }{\d x_d} \end{bmatrix}^\top \\
    = &  -2\sum_{i=1}^n \wt \sigma_i(S_x^{-1}A) \frac{a_i}{ s_{x,i} }. \qedhere
\end{align*}
\end{proof}

\begin{fact}
We have the following partial derivative:
\begin{align*}
    \frac{\d ( \wt{\sigma}_i (S_x^{-1} A) s_{x,i}^2 ) }{ \d x_l} = 2 \sum_{j=1}^n \wt{\sigma}_{i,j}^2 ( S_x^{-1} A ) \frac{ a_{j,l} }{ s_{x,j} }
\end{align*}
\end{fact}
\begin{proof}
We analyze the partial derivative term
\begin{align}\label{eq:log_partial_l}
    \frac{\partial}{\partial x_l} a_i^\top (H(x)+I_d)^{-1} a_i = & ~ a_i^\top (\frac{\partial (H(x)+I_d)^{-1}}{\partial x_l}) a_i \notag \\ 
    = & ~ a_i^\top (-(H(x)+I_d)^{-1} \frac{\partial H(x)}{\partial x_l} (H(x)+I_d)^{-1})  a_i \notag \\ 
    = & ~ a_i^\top (2(H(x)+I_d)^{-1} (\sum_{j=1}^n a_ja_j^\top \frac{a_{j,l}}{ s_{x,j}^3 })(H(x)+I_d)^{-1})a_i \notag\\
    = & ~  2\sum_{j=1}^n \frac{(a_i^\top (H(x)+I_d)^{-1}a_j)^2}{ s_{x,j}^2 } \frac{a_{j,l}}{ s_{x,j} } \notag \\
    = &~ 2 \sum_{j=1}^n \wt{\sigma}_{i,j}^2 ( S_x^{-1} A ) \frac{ a_{j,l} }{ s_{x,j} }
\end{align}
where the second step follows from Fact~\ref{fac:d_X_inverse}.
\end{proof}

\begin{fact}\label{fac:levarage_score_wt_wh}
We have
\begin{align*}
    \wt{\sigma}_i ( S_x^{-1} A ) = \wh{\sigma}_i( S_x^{-1} A )+ \sum_{j=1}^n \wt{\sigma}_{i,j}^2 ( S_x^{-1} A ) . 
\end{align*}
\end{fact}
\begin{proof}
Note that
\begin{align*}
    (H(x)+I_d)^{-1} = & ~ (H(x)+I_d)^{-1} (H(x)+I_d) (H(x)+I_d)^{-1} \\
    = & ~ (H(x)+I_d)^{-2} + \sum_{j=1}^n \frac{(H(x)+I_d)^{-1} a_ja_j^\top (H(x)+I_d)^{-1})}{ s_{x,j}^2 },
\end{align*}

therefore,
\begin{align*}
    \wt{\sigma}_i(S_x^{-1}A) 
    = & ~ a_i^\top ( H(x) + I_d )^{-1} a_i \frac{1}{ s_{x,i}^2 } \\
    = & ~ \frac{a_i^\top (H(x)+I_d)^{-2}a_i}{ s_{x,i}^2 } + \sum_{j=1}^n \frac{(a_i^\top (H(x)+I_d)^{-1}a_j)^2}{ s_{x,i}^2  s_{x,j}^2 } \\
    = & ~ \wh{\sigma}_i (S_x^{-1} A) + \sum_{j=1}^n \wt{\sigma}_{i,j}^2 (S_x^{-1} A) 
\end{align*}
Thus we complete the proof.
\end{proof}

\begin{lemma}\label{lem:F_x_is_convex}
Let $F(x)=\log \det(H(x)+I_d)$, then
\begin{align*}
    \nabla^2 F(x) \succeq & ~ 0. 
\end{align*}
Thus, $F(x)$ is convex.
\end{lemma}

\begin{proof}
Note that
\begin{align}\label{eq:log_hessian}
    \frac{\partial}{\partial x_l} \frac{\partial}{\partial x_k} F(x) = & ~ -\Big(\sum_{i=1}^n a_i^\top (H(x)+I_d)^{-1}a_i a_i \frac{\partial}{\partial x_l}(\frac{a_{i,k}}{ s_{x,i}^3 })+\frac{a_{i,k}}{ s_{x,i}^3 } \frac{\partial}{\partial x_l} a_i^\top (H(x)+I_d)^{-1} a_i \Big) \notag\\
    = & ~ 3\sum_{i=1}^n \frac{a_i^\top (H(x)+I_d)^{-1}a_i}{ s_{x,i}^2 } \frac{a_{i,k}a_{i,l}}{ s_{x,i}^2 }-\frac{a_{i,k}}{ s_{x,i}^3 } \frac{\partial}{\partial x_l} a_i^\top (H(x)+I_d)^{-1} a_i.
\end{align}

Plug in Eq.~\eqref{eq:log_partial_l} into Eq.~\eqref{eq:log_hessian}, we have
\begin{align*}
    3 \sum_{i=1}^n \wt \sigma_i(S_x^{-1}A) \frac{a_{i,k}a_{i,l}}{ s_{x,i}^2 }-2\sum_{i=1}^n \sum_{j=1}^n  \wt{\sigma}_{i,j}(S_x^{-1} A)^2 \frac{a_{i,k}a_{j,l}}{ s_{x,j} s_{x,i} },
\end{align*}

and 
\begin{align*}
    & ~ 3 \sum_{i=1}^n \wt{\sigma}_i(S_x^{-1}A) \frac{a_{i}a_{i}^\top}{ s_{x,i}^2 }-2\sum_{i=1}^n \sum_{j=1}^n \wt{\sigma}_{i,j}^2 (S_x^{-1} A) \frac{a_i a_j^\top}{ s_{x,j} s_{x,i} } \\
    =  & ~ 3 \sum_{i=1}^n ( \wh{\sigma}_{i}(S_x^{-1} A ) +\sum_{j=1}^n \wt{\sigma}_{i,j}^2(S_x^{-1} A))\frac{a_ia_i^\top}{ s_{x,i}^2 }
    - ~ 2\sum_{i=1}^n \sum_{j=1}^n \wt{\sigma}_{i,j}^2 (S_x^{-1} A) \frac{a_{i}a_{j}^\top }{ s_{x,j}  s_{x,i} } \\
    = & ~ 3\sum_{i=1}^n \wh{\sigma}_i(S_x^{-1} A) \frac{a_ia_i^\top}{ s_{x,i}^2 }+3\sum_{i=1}^n \sum_{j=1}^n \wt{\sigma}_{i,j}^2 ( S_x^{-1} A ) \frac{a_ia_i^\top}{ s_{x,i}^2 } 
    -  ~ 2\sum_{i=1}^n \sum_{j=1}^n \wt{\sigma}_{i,j}^2 (S_x^{-1} A) \cdot \frac{a_ia_j^\top}{ s_{x,j}  s_{x,i} } \\
    = & ~ 3\sum_{i=1}^n \wh{\sigma}_i(S_x^{-1} A) \frac{a_ia_i^\top}{ s_{x,i}^2 } 
    +\sum_{i=1}^n \wt{\sigma}_i(S_x^{-1}A) \frac{a_ia_i^\top}{ s_{x,i}^2 } +  ~ 2\sum_{i=1}^n \sum_{j=1}^n  \wt{\sigma}_{i,j}^2 (S_x^{-1} A) \cdot (\frac{a_ia_i^\top}{ s_{x,i}^2 }- \frac{a_ia_j^\top}{ s_{x,j}  s_{x,i} }) \\
    := & ~ B_1 + B_2 + B_3.
\end{align*}
where the first step follows from Fact~\ref{fac:levarage_score_wt_wh}, the last step follows from 
\begin{align*}
    B_1 := & ~ 3\sum_{i=1}^n \wh{\sigma}_i(S_x^{-1} A) \frac{a_ia_i^\top}{ s_{x,i}^2 } , \\
    B_2 := & ~ \sum_{i=1}^n \wt \sigma_i(S_x^{-1}A) \frac{a_ia_i^\top}{ s_{x,i}^2 } , \\
    B_3 := & ~ 2\sum_{i=1}^n \sum_{j=1}^n  \wt{\sigma}_{i,j}^2 (S_x^{-1} A) \cdot (\frac{a_ia_i^\top}{ s_{x,i}^2 }- \frac{a_ia_j^\top}{ s_{x,j}  s_{x,i} }) .
\end{align*}
It is obvious that $B_1 \succ 0$ and $B_2 \succ 0$.

It is not hard to see that
\begin{align*}
    \frac{a_i a_i^\top}{s_{x,i}^2} + \frac{a_j a_j^\top}{ s_{x,j}^2 } \succeq 2 \frac{a_ia_j^\top}{ s_{x,j}  s_{x,i} } .
\end{align*}
Thus we have $B_3 \succeq 0$ and we complete the proof.
\end{proof}

\subsection{Convexity of \texorpdfstring{$\log\det(H_{\log }(x) + I_d)$}{}: spectrahedra}

In this section, we prove that $\log\det(H_{\rm log}(x)+I_d)$ is a convex function on $x$. Combined with the bounded variance property, we provide an algorithm that samples from the SDP spectrahedron.

\begin{definition}
\label{def:ridge_projection}
We define the ridge-projection matrix $P\in \R^{n^2 \times n^2}$ as
\begin{align*}
    P :=  (S^{-1/2}\otimes S^{-1/2}){\sf A}^\top (H+I_d)^{-1} {\sf A}(S^{-1/2}\otimes S^{-1/2}).
\end{align*}
We define projection matrix $\ov{P} \in \R^{n^2 \times n^2}$ as follows
\begin{align*}
\ov{P}:=(S^{-1/2}\otimes S^{-1/2}){\sf A}^\top H^{-1} {\sf A}(S^{-1/2}\otimes S^{-1/2}).
\end{align*} 
\end{definition}

\begin{remark}
It is not hard to see that $P$ is a PSD matrix. Note that the matrix $H+I_d$ is PSD, since $H$ is PSD, therefore $(H+I_d)^{-1}$ is also PSD. Thus, the ridge-projection is PSD, as its two ``arms'' are symmetric.
\end{remark}

\begin{lemma}
\label{lem:ridge_proj_leq_proj}
Let $\ov{P} \in \R^{n^2 \times n^2}$ be the projection matrix and let $P \in \R^{n^2 \times n^2}$ be the ridge-projection matrix defined in Def.~\ref{def:ridge_projection}. Then, we have $P\preceq \ov P$.
\end{lemma}

\begin{proof}
Let $H=U\Sigma U^\top$ be its eigendecomposition, we need to show that 
\begin{align*}
    (H+I_d)^{-1} \preceq & ~ H^{-1}.
\end{align*}
Let us expand the LHS:
\begin{align*}
    (U\Sigma U^\top+I_d)^{-1} = & ~ (U(\Sigma+I_d)U^\top)^{-1} \\
    = & ~ U(\Sigma+I_d)^{-1}U^\top\\
    \preceq & ~ U\Sigma^{-1}U^\top,
\end{align*}
where the last step follows from for any non-negative number $\lambda$, $(1+\lambda)^{-1}\leq \lambda^{-1}$.
\end{proof}

The following lemma is a generalization of Theorem 4.1 of~\cite{a00}.
\begin{lemma}
Let $H_{\log}(x) \in \R^{d \times d}$ be defined as Definition~\ref{lem:sdp_log_convex}. Let $F(x) = \log \det( H_{\log} (x) + I_d )$. Then, $F(x)$ is convex in $x$.
\end{lemma}

\begin{proof}
Using standard algebra and chain rule computations, we have
\begin{align*}
\nabla^2 F(x) = 2Q(x) + R(x) - 2 T(x).
\end{align*}
For simplicity, we drop $x$. The $Q, R, T \in \R^{d \times d}$ can be defined as follows
\begin{align*}
Q_{i,j} := & ~ {\sf A} (H+I_d)^{-1} {\sf A}^\top \cdot ( S^{-1} A_i S^{-1} A_j S^{-1} \oplus S^{-1} ) , \\
R_{i,j} := & ~ {\sf A} (H+I_d)^{-1} {\sf A}^\top \cdot (S^{-1} A_i S^{-1} \oplus S^{-1} A_j S^{-1} ) , \\
T_{i,j} := & ~ {\sf A} (H+I_d)^{-1} {\sf A}^\top \cdot ( S^{-1} A_i S^{-1} \oplus S^{-1} ) \cdot {\sf A} (H + I_d)^{-1} {\sf A}^\top \cdot ( S^{-1} A_j S^{-1} \oplus S^{-1} ).
\end{align*}
Using Lemma~\ref{lem:sdp_Q_R_T}, we know that
\begin{align*}
0 \preceq Q(x) \preceq \nabla^2 F(x) \preceq 3Q(x)
\end{align*}
Thus, $F(x)$ is convex.
\end{proof}

\begin{lemma}\label{lem:sdp_Q_R_T}
For any $x$, if $S(x) \succ 0$, then we have
\begin{itemize}
    \item Part 1. $Q \succeq 0$;
    \item Part 2. $T \succeq 0$;
    \item Part 3. $T \preceq \frac{1}{2} (Q+R)$;
    \item Part 4. $\nabla^2 F(x)\succeq 0$;
    \item Part 5. $R \preceq Q$;
    \item Part 6. $\nabla^2 F(x) \preceq 3 Q(x)$.
\end{itemize}
\end{lemma}
\begin{proof}

{\bf Proof of Part 1 and 2.}
Let $\xi \in \R^d$ and $\xi \neq 0$. 

Then we have
\begin{align*}
\xi^\top Q \xi 
= & ~ \sum_{i,j} Q_{i,j} \xi_i \xi_j \\
= & ~ {\sf A} (H+I_d)^{-1} {\sf A}^\top \cdot (S^{-1} B S^{-1} B S^{-1} \oplus S^{-1}) \\
= & ~ P \cdot ( \ov{B}^2 \oplus I_d )
\end{align*}
where $B = B (\xi) = \sum_{i=1}^d \xi_i A_i$, and $\ov{B} = S^{-1/2} B S^{-1/2}$.

Similarly, we have
\begin{align*}
\xi^\top R \xi 
= & ~ {\sf A} (H+I_d)^{-1} {\sf A}^\top \cdot (S^{-1} B S^{-1} \otimes S^{-1} B S^{-1} ) \\
= & ~ P \cdot ( \ov{B} \otimes \ov{B} )
\end{align*}

We can rewrite $\xi^\top T \xi$ as follows:
\begin{align*}
\xi^\top T \xi
= & ~ {\sf A}^\top (H + I)^{-1} {\sf A} \cdot (S^{-1} B S^{-1} \oplus S^{-1} ) {\sf A} (H+I)^{-1} {\sf A}^\top \cdot ( S^{-1} B S^{-1} \oplus S^{-1} ) \\
= & ~ P \cdot ( \ov{B} \oplus I_d ) P ( \ov{B} \oplus I_d )
\end{align*}

It is obvious that $I_d, P$ and $\ov{B}^2$ are all PSD matrices. 

Using Part 3 of Lemma~\ref{lem:matrix_trace_psd} and Part 6 of Lemma~\ref{lem:matrix_otimes_oplus}, we have
\begin{align*}
\xi^\top Q \xi \geq 0 \\
\xi^\top T \xi \geq 0
\end{align*}

Since $\xi$ is arbitrary, then we know that
\begin{align*}
Q \succeq & ~ 0 \\
T \succeq & ~0
\end{align*}

{\bf Proof of Part 3.} Note that $\ov{P}$ is a projection matrix, and $P \preceq \ov{P}$ by Lemma~\ref{lem:ridge_proj_leq_proj}.
\begin{align*}
(\ov{B} \oplus I_d) P ( \ov{B} \oplus I_d )
\preceq & ~ (\ov{B} \oplus I_d) \ov{P} ( \ov{B} \oplus I_d ) \\
\preceq & ~ (\ov{B} \oplus I_d) I_d ( \ov{B} \oplus I_d ) \\
= & ~ \frac{1}{2} ( ( \ov{B}^2 \oplus I_d) + (\ov{B} \otimes \ov{B} ) )
\end{align*} 
where the last step follows from Part 2 of Lemma~\ref{lem:matrix_otimes_oplus}.

Applying Part 1 of Lemma~\ref{lem:matrix_trace_psd}, we have
\begin{align*}
\langle P, (\ov{B} \oplus I_d) P (\ov{B} \oplus I_d) \rangle \leq \frac{1}{2} \langle P, (\ov{B}^2 \oplus I_d) + (\ov{B} \otimes \ov{B}) \rangle
\end{align*}

The above equation implies the following
\begin{align*}
\xi^\top T \xi \leq \frac{1}{2} \xi^\top (Q+ R)\xi
\end{align*}

Note that, here $\xi$ is arbitrary, thus we know that
\begin{align*}
T \preceq \frac{1}{2} (Q+R)
\end{align*}

{\bf Proof of Part 4.} We have
\begin{align*}
\nabla^2 F(x) = & ~ 2 Q + R - 2 T \\
\succeq & ~ 2 Q + R - 2 \cdot \frac{1}{2} (Q+T) \\
= & ~ Q
\end{align*}

{\bf Proof of Part 5.}
Let $v_i$ be orthonormal eigenvectors of $\ov{B}$ with corresponding eigenvalues $\lambda_i$.

Then using \cite[Theorem 4.4.5]{hj91}, $\ov{B}^2 \oplus I$ has orthonormal eigenvectors $v_i \otimes v_j$ with corresponding eigenvalues $\frac{1}{2}(\lambda_i^2 + \lambda_j^2)$, while (see \cite{hj91}) $\ov{B} \otimes \ov{B}$ has the same eigenvectors $v_i \otimes v_j$, with corresponding eigenvalues $\lambda_i \lambda_j$.

It then follows from $(\lambda_i - \lambda_j)^2$ for each $i,j$ that
\begin{align*}
\ov{B}^2 \oplus I_d \succeq \ov{B} \otimes \ov{B}.
\end{align*}

Using Part 4 of Lemma~\ref{lem:matrix_trace_psd}, we know
\begin{align*}
\langle P , \ov{B}^2 \oplus I_d \rangle \geq \langle P, \ov{B} \otimes \ov{B} \rangle.
\end{align*}
Which is 
\begin{align*}
\xi^\top Q \xi \geq \xi^\top R \xi
\end{align*}
Since it's arbitrary $\xi$, then we have
\begin{align*}
Q \succeq R 
\end{align*}

{\bf Proof of Part 6.}
We have
\begin{align*}
\nabla^2 F(x) = & ~ 2 Q + R - 2 T \\
\preceq & ~ 2 Q + R \\
\preceq & ~ 3 Q
\end{align*}
where the second step follows from $T \succeq 0$, the last step follows from Part 5.
\end{proof}

\subsection{Discussion}\label{sec:assumption:discussion}

It is well-known that for popular barrier functions, such as volumetric barrier and Lee-Sidford barrier~\citep{ls14,ls19}, the function $\log \det H(x)$ is convex in $x$. However, given a PSD matrix function $H(x)$ for which $\log \det H(x)$ is convex in $x$, it is generally not true that $\log \det (H(x)+I_d)$ is convex.

\begin{fact}[Folklore]
Suppose that $H(x)$ is a positive semi-definite matrix for $x$ in the domain.
Suppose $\log \det H(x)$ is convex. It is possible $\log \det (H(x)+I)$ is not convex.
\end{fact}
\begin{proof}

If $\log \det(H(x)+I)$ is convex in $x$, then for any fixed $x\in K$, $v\in \R^d$, $\log \det (H(x+t v)+I_d)$ is convex in $t$ (for $t$ sufficiently close to $0$).

For $t$ sufficiently closely to $0$, define $f_i(t) := \lambda_i(H(x+tv))$.
Then
\begin{align*}
    \log \det H(x+tv) = \sum_{i\in [d]} \log f_i(t)
\end{align*}
and the condition that $\log\det H(x)$ is convex is equivalent to
\begin{align*}
    \frac{\d^2}{\d t^2} \log \det H(x+tv) = \sum_{i\in [d]} \frac{f_i''(t) f_i(t) - f_i'(t)^2}{f_i(t)^2} \ge 0.
\end{align*}

Now
\begin{align*}
    \log \det (H(x+tv) + I_d) = \sum_{i\in [d]} \log (f_i(t)+1).
\end{align*}
The condition that $\log \det (H(x)+I_d)$ is convex is equivalent to
\begin{align*}
    \frac{\d^2}{\d t^2} \log \det (H(x+tv)+I_d) = \sum_{i\in [d]} \frac{f_i''(t) (f_i(t)+1) - f_i'(t)^2}{(f_i(t)+1)^2} \ge 0.
\end{align*}

Suppose $d=2$, $f_1(0)=1,f_1'(0)=0,f_1''(0)=1$,
$f_2(0)=4,f_2'(0)=0,f_2''(0)=-3$.
Then
\begin{align*}
    \frac{\d^2}{\d t^2}\Big|_{t=0} \log \det H(x+tv) &= \frac{f_1''(0)}{f_1(0)} + \frac{f_2''(0)}{f_2(0)} > 0. \\
    \frac{\d^2}{\d t^2}\Big|_{t=0} \log \det (H(x+tv)+I) &= \frac{f_1''(0)}{f_1(0)+1} + \frac{f_2''(0)}{f_2(0)+1} < 0.
\end{align*}
This gives an example for which $\log \det H(x)$ is convex in a region but $\log \det (H(x)+I_d)$ is not.
\end{proof}

\section{Convexity of Regularized Volumetric Barrier Function}
\label{sec:vol}
In this section, we prove the convexity of $\log\det$ of the regularized volumetric barrier function. This is crucial, as the convexity proof for regularized Lee-Sidford barrier is identical up to replacing leverage score by Lewis weights.
\subsection{Definitions}

\begin{definition}\label{def:sigma}
Let $A \in \R^{n \times d}$. Let $b \in \R^n$.
For each $i \in [n]$, we define 
\begin{align*}
    s_{x,i} := & ~ (a_i^\top x - b_i) \\
    s_x := & ~ Ax - b
\end{align*}

For each $i\in [n]$, we define
\begin{align*}
    \sigma_{i,i}( A_x ) := & ~ a_{x,i}^\top ( A_x^\top A_x )^{-1} a_{x,i} 
\end{align*}
For each $i \in [n]$, for each $l \in [n]$
\begin{align*}
    \sigma_{i,l}( A_x ) := & ~   a_{x,i}^\top ( A_x^\top A_x )^{-1} a_{x,l} 
\end{align*}
Let $\Sigma = \sigma_{*,*}(A_x) \circ I_n$.
\end{definition}

\begin{definition}\label{def:H}
We define $H(x) \in \R^{d \times d}$ as follows
\begin{align*}
    H(x): = \underbrace{ A_x^\top }_{d \times n} \underbrace{ \Sigma( A_x ) }_{n \times n} \underbrace{ A_x }_{n \times d} 
\end{align*}
\end{definition}

\begin{definition}\label{def:P_j_P_j_k}
For each $j \in [d]$, we define $P_j(x) \in \R^{n \times n}$ as follows
\begin{align*}
    P_j(x):= \underbrace{ \frac{\d \Sigma(A_x)}{\d x_j} }_{n \times n}
\end{align*}
For each $j \in [d]$, for each $k \in [d]$, we define $P_{j,k}(x) \in \R^{n \times n}$ as follows:
\begin{align*}
    P_{j,k}(x) := \underbrace{ \frac{\d^2 \Sigma(A_x)}{\d x_k x_j} }_{n \times n}
\end{align*}
\end{definition}

In the previous sections, we mainly use notation $\sigma_{*,*}(A_x)$. For simplicity, from this section, we will use notation $Q(x)$ instead.
\begin{definition}\label{def:Q}
We define $Q(x) \in \R^{n \times n}$ as follows
\begin{align*}
    Q(x):= \underbrace{ \sigma_{*,*}(A_x) }_{n \times n}.
\end{align*}
\end{definition}

\begin{fact}\label{fac:leverage_score_is_sum_of_squares}
We have
\begin{align*}
    \sigma_{i,i}( A_x) = \langle  \sigma_{*,i}^2 ( A_x ) , {\bf 1}_n \rangle.
\end{align*}
\end{fact}
\begin{proof}
The proof is straightforward from definition.
\end{proof}

We list a number of standard calculus results.
\begin{fact}
We define the following quantities:
\begin{itemize}
    \item Let $s_x = A x -b \in \R^n$;
    \item Let $S_x = \diag(s_x) \in \R^{n \times n}$ denote the diagonal matrix by putting $s_x$ on the diagonal;
    \item Let $A_{*,j}$ denote the $j$-th column of matrix $A \in \R^{n \times d}$;
    \item Let $A_x = S_x^{-1} A $;
    \item Let $a_{x,i}^\top$ denote the $i$-th row of $A_x$ for each $i \in [n]$.
\end{itemize}
Then, we have for each $j \in [d]$
\begin{itemize}
    \item {\bf Part 1.} 
    \begin{align*} 
        \underbrace{ \frac{ \d s_{x} }{ \d x_j } }_{n \times 1} = \underbrace{ A_{*,j} }_{n \times 1}
    \end{align*}
    \item {\bf Part 2.}
    \begin{align*}
        \underbrace{ \frac{ \d s_x^{-1} }{ \d x_j } }_{n \times 1} = - \underbrace{ s_x^{-2} }_{n \times 1} \circ \underbrace{ A_{*,j} }_{n \times 1}
    \end{align*}
    \item {\bf Part 3.}
     \begin{align*}
        \underbrace{ \frac{ \d s_x^{-2} }{ \d x_j } }_{n \times 1} = - 2 \underbrace{ s_x^{-3} }_{n \times 1} \circ \underbrace{ A_{*,j} }_{n \times 1} 
    \end{align*}
    \item {\bf Part 4.}
    \begin{align*}
        \underbrace{ \frac{\d S_x^{-2}}{\d x_j} }_{n \times n} = 2 \underbrace{ \diag( - s_x^{-3} \circ A_{*,j} ) }_{n \times n} = 2 \diag( - S_x^{-3} A_{*,j} )
    \end{align*}
    \item {\bf Part 5.}
    \begin{align*}
        \underbrace{ \frac{ \d A^\top S_x^{-2} A }{ \d x_j } }_{d \times d} = 2 \underbrace{ A^\top }_{d \times n} \underbrace{ \diag( - S_x^{-3} A_{*,j} ) }_{n \times n} \underbrace{ A }_{n \times d}
    \end{align*}
    \item {\bf Part 6.}
    \begin{align*}
        \frac{ \d A_x^\top A_x }{ \d x_j } = 2 A_x^\top \diag ( - A_{x,*,j} ) A_{x}
    \end{align*}
    \item {\bf Part 7.}
    \begin{align*}
        \frac{ \d ( A_x^\top A_x )^{-1} }{ \d x_j } = 2 ( A_x^\top A_x )^{-1} \cdot A_x^\top \diag (  A_{x,*,j} ) A_{x} \cdot  ( A_x^\top A_x )^{-1} 
    \end{align*}
    \item {\bf Part 8.} For each $i \in [n]$
    \begin{align*}
        \frac{\d a_{x,i} }{\d x_j} = -  \underbrace{ A_{x,i,j} }_{ \mathrm{scalar} } \cdot \underbrace{ a_{x,i} }_{d \times 1}
    \end{align*}
    \item {\bf Part 9.} For each $i \in [n]$
    \begin{align*}
        \frac{\d a_{x,i} a_{x,i}^\top }{\d x_j} = - 2  \cdot \underbrace{ A_{x,i,j} }_{ \mathrm{scalar} } \cdot \underbrace{ a_{x,i} a_{x,i}^\top }_{d \times d}
    \end{align*}
    \item {\bf Part 10.}
    \begin{align*}
        \underbrace{ \frac{\d A_{x,i,j}}{\d x_j} }_{ \mathrm{scalar} } = - \underbrace{ A_{x,i,j}^2 }_{ \mathrm{scalar} }
    \end{align*}
    \item {\bf Part 11.}
    \begin{align*}
        \underbrace{ \frac{\d A_{x,i,j}}{\d x_k} }_{ \mathrm{scalar} } = - \underbrace{ A_{x,i,j} }_{ \mathrm{scalar} } \underbrace{ A_{x,i,k} }_{ \mathrm{scalar} }
    \end{align*}    
    \item {\bf Part 12.}
    \begin{align*}
        \frac{ \d A_{x,*,j} }{\d x_j} = - A_{x,*,j}^{\circ 2}
    \end{align*}
    \item {\bf Part 13.}
    \begin{align*}
        \frac{ \d A_{x,*,j} }{\d x_k} = - A_{x,*,j} \circ A_{x,*,k}
    \end{align*}
    \item {\bf Part 14.}
    \begin{align*}
        \underbrace{ \frac{ \d A_x }{\d x_j} }_{n \times d} = - \underbrace{ \diag( A_{x,*,j} ) }_{n \times n} \underbrace{ A_x }_{n \times d}
    \end{align*}
\end{itemize}
\end{fact}
\begin{definition}\label{def:wt_Sigma}
We define $\wt{\Sigma}(A_x) \in \R^{n \times n}$ and $\wt{\sigma}_{*,*}(A_x) \in \R^{n \times n}$ as
\begin{align*}
    \wt{\Sigma}(A_x) &:= ( A_x (H(x) + I_d)^{-1} A_x^\top  ) \circ I_n \\
    \wt{\sigma}_{*,*}(A_x) &:=  A_x (H(x) + I_d)^{-1} A_x^\top
\end{align*}
\end{definition}
\begin{definition}\label{def:F_log}
We define $F(x)$ as follows:
\begin{align*}
    F(x): = \log( \det(H(x)  +I_d) )
\end{align*}

\end{definition}
\begin{definition}\label{def:F_dx_dxt}
We define $\frac{\d }{\d x} \frac{\d F}{\d x^\top } \in \R^{d \times d}$ to be
\begin{align*}
\frac{\d }{\d x} \frac{\d F}{\d x^\top} = [\frac{\d }{\d x } \frac{\d F}{\d x^\top }]_1 + [\frac{\d }{\d x} \frac{\d F}{\d x^\top }]_2
\end{align*}
where
\begin{align*}  
    [\frac{\d }{\d x} \frac{\d F}{\d x^\top }]_{1,j,k} = & ~  + \tr[ (H(x) + I_d)^{-1} \frac{\d^2 H(x)}{\d x_j \d x_k} ] \\
    [\frac{\d }{\d x} \frac{\d F}{\d x^\top}]_{2,j,k} = & ~ - \tr[ (H(x) + I_d)^{-1} \frac{\d H(x)}{\d x_k} (H(x) + I_d)^{-1}\cdot \frac{\d H(x)}{\d x_j}  ] 
\end{align*}
\end{definition}

\subsection{Gradient of \texorpdfstring{$\sigma_{i,j}$}{}}

\begin{fact}[First derivative of leverage score]\label{fac:sigma_scalar_gradient}
We define the following quantities:
\begin{itemize}
    \item For each $j \in [d]$, let $A_{*,j} \in \R^n$ denote $j$-th column of $A$; 
    \item Let $A_x := S_x^{-1} A \in \R^{n \times d}$;
    \item Let $A_{x,*,j} \in \R^n$ denote the $j$-th column of $A_{x} \in \R^{n \times d}$;
    \item Let $\Sigma(A_x) \in \R^{n \times n}$ denote a diagonal matrix where $(i,i)$-th entry is $\sigma_{i,i}(A_x)$;
    \item Let $\sigma_{*,*}^{\circ 2} (A_x) = \sigma_{*,*}(A_x) \circ \sigma_{*,*}(A_x) \in \R^{n \times n}$;
    \item Let $\sigma_{*,i}(A_x) \in \R^n$ denote a column vector of $\sigma_{*,*}(A_x) \in \R^{n \times n}$.
\end{itemize}
Then we have for each $j \in [d]$
\begin{itemize}
    \item Part 1. For each $i \in [n]$,  
    \begin{align*}
        \frac{ \d \sigma_{i,i}( A_x ) }{ \d x_j } 
   = & ~ 2  \langle \sigma_{*,i} ( A_x ) \circ \sigma_{*,i} (A_x)  , A_{x,*,j} \rangle -  2\sigma_{i,i} (A_x) \cdot A_{x,i,j} 
    \end{align*}
    \item Part 2. For each $i \in [n]$, $l \in [n]$
    \begin{align*}
        \frac{ \d \sigma_{i,l}( A_x ) }{ \d x_j } 
   = & ~ 2  \langle \sigma_{i,*} ( A_x ) \circ \sigma_{l,*}(A_x) , A_{x,*,j} \rangle -  \sigma_{i,l}(A_x) \cdot (A_{x,i,j} + A_{x,l,j} )
   \end{align*}
   
\end{itemize}
\end{fact}
\begin{proof}

We know 
\begin{align*}
    \frac{\d \sigma_{i,i} ( A_x )}{ \d x_j }
    = & ~ \frac{ \d a_{x,i}^\top ( A_x^\top A_x )^{-1}  a_{x,i} } { \d x_j } \\
   = & ~ \frac{ \d \langle a_{x,i} a_{x,i}^\top, (A_x^\top A_x)^{-1} \rangle }{\d x_j} \\
   = & ~ \langle \frac{ \d a_{x,i} a_{x,i}^\top }{\d x_j }, (A_x^\top A_x)^{-1} \rangle + \langle a_{x,i} a_{x,i}^\top, \frac{\d  (A_x^\top A_x)^{-1} }{\d x_j} \rangle
\end{align*}
For the first term in the above, we have
\begin{align*} 
-2 A_{x,i,j} \cdot \langle a_{x,i} a_{x,i}^\top, (A_x^\top A_x)^{-1} \rangle = - 2A_{x,i,j} \cdot \sigma_{i,i}(A_x)
\end{align*}
For the second term, we have
\begin{align*}
 \langle a_{x,i} a_{x,i}^\top, \frac{\d  (A_x^\top A_x)^{-1} }{\d x_j} \rangle
 = & ~ 2 \langle a_{x,i} a_{x,i}^\top, ( A_x^\top A_x )^{-1} \cdot A_x^\top \diag (  A_{x,*,j} ) A_{x} \cdot  ( A_x^\top A_x )^{-1} \rangle \\
 = & ~ 2 a_{x,i}^\top ( A_x^\top A_x )^{-1} \cdot A_x^\top \diag (  A_{x,*,j} ) A_{x} \cdot  ( A_x^\top A_x )^{-1} a_{x,i} \\
 = & ~ 2 a_{x,i}^\top ( A_x^\top A_x )^{-1} \cdot ( \sum_{l=1}^n a_{x,l} a_{x,l}^\top A_{x,l,j} ) \cdot  ( A_x^\top A_x )^{-1} a_{x,i} \\
 = & ~  2 \sum_{l=1}^n \sigma_{l,i}(A_x)^2 A_{x,l,j} \\
 = & ~ 2 \langle \sigma_{*,i}^2(A_x), A_{x,*,j} \rangle.
\end{align*}
Similarly, we can prove for $\frac{\d \sigma_{i,l} (A_x)}{\d x_j}$.
\end{proof}

\subsection{Gradient of \texorpdfstring{$\sigma$}{}}

\begin{lemma}
Let $\sigma$ and $\Sigma$ be defined as Definition~\ref{def:sigma}. Then, we have
\begin{itemize}
    \item {\bf Part 1}
    \begin{align*}
        \frac{ \d \sigma_{*,i}(A_x) }{ \d x_j} = 2 \underbrace{ \sigma_{*,*}(A_x) }_{n \times n}  \underbrace{ ( \sigma_{*,i} \circ A_{x,*,j} ) }_{n \times 1} - 2 \underbrace{ \sigma_{*,i}(A_x) }_{n \times 1} \circ \underbrace{ A_{x,*,j} }_{n \times 1}
    \end{align*}
    \item {\bf Part 2.}
    \begin{align*}
        & ~ \frac{ \d \sigma_{*,*}(A_x) }{ \d x_j} \\
        = & ~ 2 \underbrace{ \sigma_{*,*}(A_x) }_{n \times n} \diag(A_{x,*,j}) \sigma_{*,*} (A_x)  -  \diag(A_{x,*,j}) \sigma_{*,*}(A_x)  -   \sigma_{*,*}(A_x) \diag( A_{x,*,j} )
    \end{align*}
    \item {\bf Part 3.} 
    \begin{align*}
        \underbrace{ \frac{\d \Sigma(A_x)}{\d x_j} }_{n \times n} = & ~ 2 \diag( \underbrace{ \sigma_{*,*}^{\circ 2}(A_x) }_{n \times n} \underbrace{ A_{x,*,j} }_{n \times 1} ) - 2 \diag( \underbrace{ \Sigma(A_x) }_{n \times n} \underbrace{ A_{x,*,j} }_{n \times 1} ) \\
        = & ~ 2 \diag ( ( \sigma_{*,*}^{\circ 2} (A_x) - \Sigma(A_x) ) A_{x,*,j} )
   \end{align*}
\end{itemize}
\end{lemma}
\begin{proof}
It follows from Fact~\ref{fac:sigma_scalar_gradient}.
\end{proof}

\subsection{Gradient for \texorpdfstring{$H(x)$}{}}

\begin{lemma}\label{lem:gradient_H}
Recall the definitions of the following quantities:
\begin{itemize}
    \item Let $H(x) \in \R^{d \times d}$ be defined as Definition~\ref{def:H}.
    \item For each $j \in [d]$, let $P_j(x) \in \R^{n \times n}$ be defined as Definition~\ref{def:P_j_P_j_k}.
    \item For each $j \in [d]$, for each $k \in [d]$, let $P_{j,k}(x) \in \R^{n \times n}$ be defined as Definition~\ref{def:P_j_P_j_k}.
\end{itemize} 
Then, we have
\begin{itemize}
    \item {\bf Part 1.} For each $j \in [d]$
    \begin{align*}
        \frac{\d H(x)}{\d x_j} = & ~ -  2 A_x^\top \diag( \Sigma(A_x) A_{x,*,j} ) A_x \\
        & ~ + A_x^\top P_j(x) A_x 
    \end{align*}
    \item {\bf Part 2.} For each $j \in [d]$, for each $k \in [d]$
    \begin{align*}
     \frac{\d}{ \d x_k} ( \frac{\d H(x)}{\d x_j} )
        = & ~ + 6 A_x^\top \diag(A_{x,*,k}) \Sigma(A_x) \diag(A_{x,*,j}) A_x\\
        & ~  -2 A_x^\top P_k(x) \diag( A_{x,*,j} ) A_x \\
        & ~ - 2 A_x^\top P_j(x) \diag( A_{x,*,k} ) A_x \\
        & ~ + A_x^\top P_{j,k}(x) A_x
    \end{align*}
\end{itemize}
\end{lemma}
\begin{proof}

{\bf Proof of Part 1.}
We have
\begin{align*}
    \frac{\d H(x) }{\d x_j} = & ~ \frac{ \d ( A_x^\top \Sigma(A_x) A_x ) }{ \d x_j } \\
    = & ~ ( \frac{\d A_x^\top }{\d x_j} ) \cdot \Sigma(A_x) A_x +  A_x^\top \cdot ( \frac{\d \Sigma(A_x) }{\d x_j} ) \cdot A_x +  A_x^\top \Sigma(A_x) \cdot ( \frac{\d A_x}{\d x_j} ) \\
    = & ~ -  A_x \diag( A_{x,*,j} ) \Sigma(A_x) A_x \\
    & ~ +  A_x^\top P(x)_j A_x \\
    & ~ -  A_x \Sigma(A_x) \diag( A_{x,*,j} ) A_x \\
    = & ~  -  2 A_x^\top \diag( \Sigma(A_x) A_{x,*,j} ) A_x \\
        & ~ + A_x^\top P_j(x) A_x 
\end{align*}

{\bf Proof of Part 2.}
Then, we have
\begin{align}\label{eq:hessian_H_kj}
 \frac{\d }{\d x_k} ( \frac{\d H(x)}{\d x_j} ) 
 = & ~  \frac{\d }{\d x_k} (- 2 A_x^\top \Sigma(A_x) \diag(  A_{x,*,j} ) A_x  + A_x^\top P_j(x) A_x   ) \notag\\
 = & ~  \frac{\d }{\d x_k} (- 2 A_x^\top \Sigma(A_x) \diag(  A_{x,*,j} ) A_x)  + \frac{\d }{\d x_k} (A_x^\top P_j(x) A_x )  ,
\end{align}
where the first step follows from {\bf Part 1}, the second step follows from the sum rule.

For the first term of Eq.~\eqref{eq:hessian_H_kj}, we have
\begin{align}\label{eq:hessian_H_kj_1}
    & ~ \frac{\d }{\d x_k} ( - 2 A_x^\top \Sigma(A_x) \diag(  A_{x,*,j} ) A_x) \notag \\
    = & ~ -2 \frac{\d }{\d x_k} ( A_x^\top) \Sigma(A_x) \diag(  A_{x,*,j} ) A_x - 2  A_x^\top \frac{\d }{\d x_k} (\Sigma(A_x)) \diag(  A_{x,*,j} ) A_x \notag\\
    - & ~ 2 A_x^\top \Sigma(A_x) \frac{\d }{\d x_k} (\diag(  A_{x,*,j} )) A_x - 2  A_x^\top \Sigma(A_x) \diag(  A_{x,*,j} ) \frac{\d }{\d x_k} (A_x) \notag\\
    = & ~ -2 \frac{\d }{\d x_k} ( A_x^\top) \Sigma(A_x) \diag(  A_{x,*,j} ) A_x - 2  A_x^\top \frac{\d }{\d x_k} (\Sigma(A_x)) \diag(  A_{x,*,j} ) A_x \notag\\
    - & ~ 2 A_x^\top \Sigma(A_x) \diag(  \frac{\d }{\d x_k} (A_{x,*,j} )) A_x - 2  A_x^\top \Sigma(A_x) \diag(  A_{x,*,j} ) \frac{\d }{\d x_k} (A_x) \notag\\
    = & ~ + 2 A_x^\top \diag( A_{x,*,k} ) \Sigma(A_x) \diag(A_{x,*,j}) A_x \notag\\
 & ~ - 2A_x^\top P_k(x) \diag( A_{x,*,j} ) A_x \notag\\
 & ~ + 2 A_x^\top \Sigma(A_x) \diag (A_{x,*,j} ) \diag( A_{x,*,k}) A_x \notag\\
 & ~ + 2 A_x^\top \Sigma(A_x) \diag (A_{x,*,j} ) \diag( A_{x,*,k}) A_x,
\end{align}

For the second term of Eq.~\eqref{eq:hessian_H_kj}, we have
\begin{align}\label{eq:hessian_H_kj_2}
    & ~ \frac{\d }{\d x_k} (A_x^\top P_j(x) A_x ) \notag \\
    = & ~ \frac{\d }{\d x_k} (A_x^\top) P_j(x) A_x  + A_x^\top \frac{\d }{\d x_k} (P_j(x)) A_x   + A_x^\top P_j(x) \frac{\d }{\d x_k} (A_x ) \notag\\
    = & ~ -(A_x^\top) \diag( A_{x,*,k} ) P_j(x) A_x  + A_x^\top P_{j,k}(x) A_x   - A_x^\top P_j(x) \diag( A_{x,*,k} )  A_x \notag \\
    = & ~ -(A_x^\top) \diag( A_{x,*,k} ) P_j(x) A_x  + A_x^\top P_{j,k}(x) A_x   - A_x^\top \diag( A_{x,*,k} )  P_j(x) A_x \notag \\
    = & ~ - 2 (A_x^\top) \diag( A_{x,*,k} ) P_j(x) A_x  + A_x^\top P_{j,k}(x) A_x.
\end{align}
By combining Eq.~\eqref{eq:hessian_H_kj}, Eq.~\eqref{eq:hessian_H_kj_1}, and Eq.~\eqref{eq:hessian_H_kj_2}, we have
\begin{align*}
\frac{\d }{\d x_k} ( \frac{\d H(x)}{\d x_j} ) 
 = & ~ + 2 A_x^\top \diag( A_{x,*,k} ) \Sigma(A_x) \diag(A_{x,*,j}) A_x \\
 & ~ - 2A_x^\top P_k(x) \diag( A_{x,*,j} ) A_x \\
 & ~ + 2 A_x^\top \Sigma(A_x) \diag (A_{x,*,j} ) \diag( A_{x,*,k}) A_x \\
 & ~ + 2 A_x^\top \Sigma(A_x) \diag (A_{x,*,j} ) \diag( A_{x,*,k}) A_x \\
 & ~ - 2 A_x^\top \diag(A_{x,*,k}) P_j(x) A_x \\
 & ~ + A_x^\top P_{j,k}(x) A_x \\
 = & ~ + 6 A_x^\top \diag(A_{x,*,k}) \Sigma(A_x) \diag(A_{x,*,j}) A_x\\
 & ~ - 2 A_x^\top P_k(x) \diag( A_{x,*,j} ) A_x \\
 & ~ - 2 A_x^\top P_j(x) \diag( A_{x,*,k} ) A_x \\
 & ~ + A_x^\top P_{j,k}(x) A_x.
\end{align*}
Thus, we complete the proof.
\end{proof}

\subsection{Hessian of \texorpdfstring{$H(x)$}{}}

\begin{lemma}
Recall the definitions of following quantities:
\begin{itemize}
    \item Let $H(x) := A_x^\top \Sigma( A_x ) A_x \in \R^{d \times d}$ be as definition~\ref{def:H}.
    \item Let $Q(x) := \sigma_{*,*}(A_x) \in \R^{n \times n}$ be defined as Definition~\ref{def:Q}.
\end{itemize} 
Then, we have
\begin{itemize}
    \item {\bf Part 1.} For each $j \in [d]$ and for each $k \in [d]$
    \begin{align*}
        \frac{\d^2 H(x)}{\d x_j  x_k} = C_1+C_2+ C_3 + C_4 + C_5
    \end{align*}
    where we define some local $d \times d$ size matrix variables
    \begin{itemize}
        \item $C_1 =  + 20 A_x^\top \diag(A_{x,*,j}) \Sigma(A_x) \diag(A_{x,*,k}) A_x$ 
        \item $C_2 = - 6 A_x^\top \diag( Q^{\circ 2} (x) ( A_{x,*,k} \circ A_{x,*,j} ) ) A_x$
        \item $C_3 = - 8 A_x^\top \diag(A_{x,*,k}) \diag(  Q^{\circ 2}(x) A_{x,*,j} ) A_x $
        \item $C_4 =  - 8 A_x^\top \diag(A_{x,*,j}) \diag( Q^{\circ 2}(x) A_{x,*,k} )  A_x$
        \item $C_5 =  + 8 A_x^\top ( ( Q(x) \diag(A_{x,*,k}) Q(x) \diag( A_{x,*,j} ) Q(x)  ) \circ I_n ) A_x$
    \end{itemize}
\end{itemize}
\end{lemma}
\begin{proof}

By {\bf Part 2} of Lemma~\ref{lem:gradient_H}, we can show that
\begin{align*}
    \frac{\d^2 H(x)}{\d x_j  x_k} 
    = & ~ B_1 + B_2 + B_3 + B_4
\end{align*}
where
\begin{align*}
B_1 = & ~ + 6 A_x^\top \diag(A_{x,*,k}) \Sigma(A_x) \diag(A_{x,*,j}) A_x\\
B_2 = & ~  -2 A_x^\top P_k(x) \diag( A_{x,*,j} ) A_x \\
B_3 = & ~ - 2 A_x^\top P_j(x) \diag( A_{x,*,k} ) A_x \\
B_4 = & ~ + A_x^\top P_{j,k}(x) A_x
\end{align*}
where $B_2$ can be decomposed further as
\begin{align*}
    B_2 
    = & ~ -2 A_x^\top 2 \diag( ( Q^{\circ 2}(x)  - \Sigma(A_x)  ) A_{x,*,k} ) \diag(A_{x,*,j}) A_x \\
    = & ~ -4 A_x^\top \diag( Q^{\circ 2}(x) A_{x,*,k}  - \Sigma(A_x) A_{x,*,k} ) \diag(A_{x,*,j}) A_x \\
    = & ~ -4 A_x^\top (\diag( Q^{\circ 2}(x) A_{x,*,k})  - \diag(\Sigma(A_x) A_{x,*,k} )) \diag(A_{x,*,j}) A_x \\
    = & ~ -4 A_x^\top \diag( Q^{\circ 2}(x) A_{x,*,k}) \diag(A_{x,*,j}) A_x  + 4 A_x^\top  \diag(\Sigma(A_x) A_{x,*,k} ) \diag(A_{x,*,j}) A_x\\
    = & ~ -4 A_x^\top \diag( Q^{\circ 2}(x) A_{x,*,k}) \diag(A_{x,*,j}) A_x  + 4 A_x^\top \diag(A_{x,*,k} ) \Sigma(A_x)  \diag(A_{x,*,j}) A_x,
\end{align*}
where the second step follows from simple algebra, the third step follows from the definition of $\diag(\cdot)$, the fourth step follows from simple algebra, and the last step follows from the definition of $\diag(\cdot)$. For $B_3$, consider the following:
\begin{align*}
    B_3 
    = & ~ -2 A_x^\top \cdot 2 \diag( ( Q^{\circ 2} (x) - \Sigma(A_x)  ) A_{x,*,j} ) \diag(A_{x,*,k}) A_x \\
    = & ~ -4 A_x^\top  \diag(  Q^{\circ 2} (x) A_{x,*,j} - \Sigma(A_x)   A_{x,*,j} ) \diag(A_{x,*,k}) A_x \\
    = & ~ -4 A_x^\top  (\diag(  Q^{\circ 2} (x) A_{x,*,j}) - \diag(\Sigma(A_x)   A_{x,*,j} )) \diag(A_{x,*,k}) A_x \\
    = & ~ -4 A_x^\top \diag(  Q^{\circ 2} (x) A_{x,*,j}) \diag(A_{x,*,k}) A_x +  4 A_x^\top \diag(\Sigma(A_x)   A_{x,*,j} ) \diag(A_{x,*,k}) A_x\\
    = & ~ -4 A_x^\top \diag(  Q^{\circ 2} (x) A_{x,*,j}) \diag(A_{x,*,k}) A_x +  4 A_x^\top \diag( A_{x,*,j} ) \Sigma(A_x)  \diag(A_{x,*,k}) A_x,
\end{align*}
where the second step follows from simple algebra, the third step follows from the definition of $\diag(\cdot)$, the fourth step follows from simple algebra, and the last step follows from the definition of $\diag(\cdot)$.

Finally for $B_4$,
\begin{align*}
B_4 = & ~ A_x^\top ( \\
        & ~ + 8 ( Q(x) \diag(A_{x,*,k}) Q(x) \diag( A_{x,*,j} ) Q(x)  ) \circ I_n \\
         & ~ - 6 \diag ( Q^{\circ 2}(x) \cdot ( A_{x,*,k} \circ A_{x,*,j} ) ) \\
         & ~ - 4 \diag( Q^{\circ 2}(x) \cdot A_{x,*,j}  ) \diag(A_{x,*,k}) \\
         & ~ - 4 \diag( Q^{\circ 2}(x) \cdot A_{x,*,k}  ) \diag(A_{x,*,j}) \\
         & ~ + 6 \Sigma(A_x) \diag( A_{x,*,j} ) \diag(A_{x,*,k})\\
         & ~ ) A_x
\end{align*}

Eventually, we can show Hessian is $C_1+ \cdots + C_5$. The reason is following: 

For the term $C_1$, we have
\begin{align*}
C_1 = & ~ B_1 + B_{2,2} + B_{3,2} + B_{4,5} \\
= & ~ (6 + 4 + 4 + 6) \cdot A_x^\top \diag(A_{x,*,j}) \Sigma(A_x) \diag(A_{x,*,i}) \\
= & ~ 20 \cdot A_x^\top \diag(A_{x,*,j}) \Sigma(A_x) \diag(A_{x,*,i}) 
\end{align*}
where $B_{2,2}$ is the second term of $B_2$, $B_{3,2}$ is the second term of $B_3$ and $B_{4,5}$ is the last term of $B_4$.

For the term $C_2$, we have
\begin{align*}
    C_2 = B_{4,2}
\end{align*}

For the term $C_3$, we have
\begin{align*}
    C_3 = & ~ B_{3,1} + B_{4,3} \\
    = & ~ -(4 + 4) \cdot A_x^\top \diag(A_{x,*,k}) \diag(  Q^{\circ 2}(x) A_{x,*,j} ) A_x \\
    = & ~ - 8 \cdot A_x^\top \diag(A_{x,*,k}) \diag(  Q^{\circ 2}(x) A_{x,*,j} ) A_x
\end{align*}

For the term $C_4$, we have
\begin{align*}
    C_4 
    = & ~ B_{2,1} + B_{4,4}\\
    = & ~ -4 A_x^\top \diag( Q^{\circ 2}(x) A_{x,*,k}) \diag(A_{x,*,j}) A_x - 4 A_x^\top \diag( Q^{\circ 2}(x) \cdot A_{x,*,k}  ) \diag(A_{x,*,j})A_x\\
    = & ~ -8 A_x^\top \diag( Q^{\circ 2}(x) A_{x,*,k}) \diag(A_{x,*,j}) A_x.
\end{align*}

For the term $C_5$, we have
\begin{align*}
    C_5 = & ~ B_{4,1}. \qedhere
\end{align*}
\end{proof}
\subsection{Gradient and Hessian of \texorpdfstring{$F(x)$}{} }

\begin{lemma}
Let $F(x) \in \R$ be defined as Definition~\ref{def:F_log}. Then, we have
\begin{itemize}
\item {\bf Part 1.}
    \begin{align*}
        \frac{\d F(x)}{\d x_j} = \tr[ (H(x) + I_d)^{-1} \cdot \frac{\d H(x)}{\d x_j}  ]
    \end{align*}
    \item {\bf Part 2}
    \begin{align*}
       \frac{\d}{\d x_k} \frac{\d F(x)}{\d x_j} = 
       & ~ + \tr[ (H(x) + I_d)^{-1} \frac{\d^2 H}{\d x_j \d x_k} ] \\
       & ~ - \tr[ (H(x) + I_d)^{-1} \frac{\d H(x)}{\d x_k} (H(x) + I_d)^{-1}\cdot \frac{\d H(x)}{\d x_j}  ] \\
    \end{align*}
\end{itemize}
\end{lemma}
\begin{proof}

{\bf Proof of Part 1.} We have
\begin{align*}
    \frac{\d F(x)}{\d x_j} 
    = & ~ \frac{\d \log( \det(H(x)  +I_d) )}{\d x_j} \\
    = & ~ \tr[ (H(x)  +I_d)^{-1} \frac{\d (H(x)  +I_d)}{\d x_j} ]\\
    = & ~ \tr[ (H(x)  +I+d)^{-1} (\frac{\d H(x) }{\d x_j}  + \frac{\d I_d}{\d x_j}) ]\\ 
    = & ~ \tr[ (H(x) + I_d)^{-1} \cdot \frac{\d H(x)}{\d x_j}  ],
\end{align*}
where the first step follows from the definition of $F(x)$ (see Definition~\ref{def:F_log}), 
the third step follows from the basic calculus rule, and the last step follows from $\frac{\d I_d}{\d x_j} = 0$.

{\bf Proof of Part 2.} We have 
\begin{align*}
       & ~ \frac{\d}{\d x_k} \frac{\d F(x)}{\d x_j} \\
       = & ~ \frac{\d}{\d x_k} (\tr[ (H(x) + I_d)^{-1} \cdot \frac{\d H(x)}{\d x_j}  ]) \\
       = & ~ \tr[ \frac{\d}{\d x_k} ((H(x) + I_d)^{-1}) \cdot \frac{\d H(x)}{\d x_j}  ] + \tr[ (H(x) + I_d)^{-1} \cdot \frac{\d}{\d x_k} (\frac{\d H(x)}{\d x_j} ) ] \\
       = & ~ \tr[ -(H(x) + I_d)^{-1} \frac{\d}{\d x_k}(H(x)) (H(x) + I_d)^{-1}  \cdot \frac{\d H(x)}{\d x_j}  ] + \tr[ (H(x) + I_d)^{-1} \cdot \frac{\d}{\d x_k} (\frac{\d H(x)}{\d x_j} ) ] \\
       = & ~ + \tr[ (H(x) + I_d)^{-1} \frac{\d^2 H}{\d x_j \d x_k} ] - \tr[ (H(x) + I_d)^{-1} \frac{\d H(x)}{\d x_k} (H(x) + I_d)^{-1}\cdot \frac{\d H(x)}{\d x_j}  ],
    \end{align*}
    where the first step follows from {\bf Part 1}, the second step follows from the product rule, and the last step follows from simple algebra.
\end{proof}

Using standard algebraic tools in literature \citep{ls19}, we can show that,
\begin{lemma}
We can rewrite it as follows 
\begin{align*}
     \frac{\d }{\d x} \frac{\d F}{\d x} = & ~ 20 H_1 - 6 H_2 - 8 H_3 - 8 H_4 + 8 H_5 \\
     & ~ -4 G_1 + 8 G_2 + 8 G_3 - 16G_4
\end{align*}
and 
\begin{align*}
    \frac{\d}{\d x} \frac{\d F}{\d x} \succ & ~ 0.
\end{align*}
Thus, the function $F$ is convex in $x$.
\end{lemma}

\subsection{Quantity I for Hessian of \texorpdfstring{$F$}{}}

\begin{lemma}
We have
\begin{align*}
    [\frac{\d }{\d x } \frac{\d F}{\d x^\top} ]_1 = 20 H_1 - 6 H_2 - 8 H_3 - 8 H_4 + 8 H_5
\end{align*}
where
\begin{itemize}
    \item $H_1 = A_x^\top \wt{\Sigma}(A_x) \Sigma(A_x) A_x$ 
    \item $H_2 = A_x^\top \diag( Q^{\circ 2}(x) \wt{\Sigma}(A_x) {\bf 1}_n ) A_x$
    \item $H_3 = A_x^\top Q^{\circ 2}(x) \wt{\Sigma}(A_x) A_x$
    \item $H_4 = A_x^\top  \wt{\Sigma}(A_x) Q^{\circ 2}(x) A_x$
    \item $H_5 = A_x^\top ( Q(x) \wt{\Sigma}(A_x) Q(x) ) \circ Q(x) ) A_x$
\end{itemize}
\end{lemma}
\begin{proof}
It comes from $P_{i,j}$ and the following lemma.
\end{proof}

\begin{lemma}\label{lem:H_1}
Let $H_1 \in \R^{d \times d}$ be Hessian where each entry is
\begin{align*}
    (H_1)_{j,k} = \tr[ (H(x) + I_d)^{-1} A_x^\top \diag(A_{x,*,j}) \Sigma(A_x) \diag(A_{x,*,k}) A_x  ]
\end{align*}
Then, we have
\begin{align*}
    H_1 = A_x^\top \wt{\Sigma}(A_x) \Sigma(A_x) A_x
\end{align*}
\end{lemma}
\begin{proof}
We can rewrite $(H_1)_{j,k}$ as follows
\begin{align*}
    (H_1)_{j,k} = & ~ \tr[ (H(x) + I_d)^{-1} A_x^\top \diag(A_{x,*,j}) \Sigma(A_x) \diag(A_{x,*,k}) A_x  ] \\
    = & ~ \tr[ A_x (H(x) + I_d)^{-1} A_x^\top \diag(A_{x,*,j}) \Sigma(A_x) \diag(A_{x,*,k}) ] \\
    = & ~ \tr[ \wt{\sigma}_{*,*} (A_x) \diag(A_{x,*,j}) \Sigma(A_x) \diag(A_{x,*,k}) ] \\
    = & ~ A_{x,*,j}^\top  \wt{\Sigma}(A_x) \Sigma(A_x)  A_{x,*,k},
\end{align*}
where the first step follows from the Lemma statement, the second step follows from the fact that all of $(H(x) + I_d)^{-1}$, $\diag(A_{x,*,j})$, $\Sigma(A_x)$, and $\diag(A_{x,*,k})$ are diagonal matrices, the third step follows from the definition of $\wt{\sigma}_{*,*}$ (see Definition~\ref{def:wt_Sigma}), and the last step follows from the definition of $\wt{\Sigma}(A_x)$ (see Definition~\ref{def:wt_Sigma}).

Thus, we have
\begin{align*}
    H_1 = & ~ A_x^\top \wt{\Sigma}(A_x) \Sigma(A_x) A_x. \qedhere
\end{align*}
\end{proof}

\begin{lemma}\label{lem:H_2}
Let $H_2$ be defined as
\begin{align*}
    (H_2)_{j,k} = \tr[ (H(x) + I_d)^{-1} A_x^\top \diag( Q^{\circ 2} (x) ( A_{x,*,k} \circ A_{x,*,j} ) ) A_x ]
\end{align*}
Then, we have 
\begin{align*}
    H_2 = A_x^\top \diag( Q^{\circ 2}(x) \wt{\Sigma}(A_x) {\bf 1}_n ) A_x
\end{align*}
\end{lemma}
\begin{proof}
We have
 \begin{align*}
    (H_2)_{j,k} = & ~ \tr[ (H(x) + I_d)^{-1} A_x^\top \diag( Q^{\circ 2} (x) ( A_{x,*,k} \circ A_{x,*,j} ) ) A_x ] \\
    = & ~ \tr[ \wt{\Sigma}(A_x) \diag( Q^{\circ 2} (x) ( A_{x,*,k} \circ A_{x,*,j} ) ) ] \\
    = & ~ {\bf 1}_n^\top \wt{\Sigma}(A_x) Q^{\circ 2}(x) ( A_{x,*,k} \circ A_{x,*,j} ) \\
     = & ~ A_{x,*,k}^\top \diag( Q^{\circ 2}(x) \wt{\Sigma}(A_x) {\bf 1}_n )  A_{x,*j}  
\end{align*}
where the first step follows from the Lemma statement, the second step follows from the definition of $\wt{\Sigma}(A_x)$ (see Definition~\ref{def:wt_Sigma}), the third step follows from the definition of ${\bf 1}_n^\top$,  and the last step follows from the definition of $\Sigma(A_x)$.

Thus, we have
\begin{align*}
    H_2 = & ~ A_x^\top \diag( Q^{\circ 2}(x) \wt{\Sigma}(A_x) {\bf 1}_n ) A_x. \qedhere
\end{align*}
\end{proof}

\begin{lemma}\label{lem:H_3}
 Let $H_3$ be defined as
 \begin{align*}
    (H_{3})_{j,k} = \tr[ (H(x) + I_d)^{-1} A_x^\top \diag( A_{x,*,k}) \diag( Q^{\circ 2} (x) (   A_{x,*,j} ) ) A_x ]
 \end{align*}
 Then, we 
 \begin{align*}
     H_3 = A_{x}^\top Q^{\circ 2}(x) \wt{\Sigma}(A_x)  A_x
 \end{align*}
\end{lemma}
\begin{proof}
Then, we have
\begin{align*}
     (H_{3})_{j,k} = & ~ \tr[ (H(x) + I_d)^{-1} A_x^\top \diag( A_{x,*,k}) \diag( Q^{\circ 2} (x) (   A_{x,*,j} ) ) A_x ] \\
     = & ~ \tr[ \wt{\Sigma}(A_x)  \diag( A_{x,*,k}) \diag( Q^{\circ 2} (x) (   A_{x,*,j} ) ) ] \\
     = & ~ A_{x,*,k}^\top \wt{\Sigma}(A_x) Q^{\circ 2} (x)    A_{x,*,j},
\end{align*}
where the first step follows from the Lemma statement, the second step follows from the definition of $\wt{\Sigma}(A_x)$ (see Definition~\ref{def:wt_Sigma}), and the last step follows from the property of $\diag( \cdot )$.

Thus, we have
\begin{align*}
    H_3 = & ~ A_{x}^\top Q^{\circ 2}(x) \wt{\Sigma}(A_x)   A_x. \qedhere
\end{align*}
\end{proof}

\begin{lemma}\label{lem:H_4}
 Let $H_4$ be defined as
 \begin{align*}
    (H_{4})_{j,k} = \tr[ (H(x) + I_d)^{-1} A_x^\top \diag( A_{x,*,j}) \diag( Q^{\circ 2} (x) (   A_{x,*,k} ) ) A_x ]
 \end{align*}
 Then, we 
 \begin{align*}
     H_4 = A_{x}^\top \wt{\Sigma}(A_x) Q^{\circ 2}(x) A_x
 \end{align*}
\end{lemma}
\begin{proof}
This proof is very similar to Lemma~\ref{lem:H_3}, so we omit the details here.
\end{proof}

\begin{lemma}\label{lem:H_5}
Let $H_5$ be defined as
\begin{align*}
    (H_5)_{j,k} = \tr[ (H(x)+I_d)^{-1} A_x^\top ( Q(x) \diag(A_{x,*,k}) Q(x) \diag( A_{x,*,j} ) Q(x)  ) \circ I_n ) A_x ]
\end{align*}
Then, we have
\begin{align*}
    H_5 = A_x^\top ( Q(x) \wt{\Sigma}(A_x) Q(x) ) \circ Q(x) ) A_x
\end{align*}
\end{lemma}
\begin{proof}
We have
\begin{align*}
    (H_5)_{j,k} = & ~ \tr[ (H(x)+I_d)^{-1} A_x^\top ( Q(x) \diag(A_{x,*,k}) Q(x) \diag( A_{x,*,j} ) Q(x)  ) \circ I_n ) A_x ] \\
    = & ~ \tr[ \wt{\Sigma}(A_x) ( Q(x) \diag(A_{x,*,k}) Q(x) \diag( A_{x,*,j} ) Q(x)  ) \circ I_n  ] \\
    = & ~ \sum_{l=1}^n \wt{\Sigma}_{l,l}(A_x) Q_{*,l}(x)^\top \diag(A_{x,*,k}) Q(x) \diag( A_{x,*,j})  Q_{*,l}(x) \\
    = & ~ A_{x,*,k}^\top \sum_{l=1}^n \wt{\Sigma}_{l,l}(A_x) \diag ( Q_{*,l}(x) ) Q(x) \diag ( Q_{*,l}(x) ) A_{x,*,j} \\
    = & ~ A_{x,*,k}^\top ( \sum_{l=1}^n \wt{\Sigma}_{l,l}(A_x)  ( Q_{*,l}(x) Q_{*,l}(x)^\top \circ Q(x) ) ) A_{x,*,j} \\
    = & ~ A_{x,*,k}^\top ( ( Q(x) \wt{\Sigma}(A_x) Q(x) ) \circ Q(x) ) A_{x,*,j},
\end{align*}
where the first step follows from the Lemma statement, the second step follows from the definition of $\wt{\Sigma}(A_x)$ (see Definition~\ref{def:wt_Sigma}), the third step follows from the definition of $\tr[\cdot]$, and the last step follows from the fact that $\wt{\Sigma}(A_x)$ is a diagonal matrix (see Definition~\ref{def:wt_Sigma}).

Thus, we have
\begin{align*}
    H_5 = & ~ A_x^\top ( Q(x) \wt{\Sigma}(A_x) Q(x) ) \circ Q(x) ) A_x. \qedhere
\end{align*}
\end{proof}

\subsection{Quantity II for Hessian of \texorpdfstring{$F$}{}}

Recall that
\begin{align*}
\frac{\d H}{\d x_j} = & ~ - 2 A_x^\top \diag( \Sigma(A_x) A_{x,*,j} ) A_x \\
& ~ +  A_x^\top \frac{\d \Sigma(A_x)}{\d x_j} A_x \\
= & ~ - 2 A_x^\top \diag( \Sigma(A_x) A_{x,*,j} ) A_x \\
& ~ + 2 A_x^\top \diag ( ( Q^{\circ 2}(x) - \Sigma(A_x) ) A_{x,*,j} ) A_x \\
= & ~ A_x^\top \diag ( 2 Q^{\circ 2}(x) - 4 \Sigma(A_x) ) A_{x,*,j} ) A_x 
\end{align*}

\begin{lemma}
For the second item, we have
\begin{align*}
    [\frac{\d }{\d x} \frac{\d F}{\d x} ]_2 = -4 G_1 + 8 G_2 + 8 G_3 - 16G_4
\end{align*}
\begin{itemize}
    \item $G_1=  A_x^\top Q^{\circ 2}(x) \wt{\Sigma}(A_x)^2 Q^{\circ 2}(x) A_x$
    \item $G_2 = A_x^\top Q^{\circ 2}(x) \wt{\Sigma}(A_x)^2 \Sigma(A_x) A_x$
    \item  $G_2 = A_x^\top \Sigma(A_x) \wt{\Sigma}(A_x)^2  Q^{\circ 2}(x)  A_x$
    \item  $G_4 = A_x^\top \Sigma(A_x) \wt{\Sigma}(A_x)^2 \Sigma(A_x) A_x$
\end{itemize} 
\end{lemma}
\begin{proof}
The proof follows by combining the following lemmas.
\end{proof}

\begin{lemma}
Let $(G_1)_{j,k}$
\begin{align*}
    (G_1)_{j,k} = & ~ \tr[ (H+I_d)^{-1}  A_x^\top \diag( Q^{\circ 2}(x) A_{x,*,j} )  A_x (H+I_d)^{-1} A_x^\top \diag( Q^{\circ 2}(x) A_{x,*,k} ) A_x ] 
\end{align*}
Then, we have 
\begin{align*}
    G_1 = A_x^\top Q^{\circ 2}(x) \wt{\Sigma}(A_x)^2 Q^{\circ 2}(x) A_x
\end{align*}
\end{lemma}
\begin{proof}
We can show
\begin{align*}
    (G_1)_{j,k} = & ~ \tr[ (H+I_d)^{-1}  A_x^\top \diag( Q^{\circ 2}(x) A_{x,*,j} )  A_x (H+I_d)^{-1} A_x^\top \diag( Q^{\circ 2}(x) A_{x,*,k} ) A_x ] \\
    = & ~ \tr[ \wt{\Sigma}(A_x) \diag(Q^{\circ 2} A_{x,*,j}) A_x (H+I_d)^{-1} A_x^\top \diag(Q^{\circ 2}(x) A_{x,*,k}) ] \\
    = & ~ \tr[ \wt{\Sigma}(A_x) \diag(Q^{\circ 2} A_{x,*,j}) \wt{\Sigma}(A_x) \diag(Q^{\circ 2}(x) A_{x,*,k}) ] \\
    = & ~ A_{x,*,j}^\top Q^{\circ 2}(x) \wt{\Sigma}(A_x)^2 Q^{\circ 2}(x) A_{x,*,k}
\end{align*}
Thus, we have
\begin{align*}
    G_1 = & ~ A_x^\top Q^{\circ 2}(x) \wt{\Sigma}(A_x)^2 Q^{\circ 2}(x) A_x. \qedhere
\end{align*}  
\end{proof}

The proofs of $G_2,G_3,G_4$ are similar to $G_1$. So we omit the details here.
\begin{lemma}
Let $(G_2)_{j,k}$
\begin{align*}
    (G_2)_{j,k} = & ~ \tr[ (H+I_d)^{-1}  A_x^\top \diag( Q^{\circ 2}(x) A_{x,*,j} )  A_x (H+I_d)^{-1} A_x^\top \diag( \Sigma(A_x) A_{x,*,k} ) A_x ] 
\end{align*}
Then, we have 
\begin{align*}
    G_2 = A_x^\top Q^{\circ 2}(x) \wt{\Sigma}(A_x)^2 \Sigma(A_x) A_x.
\end{align*}
\end{lemma}

\begin{lemma}
Let $(G_3)_{j,k}$
\begin{align*}
    (G_3)_{j,k} = & ~ \tr[ (H+I_d)^{-1}  A_x^\top \diag( \Sigma(A_x) A_{x,*,j} )  A_x (H+I_d)^{-1} A_x^\top \diag( Q^{\circ 2}(x) A_{x,*,k} ) A_x ] 
\end{align*}
Then, we have 
\begin{align*}
    G_3 = A_x^\top \Sigma(A_x) \wt{\Sigma}(A_x)^2 Q^{\circ 2}(x) A_x.
\end{align*}
\end{lemma}

\begin{lemma}
Let $(G_4)_{j,k}$
\begin{align*}
    (G_4)_{j,k} = & ~ \tr[ (H+I_d)^{-1}  A_x^\top \diag( \Sigma(A_x) A_{x,*,j} )  A_x (H+I_d)^{-1} A_x^\top \diag(  \Sigma(A_x) A_{x,*,k} ) A_x ] 
\end{align*}
Then, we have 
\begin{align*}
    G_4 = A_x^\top \Sigma(A_x) \wt{\Sigma}(A_x)^2 \Sigma( A_x) A_x.
\end{align*}
\end{lemma}

\ifdefined\isarxiv
\else
\newpage
\input{checklist}
\fi



\end{document}